\documentclass{aa}
\usepackage{upgreek}
\usepackage{graphicx}
\usepackage[varg]{txfonts}
\usepackage{natbib}
\usepackage{float}
\usepackage{soul}
\newlength{\linwx}
\usepackage{color}

\begin{document}

\title{The eccentricity distribution of giant planets and their relation to super-Earths in the pebble accretion scenario}

%Eccentricity of giant planets and the super-Earth cold Jupiter relation in the pebble accretion scenario

%Growth of giant planets from a small ring of planetary embryos via pebble accretion I: rings at ice lines}
%  \subtitle{empty}
%%
\author{
Bertram Bitsch \inst{1}, Trifon Trifonov \inst{1}, Andre Izidoro \inst{2,3}
}
\offprints{B. Bitsch,\\ \email{bitsch@mpia.de}}
\institute{
Max-Planck-Institut f\"ur Astronomie, K\"onigstuhl 17, 69117 Heidelberg, Germany
\and
Department of Earth, Environmental and Planetary Sciences, MS 126, Rice
 University, Houston, TX 77005, USA
 \and
 UNESP, Univ. Estadual Paulista - Grupo de Din\`amica Orbital Planetologia, Guaratinguet\`a, CEP 12.516-410, S\~ao Paulo, Brazil
}
\abstract{Observations of the population of cold Jupiter planets ($r>$1 AU) show that nearly all of these planets orbit their host star on eccentric orbits. For planets up to a few Jupiter masses, eccentric orbits are thought to be the outcome of planet-planet scattering events taking place after gas dispersal. We simulated the growth of planets via pebble and gas accretion as well as the migration of multiple planetary embryos in their gas disc. We then followed the long-term dynamical evolution of our formed planetary system up to 100 Myr after gas disc dispersal. We investigated the importance of the initial number of protoplanetary embryos and different damping rates of eccentricity and inclination during the gas phase for the final configuration of our planetary systems. We constrained our model by comparing the final dynamical structure of our simulated planetary systems to that of observed exoplanet systems. Our results show that the initial number of planetary embryos has only a minor impact on the final orbital eccentricity distribution of the giant planets, as long as the damping of eccentricity and inclination is efficient. If the damping is inefficient (slow), systems with a larger initial number of embryos harbor larger average eccentricities. In addition, for slow damping rates, we observe that scattering events are already common during the gas disc phase and that the giant planets that formed in these simulations match the observed giant planet eccentricity distribution best. These simulations also show that massive giant planets (above Jupiter mass) on eccentric orbits are less likely to host inner super-Earths as they get lost during the scattering phase, while systems with less massive giant planets on nearly circular orbits should harbor systems of inner super-Earths. Finally, our simulations predict that giant planets are not single, on average, but they live in multi-planet systems.
}
\authorrunning{Bitsch et al.}\titlerunning{Eccentric giant planets and their super-Earths}\maketitle

\section{Introduction}
\label{sec:Introduction}

Since the discovery of the first close-in giant planet around a main sequence star \citep{1995Natur.378..355M}, many more exoplanets have been detected. These detected planets come in the following different categories: (i) terrestrial planets similar in mass to those as in our inner Solar System, (ii) close-in super-Earths and mini-Neptunes with masses of several Earth masses, and (iii) short period Jovian-mass planets  (hot Jupiters, r<0.1 au), or (iv) more distant giant planets (warm ($r<$1.0 AU) and cold Jupiters ($r>$1.0 au).

Based on the observational evidence, these different types of planets also show different occurrence rates. For instance, close-in super-Earths seem to be most common, with an occurrence rate of 30-50\%, in inner systems \citep{2011arXiv1109.2497M, 2013ApJ...766...81F, 2018AJ....156...24M}, while cold Jupiters seem to exist around 10-20\% of solar-type stars \citep{J2010}. Recent observations have argued for a correlation between inner super-Earths and cold gas giants, where up to 90\% of cold gas giants should have inner super-Earth systems \citep{2018arXiv180608799B, 2018arXiv180502660Z}. However, this trend is under debate \citep{2018arXiv180408329B}. If such a correlation exists, these inner super-Earths should have formed interior to the orbit of the gas giants \citep{2015ApJ...800L..22I}.

The observed giant planets seem to show significant eccentricity (e.g. \citealt{2013ApJ...767L..24D, 2018arXiv180206794B}), although the planet eccentricity distribution might suffer from observational biases. Recent observations \citep{2010ApJ...709..168A, 2013ApJS..208....2W, 2015A&A...577A.103K, 2018MNRAS.480.2846B, 2019MNRAS.489..738H, 2019MNRAS.484.4230W} have claimed that two giant planets in circular orbits could mimic the radial velocity (RV) signature of a single eccentric giant planet if not enough RV measurements have been used to determine the true orbital configuration of the system. In addition, simulations have shown that the eccentricities of multi-planet systems can oscillate in time due to secular perturbations, with quite large variations at any given moment in time (e.g. \citealt{2009ApJ...699L..88R, 2014A&A...568A..64T, 2017A&A...598A..70S, 2019A&A...631A.136L})

The observed high eccentricities of these giant planets can be explained by the following two mechanisms: (i) via gravitational planet-disc interactions \citep{2001A&A...366..263P, 2006A&A...447..369K, 2013A&A...555A.124B} or (ii) via planet-planet scattering events after gas disc dispersal \citep{2008ApJ...686..603J, 2009ApJ...699L..88R, 2017A&A...598A..70S}. The first mechanism is related to the fact that very massive planets (above 5 Jupiter masses) open very deep gaps that deplete the Lindblad resonances responsible for eccentricity damping, resulting in an increase in the orbital eccentricity, which can even happen for single giant planets. The second mechanism is based on gravitational interactions between multiple planets.

Recent studies have found that eccentric giant planets are more frequently observed around metal rich stars \citep{2013ApJ...767L..24D, 2018arXiv180206794B}. In the core accretion paradigm for planet formation \citep{1996Icar..124...62P}, this can be explained by the larger amount of building blocks available, which enhances the formation of pebbles, planetesimals \citep{2010ApJ...722L.220B, 2017A&A...606A..80Y} and thus planetary cores via the accretion of planetesimals \citep{1996Icar..124...62P} or pebbles \citep{2010MNRAS.404..475J, 2010A&A...520A..43O, 2012A&A...544A..32L}. As more planets are formed, mutual gravitational interactions among multiple giant planets are more likely to lead to dynamical instabilities and scattering events that pump planetary eccentricities to observed levels.

Planet-planet scattering simulations dedicated to explain the observed giant planet eccentricity distribution have typically invoked fully formed giant planets from the beginning. Traditionally, these planets are artificially put close to each other to favour a prompt onset of the dynamical instability once the simulation starts \citep{2008ApJ...686..603J, 2009ApJ...699L..88R} or have only type-II migration of already fully formed giant planet \citep{2017A&A...598A..70S}. These simulations are quite successful in reproducing the giant planet eccentricity distribution. In addition, all these simulations required that at least three planets are needed to trigger scattering events, if no other sources to disrupt the system are present.

In this work, we aim to reproduce the observed giant planet eccentricity distribution by including the initial step of modelling the growth and migration from planetary embryos (of around Moon mass) to full planetary systems. We follow the planetary evolution model of \citet{2019arXiv190208772I} and \citet{2019A&A...623A..88B}. This model includes growth by pebble \citep{2014A&A...572A.107L} and gas accretion \citep{2014ApJ...786...21P, 2010MNRAS.405.1227M}, migration in the type-I \citep{2011MNRAS.410..293P} and type-II \citep{2018arXiv180511101K} regime as well as an evolving disc model \citep{2015A&A...575A..28B}. We also vary the initial number of planetary embryos that can grow and migrate to study their influence on the final planetary systems. In addition, we investigate the influence of the damping rates in the type-II regime as free parameter, motivated by the fact that damping rates have been derived in hydrodynamical simulations only for single planets, making it unclear what the exact damping rates are for multiple gap opening planets. Finally, we test two different scenarios for the growth of the giant planets. In the first one, giants planets are only allowed to grow up to one Jupiter-mass. In the second set, gas accretion onto gas giants is only limited by the gas disc flow and it only shuts down at the time of gas disc dispersal, allowing the growth of planets of up to several Jupiter masses.

The code used in this work was already introduced in our previous works, where we studied the formation of super-Earth systems \citep{2019arXiv190208772I} and gas giants \citep{2019A&A...623A..88B}. However, in these previous works we did not touch upon the eccentricity distribution of the giant planets and the super-Earth - cold Jupiter relation. 

Our work is structured as follows. In Section~\ref{sec:methods}, we summarise the numerical methods used in this work. In Section~\ref{sec:individual}, we show the evolution of individual systems and in Section~\ref{sec:statistics} we investigate the link to the eccentricity distribution of Jupiter mass planets. In Section~\ref{sec:growth}, we investigate how faster growth and thus more massive planets influence the outcomes of our simulations. In Section~\ref{sec:super}, we comment on the implications of our simulations in respect to the super-Earth - cold Jupiter relation found in observations. We then discuss further implications of our results and their shortcomings in Section~\ref{sec:discussion}, while we summarise in Section~\ref{sec:summary}.

%{\bf super earth fraction where no warm jupiters have formed?}

%\begin{itemize}
% \item giant planet detection, occurence rates of giants
% \item eccentricity distribution of planets: observations: metallicity correlation: Buchhave and Dawson paper, bias in eccentricity distribution: Wittenmyer paper, also eccentricity of planets is a momentary measure, e.g. long term oscillations can change the eccentricity of the planets in time.
% \item causes of eccentricity: planet-disc interactions: large planets, driving of eccentricity  (Bitsch 2013, Duffel paper, Kley \& Dirksen paper etc.)
% \item causes of eccentricity: scattering after the gas disc phase: cite Juric \& Tremaine, Sotiradis paper, Raymond work. Say what they did: fully formed planet, just put them in, sometimes with and sometimes without migration
% \item what we want to do in this work: pebble and gas accretion framework: model growth and migration in the disc to see if we can match the eccentricity distribution. As hydro simulations of multiple planets are scarse and we explore different damping rates and what that could mean for the eccentricity distribution of giant exoplanets. We also test different initial number of seeds
% \item structure of papers
%\end{itemize}

\section{Methods}
\label{sec:methods}

In this section we summarise the used methods and state the parameters of the model that are used in this work. The code is described in more detail in our previous work \citep{2019arXiv190208772I,  2019A&A...623A..88B}. In addition, we also discuss the initial conditions of our simulations. Our planet formation model features pebble accretion \citep{2014A&A...572A.107L}, gas accretion \citep{2014ApJ...786...21P, 2010MNRAS.405.1227M}, type-I and type-II migration \citep{2011MNRAS.410..293P, 2018arXiv180511101K} as well as a simple accretion disc model with decaying $\dot{M}$ \citep{2015A&A...575A..28B}. We follow the planetary disc model outlined in \citet{2015A&A...575A..28B} and used in the N-body planet formation simulations of \citet{2019arXiv190208772I} and \citet{2019A&A...623A..88B}. The disc lifetime is in total 5 Myr, and the initial disc age is 2 Myr as in \citet{2019A&A...623A..88B}, meaning that the planets have 3 Myr in a gas disc environment to grow and migrate via interactions with the gas disc. In the last 100kyr we employ an exponential decay of the gas disc surface density to mimic the disc dispersal. We set the inner edge of the gas disc to be at 0.1 AU, in line with recent hydrodynamical models \citep{2019A&A...630A.147F}. Collisions between planetary embryos are treated as perfect mergers between the bodies.

We want to stress here that our simulations are designed to form giant planets, by choosing disc properties and pebble fluxes that allow the efficient growth of giant planets, motivated by our earlier work \citep{2019A&A...623A..88B}. We can thus not make any statements about giant planet occurrence rates.

\subsection{Pebble and gas accretion}

Pebble accretion is modelled following the approach in \citet{Johansen2015}. This pebble accretion approach takes into account the changes of the pebble accretion rates for planetary embryos on eccentric and inclined orbits \citep{Johansen2015}. This prescription reduces the pebble accretion rates for planetary embryos on eccentric and inclined orbits. The accretion rate of the planetary core $\dot{M}_{\rm core}$ is directly proportional to the pebble surface density $\Sigma_{\rm peb}$. 

Pebbles in the protoplanetary disc settle towards the midplane depending on their size, parameterised in this work by the dimensionless Stokes number $\tau_{\rm f}$, and depending on the level of turbulence in the protoplanetary disc described through the disc's viscosity $\alpha_{\rm disc}$. Using the $\alpha$ prescription, \citet{2007Icar..192..588Y} derived the pebble scale height $H_{\rm peb}$ using the gas scale height $H_{\rm g}$ via
\begin{equation}
 H_{\rm peb} = H_{\rm g} \sqrt{\alpha_{\rm disc} / \tau_{\rm f}} \ .
\end{equation}
Here $\alpha_{\rm disc}$ corresponds to the $\alpha$-viscosity value inside the gas disc. Typically the pebbles in our simulations have Stokes numbers of 0.05-0.1, which is calculated by equating the drift time-scale with the growth time-scale \citep{2012A&A...539A.148B, 2012A&A...544A..32L, 2015A&A...582A.112B}. That yields a value of $H_{\rm peb}/H_{\rm g}\sim$0.1, in agreement with observations of protoplanetary discs \citep{2016ApJ...816...25P}. In order to be compareable to our previous works \citep{2019A&A...623A..88B} we use an $\alpha_{\rm disc}$ value of 0.0054 for the pebble scale height\footnote{This $\alpha$ value is also used to derive the disc structure in the disc model of \citet{2015A&A...575A..28B} that we use here.}, which is different compared to the value used for the migration (see below). A lower $\alpha$ value used for the pebble scale height will result in a denser midplane layer of pebbles, with faster pebble accretion rates onto the planets as a consequence \citep{2014A&A...572A.107L, 2019A&A...623A..88B}.

The pebble surface density $\Sigma_{\rm peb} (r_{\rm P})$ at the planets location can be calculated from the pebble flux $\dot{M}_{\rm peb}$ via
\begin{equation}
 \label{eq:SigmaPeb}
 \Sigma_{\rm peb} (r_{\rm P}) = \sqrt{\frac{2 S_{\rm peb} \dot{M}_{\rm peb} \Sigma_{\rm g}(r_{\rm P}) }{\sqrt{3} \pi \epsilon_{\rm P} r_{\rm P} v_{\rm K}}} \ ,
\end{equation}
where $r_{\rm P}$ denotes the semi-major axis of the planet, $v_{\rm K}$ the Keplerian velocity, and $\Sigma_{\rm g} (r_{\rm P})$ stands for the gas surface density at the planets locations. The pebble flux $\dot{M}_{\rm peb}$ is calculated self consistently through an equilibrium between dust growth and drift \citep{2012A&A...539A.148B, 2014A&A...572A.107L, 2018A&A...609C...2B}, where these simulations predict a decrease in the pebble flux in time. Here $S_{\rm peb}$ describes the scaling factor to the pebble flux $\dot{M}_{\rm peb}$ to test the influence of different pebble fluxes, where we use $S_{\rm peb}=$2.5 and $S_{\rm peb}=$5.0, which is similar to our previous works \citep{2019A&A...623A..88B, 2019arXiv190208772I} and as explained in \citep{2018A&A...609C...2B}. Using $S_{\rm peb}$=2.5 a total of 175 Earth masses of pebbles drift through the disc in the 3 Myr of integration, which is doubled for $S_{\rm peb}$=5.0. The pebble sticking efficiency can be taken as $\epsilon_{\rm P} =0.5$ under the assumption of near-perfect sticking \citep{2014A&A...572A.107L}. 

The Stokes number of the pebbles can be related to the pebble surface density $\Sigma_{\rm peb}$ and gas surface density $\Sigma_{\rm g}$ through the following relation
\begin{equation}
 \tau_{\rm f} = \frac{\sqrt{3}}{8} \frac{\epsilon_{\rm P}}{\eta} \frac{\Sigma_{\rm peb}(r_{\rm P})}{\Sigma_{\rm g} (r_{\rm P})} \ .
\end{equation}
Here $\eta$ represents a measurement of the sub-Keplerianity of the gas velocity.

At higher temperatures, water sublimates and we fix the radius of the pebbles to 1 mm for $T>$170K, corresponding to typical chondrule sizes \citep{2015Icar..258..418M, 2016A&A...591A..72I}. This is consistent with the assumptions made in \citet{2015Icar..258..418M} to explain the dichotomy between the terrestrial planets and the gas giants in the solar system. Additionally, we reduce the pebble flux $\dot{M}_{\rm peb}$ to half its nominal value to account for water loss. In our disc model, the water ice line is located at $\approx$1 AU at the beginning of our simulations, but moves even further inwards in time as the disc evolves \citep{2015A&A...575A..28B}.

As a planet grows, it starts to push away material from its orbit, generating a partial gap in the protoplanetary disc, where the planet generates an inversion in the radial pressure gradient of the disc exterior its orbit, halting the inward drift of pebbles \citep{2006A&A...453.1129P, 2012A&A...546A..18M, 2014A&A...572A..35L, 2018arXiv180102341B, 2018A&A...615A.110A}. This planetary mass is referred to as the pebble isolation mass. The pebble isolation mass in itself is a function of the local properties of the protoplanetary discs, namely the disc's viscosity $\nu$, aspect ratio $H/r$ and radial pressure gradient $\partial \ln P/\partial\ln r$ as well as of the Stokes number of the particles, which can diffuse through the partial gap in the disc generated by the planet \citep{2018arXiv180102341B}. We follow here the fit of \citet{2018arXiv180102341B}, who gave the pebble isolation mass including diffusion of small pebbles as
\begin{equation}
\label{eq:MisowD}
  M_{\rm iso} = 25 f_{\rm fit} {\rm M}_{\rm E} + \frac{\Pi_{\rm crit}}{\lambda} {\rm M}_{\rm E} \ ,
\end{equation}
with $\lambda \approx 0.00476 / f_{\rm fit}$, $\Pi_{\rm crit} = \frac{\alpha_{\rm disc}}{2\tau_{\rm f}}$, and
\begin{equation}
\label{eq:ffit}
 f_{\rm fit} = \left[\frac{H/r}{0.05}\right]^3 \left[ 0.34 \left(\frac{\log(\alpha_3)}{\log(\alpha_{\rm disc})}\right)^4 + 0.66 \right] \left[1-\frac{\frac{\partial\ln P}{\partial\ln r } +2.5}{6} \right] \ ,
\end{equation}
where $\alpha_3 = 0.001$. 

As all planets accrete pebbles at the same time, the innermost planets will feel a reduced pebble flux, compared to the case of single planets in the disc. The pebbles accreted by the outer planets are subtracted from the pebble flux that arrives at the inner planets. Once a planet reaches pebble isolation mass, we set the pebble flux to zero for all the inner planets, stopping their growth by pebble accretion. We calculate the pebble accretion rate directly from the orbital position of the planetary embryo, so a planetary embryo on an eccentric orbit that is briefly exterior to an embryo that has reached pebble isolation mass could still accrete pebbles. However, the pebble accretion rates in our model decrease strongly if the planetary eccentricity increases, preventing pebble accretion of very eccentric planetary embryos \citep{Johansen2015}.

After the planet has reached pebble isolation mass \citep{2014A&A...572A..35L, 2018arXiv180102341B, 2018A&A...615A.110A}, a gaseous envelope can contract \citep{2014A&A...572A..35L}. Once the envelope becomes as massive as the planetary core, the planet can undergo runaway gas accretion \citep{1996Icar..124...62P}. We follow here the approaches of our previous works \citep{2015A&A...582A.112B, 2019A&A...623A..88B}. Even though gas accretion rates are heavily debated in the literature \citep{2009MNRAS.393...49A, 2010MNRAS.405.1227M, 2013ApJ...778...77D, 2019A&A...632A.118S} and span several orders of magnitude in range, we keep these rates as in our previous works to allow a better comparison.

After the planet has reached pebble isolation mass, it can contract its envelope, where the calculate the envelope contraction via
\begin{eqnarray}
\label{eq:Mdotenv}
 \dot{M}_{\rm gas} &= 0.000175 f^{-2} \left(\frac{\kappa_{\rm env}}{1{\rm cm}^2/{\rm g}}\right)^{-1} \left( \frac{\rho_{\rm c}}{5.5 {\rm g}/{\rm cm}^3} \right)^{-1/6} \left( \frac{M_{\rm c}}{{\rm M}_{\rm E}} \right)^{11/3} \nonumber \\ 
 &\left(\frac{M_{\rm env}}{{\rm M}_{\rm E}}\right)^{-1} \left( \frac{T}{81 {\rm K}} \right)^{-0.5} \frac{{\rm M}_{\rm E}}{{\rm Myr}} \ .
\end{eqnarray}
Here $f$ is a factor to change the accretion rate in order to match numerical and analytical results, which is normally set to $f=0.2$ \citep{2014ApJ...786...21P}. For the opacity in the envelope we use the fixed value of $\kappa_{\rm env} = 0.05{\rm cm}^2/{\rm g}$, which is very similar to the values used in the study by \citet{2008Icar..194..368M}. Lower and higher envelope opacities would result in higher and lower envelope contraction rates. As soon as $M_{\rm core}=M_{\rm env}$, the envelope contraction ends and rapid gas accretion can start.

For the rapid gas accretion we follow \citet{2010MNRAS.405.1227M}, who calculated the gas accretion rates via 3D hydrodynamical simulations in shearing boxes. The accretion rates are divided into two branches
\begin{equation}
 \dot{M}_{\rm gas,low} = 0.83 \Omega_{\rm K} \Sigma_{\rm g} H^2 \left( \frac{r_{\rm H}}{H} \right)^{9/2}
\end{equation}
and
\begin{equation}
 \dot{M}_{\rm gas,high} = 0.14 \Omega_{\rm K} \Sigma_{\rm g} H^2 \ ,
\end{equation}
where $r_{\rm H}$ denotes the planetary hill radius. These two branches divide low mass and high mass planets, where the effective accretion rate is the minimum of these two rates. In our approach, the gap opening does not affect the gas accretion rate, because the limiting factor of the gas accretion rate is what the disc can provide to the planet, which we set to $80\%$ of the $\dot{M}$ value of the disc, because gas accretion is not 100\% efficient \citep{2006ApJ...641..526L}.

\subsection{Planetary migration}

Planetary migration in the type-I migration regime is modelled using the equations from \citet{2011MNRAS.410..293P}. This prescription includes the Lindblad torques as well as the barotropic and entropy related corotation torque. Recently \citet{2017MNRAS.471.4917J} introduced a new torque formula, which includes the same effects, but should be more accurate, because it is based on 3D hydrodynamical simulations in contrast to \citet{2011MNRAS.410..293P}, which was based on 2D simulations. However, the changes compared to the \citet{2011MNRAS.410..293P} torque formula seem quite small in the pebble accretion scenario \citep{2020arXiv200400874B}. 

The accretion of material onto the planet can change the gas dynamics around it, leading to a thermal torque \citep{2014MNRAS.440..683L, 2015Natur.520...63B}, which can, if the accretion rates are large, lead to outward migration \citep{2015Natur.520...63B}. In addition, this effect could also increase the planetary eccentricity \citep{2017A&A...606A.114C}. However, \citet{2020arXiv200400874B} showed that these effects only become very important if the accretion rates onto the planet are very large. In fact, we do not reach the accretion rates needed for the thermal torque to become positive and thus ignore its effects in this work.

Planets that become very massive and start to open deep gaps in the protoplanetary disc, change their migration regime to type-II. \citet{2018arXiv180511101K} relate the type-II migration time-scale to the type-I migration time-scale (which we calculate as explained above) in the following way
\begin{equation}
\label{eq:migII}
 \tau_{\rm mig II} = \frac{\Sigma_{\rm up}}{\Sigma_{\rm min}} \tau_{\rm mig I} \ ,
\end{equation}
where $\Sigma_{\rm up}$ corresponds to the unperturbed gas surface density and $\Sigma_{\rm min}$ to the minimal gas surface density at the bottom of the gap generated by the planet. The ratio $\Sigma_{\rm up}/\Sigma_{\rm min}$ can be expressed through \citep{2013ApJ...769...41D, 2014ApJ...782...88F, 2015MNRAS.448..994K}
\begin{equation}
\label{eq:Kgapopen}
 \frac{\Sigma_{\rm up}}{\Sigma_{\rm min}} = 1 + 0.04 K_{\rm mig} \ ,
\end{equation}
where
\begin{equation}
 K_{\rm mig} = \left( \frac{M_{\rm P}}{{\rm M}_\odot} \right)^2 \left( \frac{H}{r} \right)^{-5} \alpha_{\rm mig}^{-1} \ .
\end{equation}
We use here for the migration $\alpha_{\rm mig}=10^{-4}$, as this viscosity value was found in \citet{2019A&A...623A..88B} to allow giant planets to stay exterior to 1 AU. This choice of low viscosity in the midplane of the protoplanetary disc is motivated by the fact that recent disc observations point to low levels of turbulence \citep{2018ApJ...856..117F, 2018ApJ...869L..46D}. In addition, new evolution models of protoplanetary discs, where most of the angular momentum is transported away through disc winds instead of a large bulk viscosity, indicate a low midplane turbulence \citep{2016arXiv160900437S, 2016ApJ...821...80B, 2019ApJ...879...98C}. Finally, it should be noticed that detailed studies of planet migration in 3D discs driven by disc winds are still lacking, even though 2D approaches have been made \citep{2020A&A...633A...4K} indicating the importance of disc winds for planet migration that should be included in future models. Our migration prescription results in slow inward migration for massive planets due to the low viscosity, preventing large scale migration.

This type-II migration prescription is only valid in the case of low viscosities. At higher viscosities (e.g. $\alpha_{\rm mig} >0.001$), the entropy driven corotation torque could operate \citep{2011MNRAS.410..293P}, which could allow outward migration in certain regions of the protoplanetary discs \citep{2015A&A...575A..28B}. Applying in the case of high viscosity the migration prescription of eq.~\ref{eq:migII} would lead to an unphysical outward migration of giant planets. At low viscosities, as we use in our simulations, the entropy driven corotation torque saturates \citep{2008ApJ...672.1054B, 2011MNRAS.410..293P}, so that planets in our simulations always migrate inwards.

\subsection{Damping of eccentricity and inclination}

Damping of the orbital eccentricity and inclination during the gas disc phase tends to increase the pebble accretion efficiency and also to avoid orbital crossing and instabilities during the gas disc phase. For small planets undergoing type-I migration, we use the type-I damping rates of \citet{2008A&A...482..677C}. These damping rates are then applied to each low mass planets. Small mass planets only perturb the gas disc slightly, so that the damping formulae are still valid even if multiple small mass planets are present \citep{2013A&A...558A.105P}. The exact implementation of the damping formulae for small planets in our code is given in \citet{2019arXiv190208772I}. 

As soon as the planets start to open a gap in the disc, they start to migrate in type-II migration, where also the damping rates onto the planet are different. Changes of eccentricity and inclination of giant gap opening planets have been studied in the past \citep{2001A&A...366..263P, 2006A&A...447..369K}, where \citet{2013A&A...555A.124B} derived damping rates for eccentricity and inclination based on 3D isothermal hydrodynamical simulation. However, all these works have only considered single giant planets in discs, so the effects of damping induced by the gas disc if multiple giant planets are present have not been studied, which is why we agnostically vary the damping rates for giant planets in our simulations.

The classical K-damping prescription \citep{2002ApJ...567..596L} relate the damping rates to the type-II migration rates through
\begin{equation}
\label{eq:typeIIdamp}
 \dot{e}/e = - K |\dot{a}/a| \quad ; \quad \dot{i}/i = - K |\dot{a}/a| \ .
\end{equation}
In the following, we use this damping prescription (eq.~\ref{eq:typeIIdamp}) rather than the damping prescription of \citet{2013A&A...555A.124B}, because the K-damping prescription makes it easier to vary the damping rates. We use here $K$=5, 50, 500, and 5000. A large $K$ value implies a fast damping rate, meaning that any orbital eccentricity and inclinations that planets eventually get during the gas disc phase will be quickly damped to low values. We then explore how different damping values influence the formation of giant planet systems. As soon as the planet reaches a gap depth of 10\% of the uperturbed gas surface density, we apply type-II damping (eq.~\ref{eq:typeIIdamp}) without any smoothing function from type-I damping towards the type-II damping rate.

We use in the following for $|\dot{a}/a|$ (in eq.~\ref{eq:typeIIdamp}) not the migration rates from \citet{2018arXiv180511101K}, but instead use the classical type-II migration rate $\tau_{\rm visc} = r_{\rm P}^2 / \nu$ as soon as the gas surface density inside the opening gap reduces to 10$\%$ of the unperturbed gas surface density. Here we calculate $\nu = \alpha_{\rm disc} H^2 \Omega$, with $\alpha_{\rm disc} =0.0054$ corresponding to the $\alpha$ value used to compute the thermal structure of the protoplanetary disc \citep{2015A&A...575A..28B}. In addition, we include the effect of the mass inertia, which slows down type-II migration once the planet becomes more massive than the disc (e.g. \citealt{2013arXiv1312.4293B}). This implementation of the damping rates ensures that the damping rates are easier to control in our simulations, because they just depend on the disc's viscosity. The dependency on the planetary mass due to the mass inertia only plays a small role in our simulations with the large pebble flux. 

\begin{figure}
 \centering
 \includegraphics[scale=0.7]{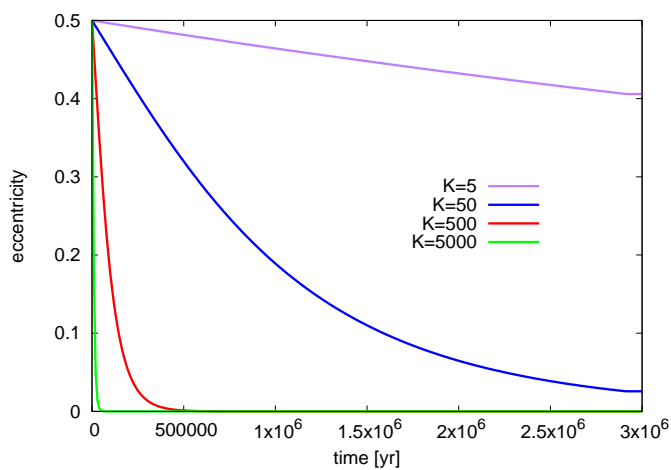}  
 \caption{Change of the eccentricity of a 1 Jupiter mass planet as function of time for the different damping parameters. The $K$ damping values are derived using the classical type-II migration timescales. The flattening of the damping at the end of the disc lifetime is due to the exponential decay of the gas disc surface in the last 100 kyr, which reduces the gas surface density and thus the damping effects.
   \label{fig:Damping}
   }
\end{figure}

In Fig.~\ref{fig:Damping}, we show how the damping rates influence the evolution of eccentricity of a single Jupiter mass planet without growth as a function of time. The disc structure and 3 Myr gas disc lifetime is the same as in all our simulations (see above).

\subsection{Initial conditions and simulation parameters}

Initially the planetary embryos embedded in the disc have $\sim$0.01 Earth masses, which corresponds to the pebble transition mass, where the accretion in the Hill regime becomes dominant \citep{2012A&A...544A..32L}. The initial eccentricities and inclinations of the planetary embryo are randomly selected from uniform distributions and are 0.001-0.01 and 0.01-0.5 degrees, respectively. This is also identical to our previous simulations \citep{2019A&A...623A..88B}.

The embryos are distributed radially between $\approx$3 and 17 AU, with equal radial spacing as in our previous work \citep{2019A&A...623A..88B}. In addition embryos starting interior to the water ice line are normally outgrown by their counterparts exterior to the water ice line \citep{2019arXiv190208772I}. We test three different configuration with initially 15, 30 or 60 planetary embryos. As the total radial distance over which we spread the embryos is constant, the initial distances between the embryos varies for the different configurations.

After the gas disc phase of 3 Myr, we evolve the system until 100 Myr to study its long-term dynamical stability after gas dispersal. This effect is very important, as instabilities in the inner systems can occur several 10 Myrs after gas disc dissipation, shaping the structure of the inner systems \citep{2017MNRAS.470.1750I, 2019arXiv190208772I, 2019arXiv190208694L}.

As mentioned before, we test four different $K$-damping scenarios. For each $K$ value we run 50 simulations, where we slightly vary the initial conditions regarding the initial planetary embryos mass, as well as the initial eccentricity and inclination and the orbital elements of the planets. In the simulations with $S_{\rm peb}$=2.5 we limit the growth of the planet by gas accretion to 1 Jupiter mass to investigate a scenario for Jupiter mass planets. Most of the giant planets formed in these simulations reach 1 Jupiter mass only at the end of the disc's lifetime. We relax this condition for simulations with $S_{\rm peb}$=5.0, where planets can grow to larger masses by accretion of pebbles and gas. For simulations with $S_{\rm peb}$=5.0 we constrain ourselves to simulations with 30 initial planetary embryos. In addition we do not present the results of simulations with 60 planetary embryos, $S_{\rm peb}=2.5$ and $K$=5. In total, we present here the results of 750 N-body simulations.

\section{Individual systems}
\label{sec:individual}

In this section we discuss the results of two selected simulations that span different $K$ damping values. We show an additional three planetary systems in appendix~\ref{ap:indi}. All the simulations presented in this section use $S_{\rm peb}$=2.5, where the growth by gas accretion is limited to 1 Jupiter mass, but planets can grow more massive through mutual collisions.

\begin{figure*}
 \centering
 \includegraphics[scale=0.7]{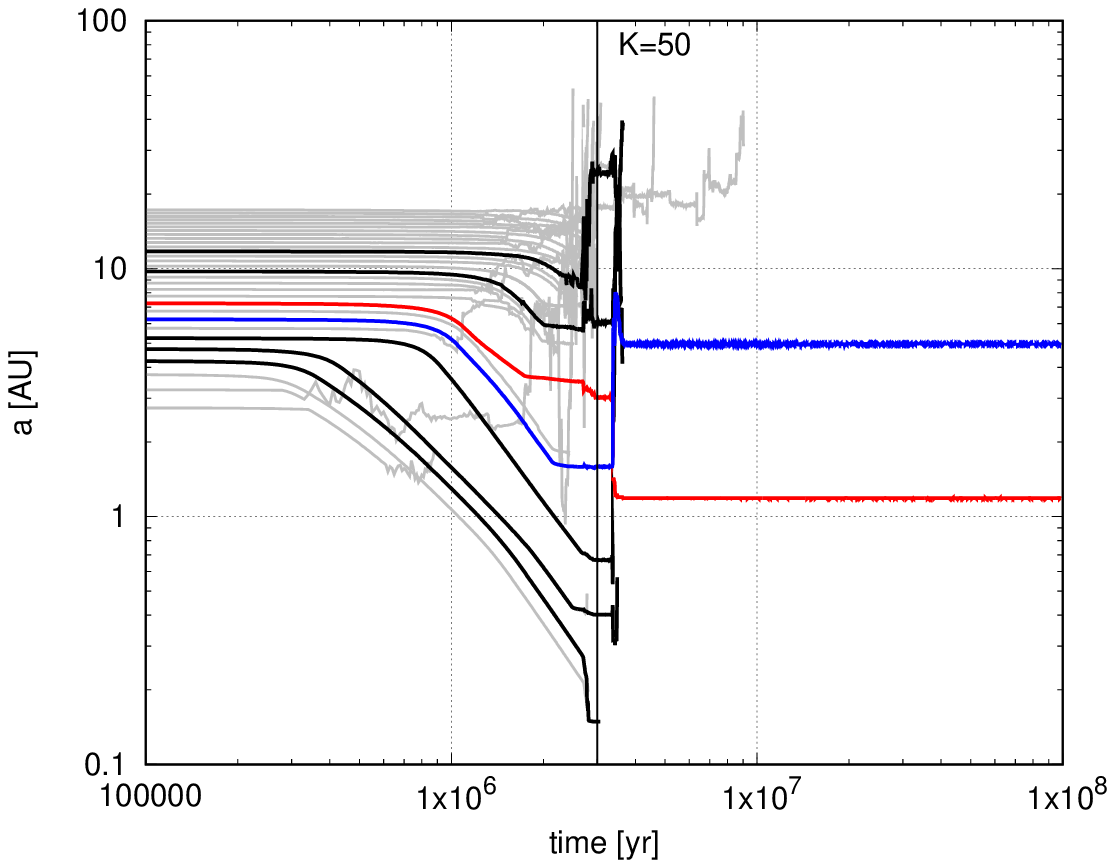}
 \includegraphics[scale=0.7]{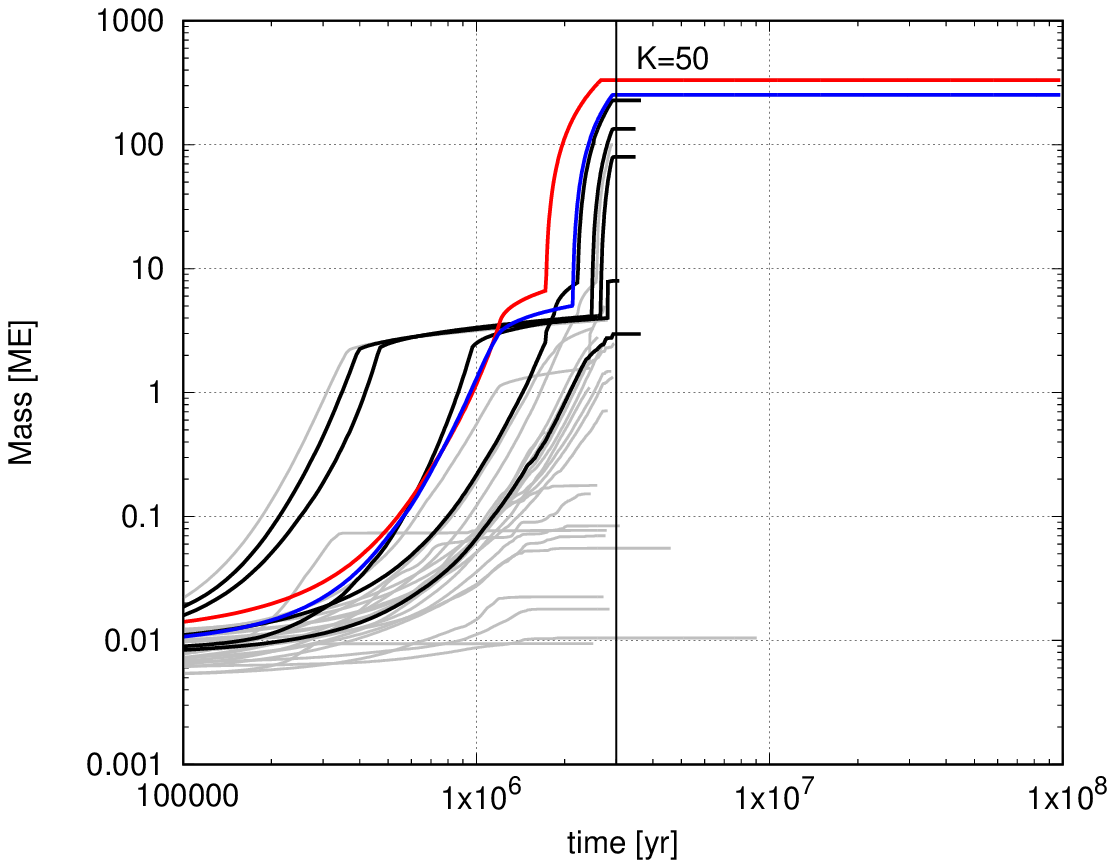}
 \includegraphics[scale=0.7]{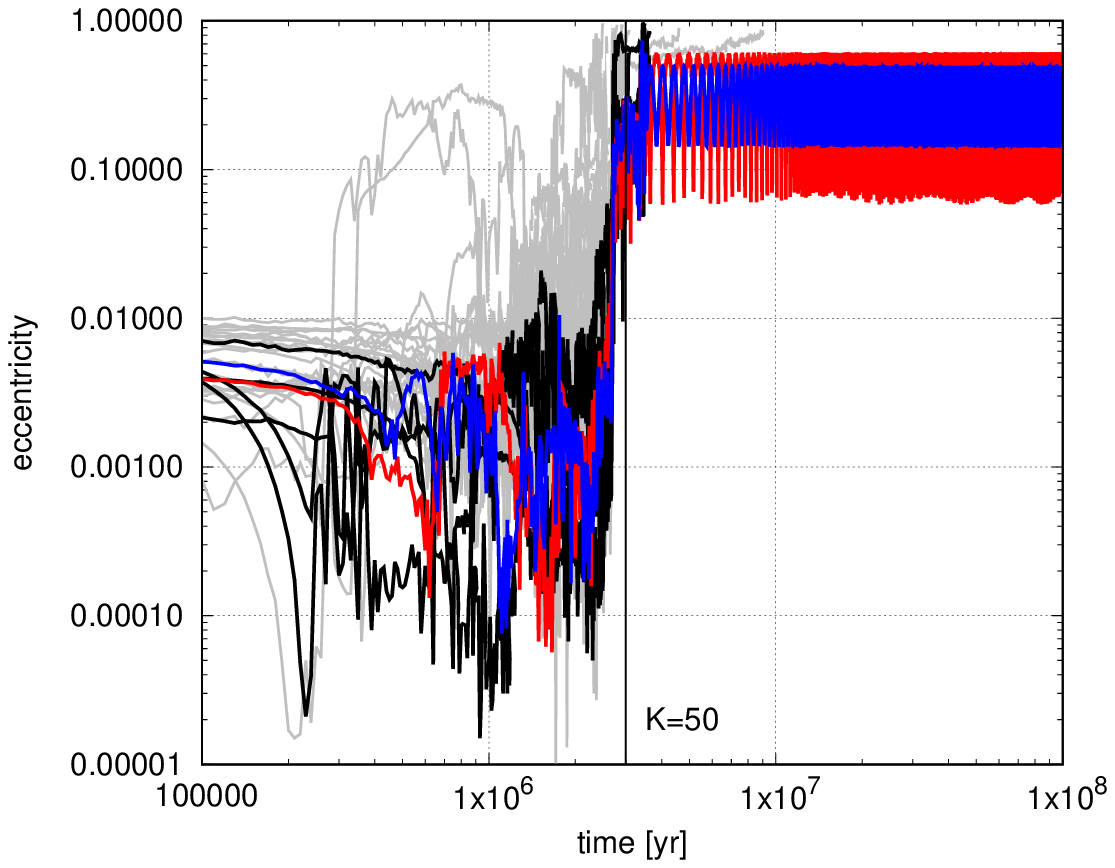}
 \includegraphics[scale=0.7]{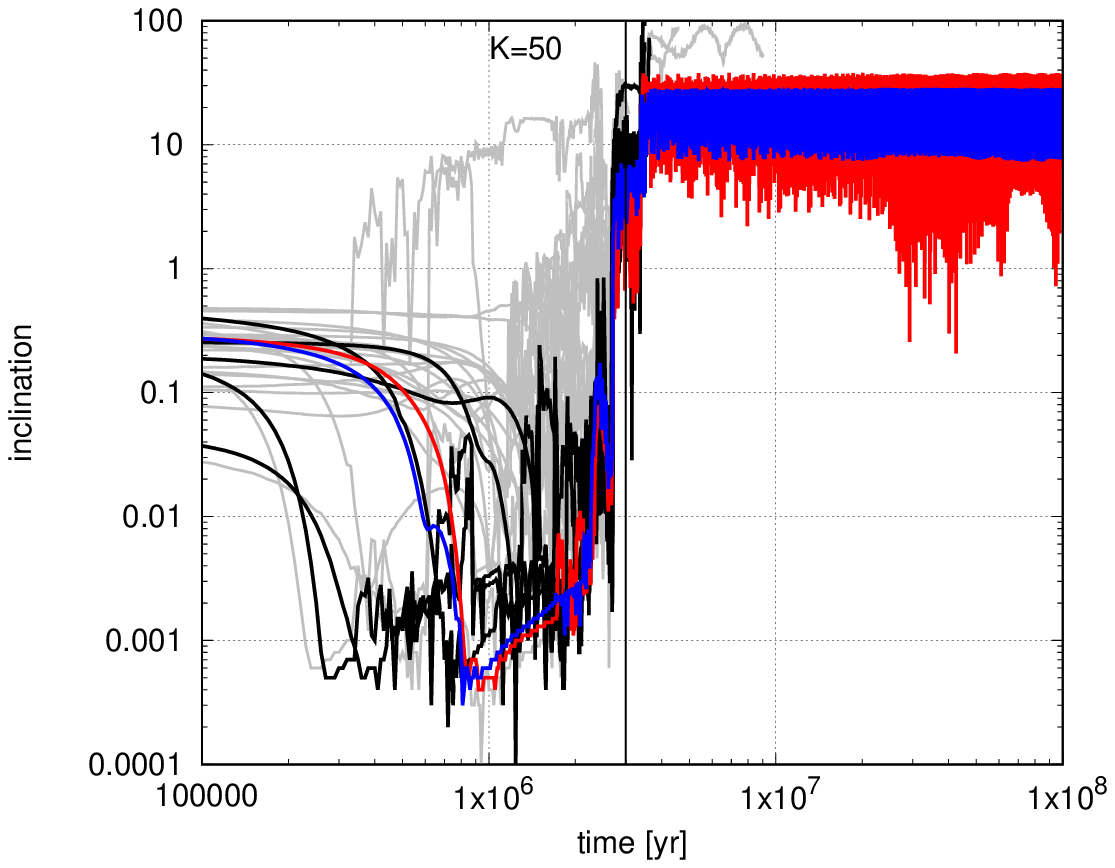}   
 \caption{Evolution of a system with a damping factor of $K=$50. Semi major axis (top left), planetary mass (top right), eccentricity (bottom left) and inclination (bottom right) of 30 planetary embryos as function of time. The gas disc lifetime is 3 Myr after injection of the planetary embryos, where the end of the gas disc lifetime is marked by the vertical black line. The grey lines represent either small mass bodies, or bodies that are ejected during the lifetime of the gas disc. The black lines represent massive planets that are scattered away after the gas disc phase. The coloured lines represent the two surviving planets.
   \label{fig:30bodyK50}
   }
\end{figure*}

In Fig.~\ref{fig:30bodyK50} we show the evolution of semi-major axis, planetary mass, eccentricity and inclination of a set of 30 planetary embryos using $K=$50 as a function of time. The grey lines represent small bodies, while the black line represents larger bodies that are scattered away after the gas disc phase and the coloured lines mark the surviving planets.

Initially the planetary embryos start growing by accreting pebbles, which can be seen by the smooth increase in planetary mass at the beginning of the simulation during the gas disc phase. As soon as the planets reach the pebble isolation mass, they stop accreting pebbles and undergo a phase of envelope contraction before runaway gas accretion can start. Three planetary embryos reach pebble isolation mass quite fast, after only $\approx$ 500 kyr, but they remain very small until the end of the disc's life time. This is caused by the fact that they finish their assembly in the inner regions of the disc, where the pebble isolation mass is small \citep{2015A&A...582A.112B, 2018arXiv180102341B} and in agreement with the typical masses of super-Earths \citep{2019ApJ...874...91W, 2019A&A...630A..51B}, preventing them from accreting a gaseous envelope. The envelope contraction phase is a strong function of planetary mass and the opacity in the planetary envelope \citep{2000ApJ...537.1013I, 2008Icar..194..368M, 2014ApJ...786...21P, 2017A&A...606A.146L}.

Only the planets accreting pebbles efficiently in the outer disc reach planetary core masses (of a few Earth masses) that allow a fast enough envelope contraction for them to reach runaway gas accretion during the lifetime of the protoplanetary disc. This phenomenon was already observed in our previous works for single bodies \citep{2015A&A...582A.112B, 2018MNRAS.474..886N} and in our N-body framework \citep{2019A&A...623A..88B}.

During the gas phase of the disc about 5 planets with Saturnian masses (and larger) are formed. However, towards the end of the gas disc lifetime, the eccentricity and inclination are not damped efficiently any more due to the low gas density and the system undergoes a dynamical instability shortly after gas disc dissipation. As a consequence, only the two most massive gas giants survived on very eccentric and inclined orbits. The inclination and eccentricity of both planets oscillate in time. In particular the eccentricity of the inner gas giant (marked in red in Fig.~\ref{fig:30bodyK50}) varies between $\sim$0.07 and $\sim$0.6. This behaviour has important consequences for matching the eccentricity distribution of the observed giant planets (see below). This behaviour has also been observed in many N-body simulations that deal with giant planets (e.g. \citealt{2008MNRAS.384..663R, 2009ApJ...699L..88R}).

\begin{figure*}
 \centering
 \includegraphics[scale=0.7]{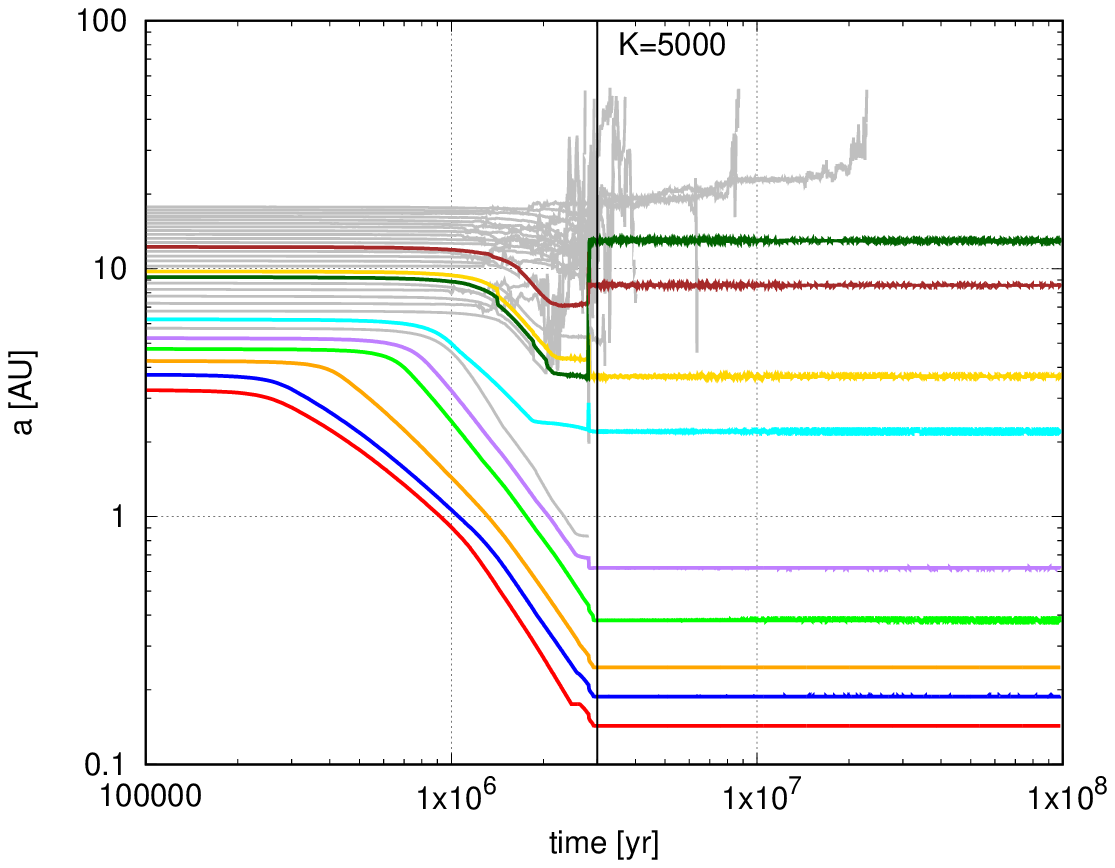}
 \includegraphics[scale=0.7]{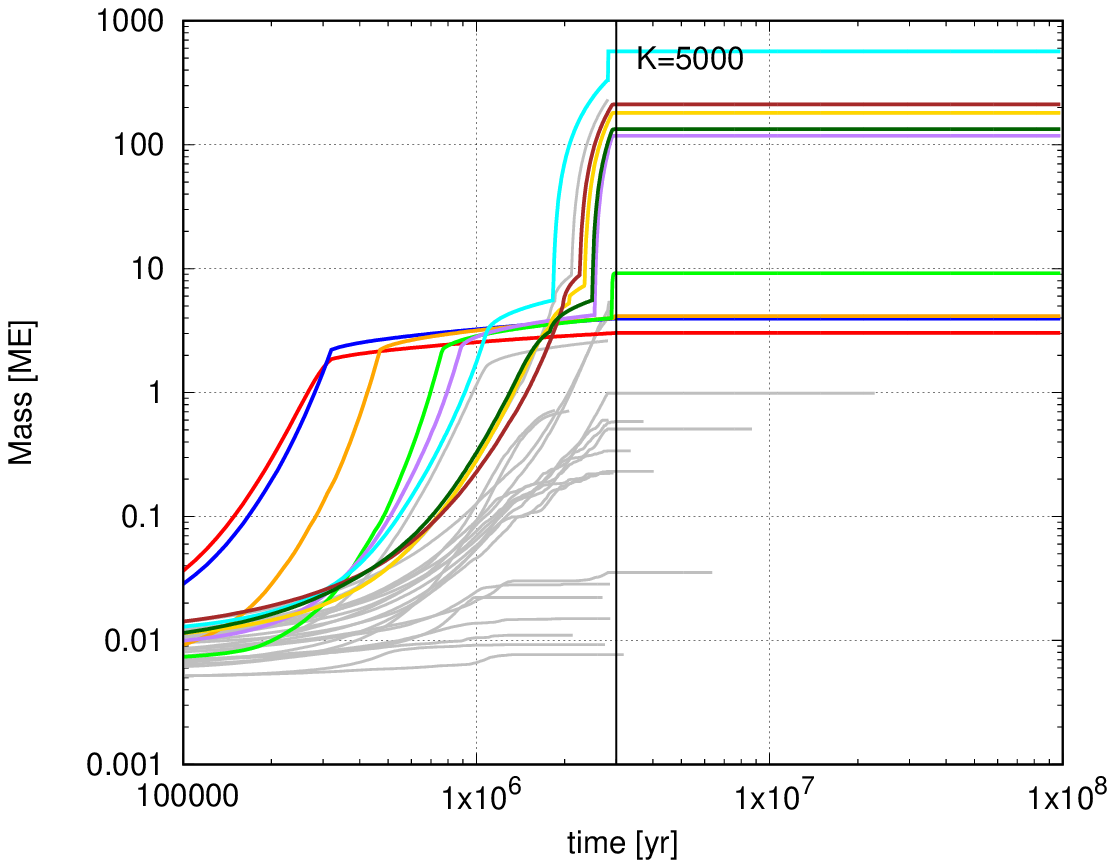}
 \includegraphics[scale=0.7]{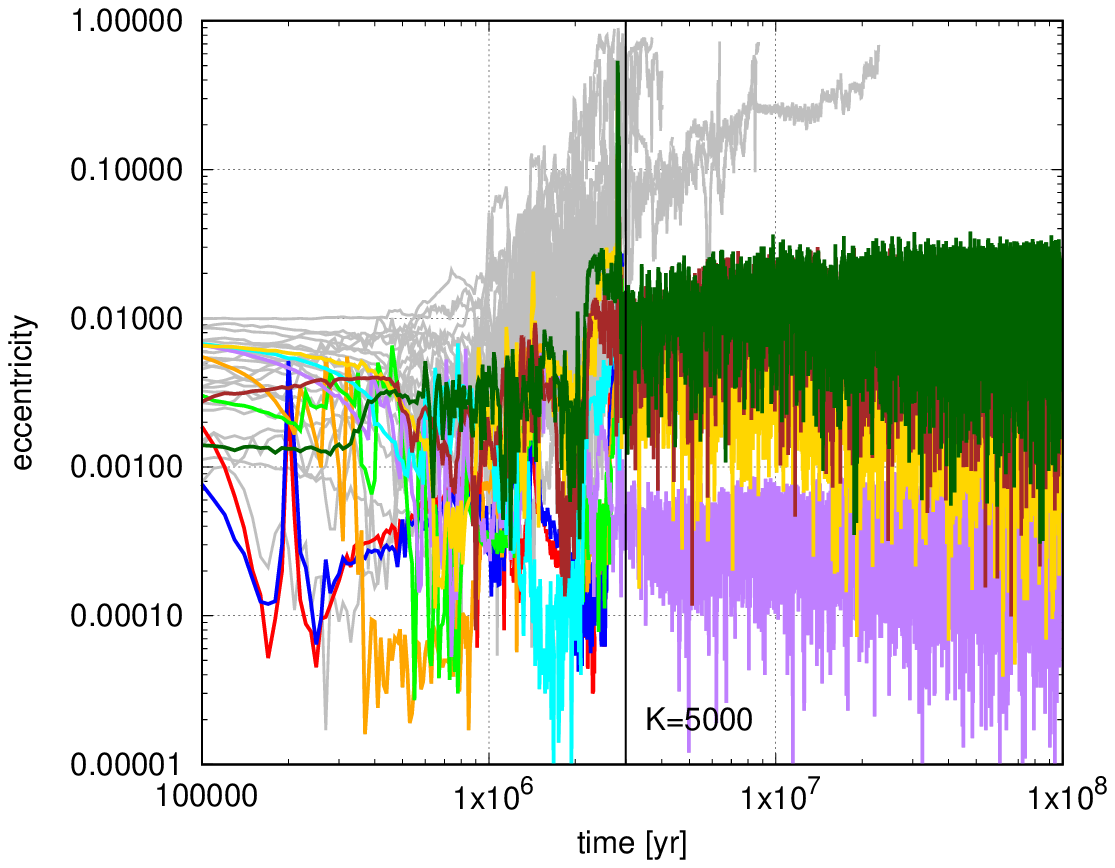}
 \includegraphics[scale=0.7]{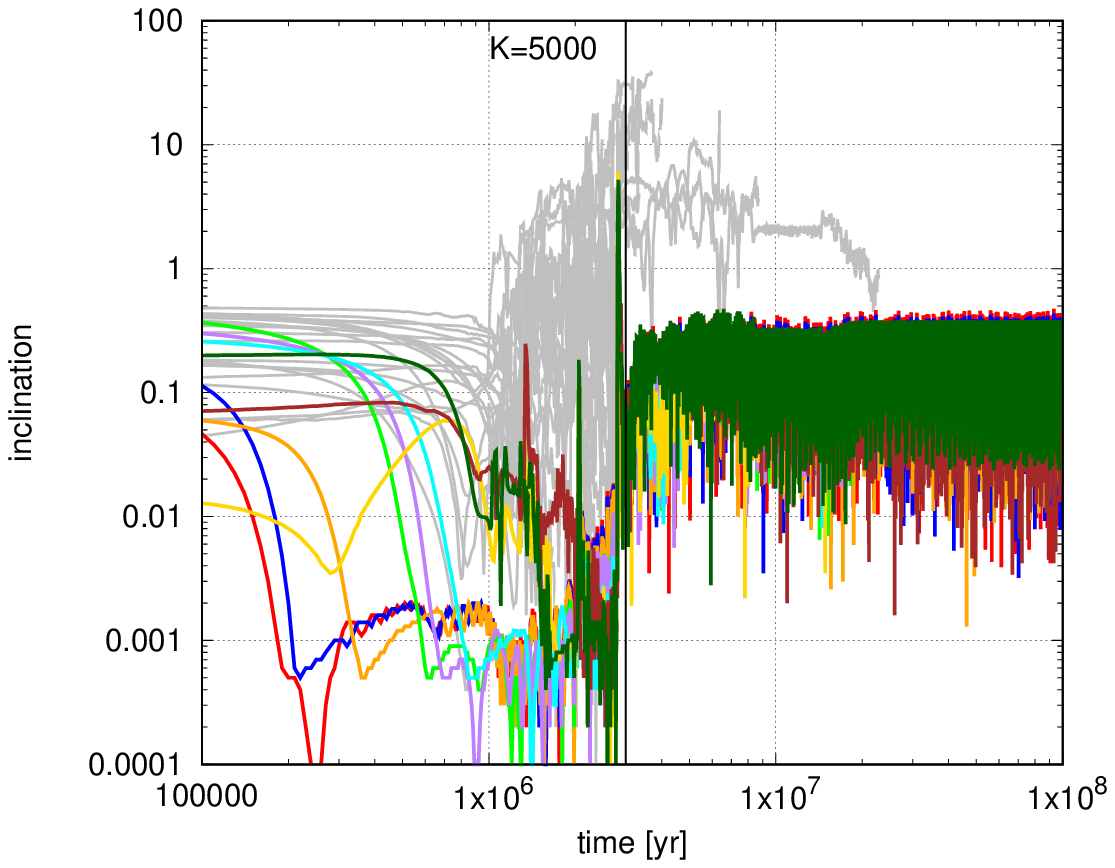}   
 \caption{Evolution of a system of 30 initial planetary embryos with a damping factor of $K$=5000. The plots and lines have the same meaning as in Fig.~\ref{fig:30bodyK50}. In this case, no giant instability after the gas disc phase happened within 100 Myr and a system with inner super-Earths and several outer gas giants survived.
   \label{fig:30bodyK5000}
   }
\end{figure*}

We show in Fig.~\ref{fig:30bodyK5000} a simulations with $K$=5000. This large $K$ value prevents the build up of eccentricities and inclinations during the gas disc phase for the growing planetary embryos. As some planets grow more and more, the eccentricities of the small bodies (below 1 Earth masses), increase dramatically towards the end of the gas disc lifetime. This then results with the ejection of the small bodies at this stage. Additionally, a collision between two gas giants occurs at the end of the gas disc lifetime, but the damping is so efficient that the eccentricity of the giants is damped almost immediately.

In the end a planetary system with several inner super-Earths and several outer gas giants forms. This result is similar to the results of \citet{2019A&A...623A..88B}, where a different damping formalism was used \citep{2013A&A...555A.124B}, but effectively it corresponded to a large $K$ value. The resulting system confirms that inner super-Earths ($r<$1.0 AU) and cold Jupiters can form within the same system, as predicted by observation \citep{2018arXiv180608799B, 2018arXiv180502660Z}. All planets feature low eccentricities and the system is nearly co-planar. We discuss this further in section~\ref{sec:discussion}.

We note that the here presented simulations are just examples of the full set of simulations. Also simulations with small $K$ can harbor systems of multiple giant planets (e.g. appendix~\ref{ap:structure}) and not all systems with small $K$ undergo dynamical rearrangements. Reciprocally, also systems with a large $K$ value can undergo dynamical instabilities. We thus discuss in the next section about the statistics of our simulations.

\section{Statistics and comparison to observations}
\label{sec:statistics}

In this section, we discuss the period ratio of adjacent planets, their orbital separation and the eccentricity distribution. We also discus the number of survived planets in our simulations at the end of the gas disc lifetime and after 100 Myr of integration. In this section we focus on the simulations using $S_{\rm peb}$=2.5 with planets that can grow up to Jupiter mass by gas accretion. We show the results of our simulations with $S_{\rm peb}$=5.0 in section~\ref{sec:growth}.

In our comparison to the observations we define giant planets as planets with masses above 0.5 Jupiter masses. The influence of our results is not affected if we were to use a cut at around Saturn mass (about $1/3$ the mass of Jupiter), as this would increase the number of giant planets by a maximum of 5\% for simulations with $K\geq$500, but much less for simulations with lower $K$. In addition, the influence is also smaller, if more initial planetary embryos were used, because the scattering effects in simulations with more initial planetary embryos are stronger (see below), where the lower mass bodies are removed efficiently. We want to stress here that our simulations are designed to form giant planets, so it is no surprise that all our simulations form giant planets. In addition this prevents us from saying anything about giant planet occurrence rates.

\subsection{Evolution of the planetary systems}

\begin{figure}
 \centering
 \includegraphics[scale=0.67]{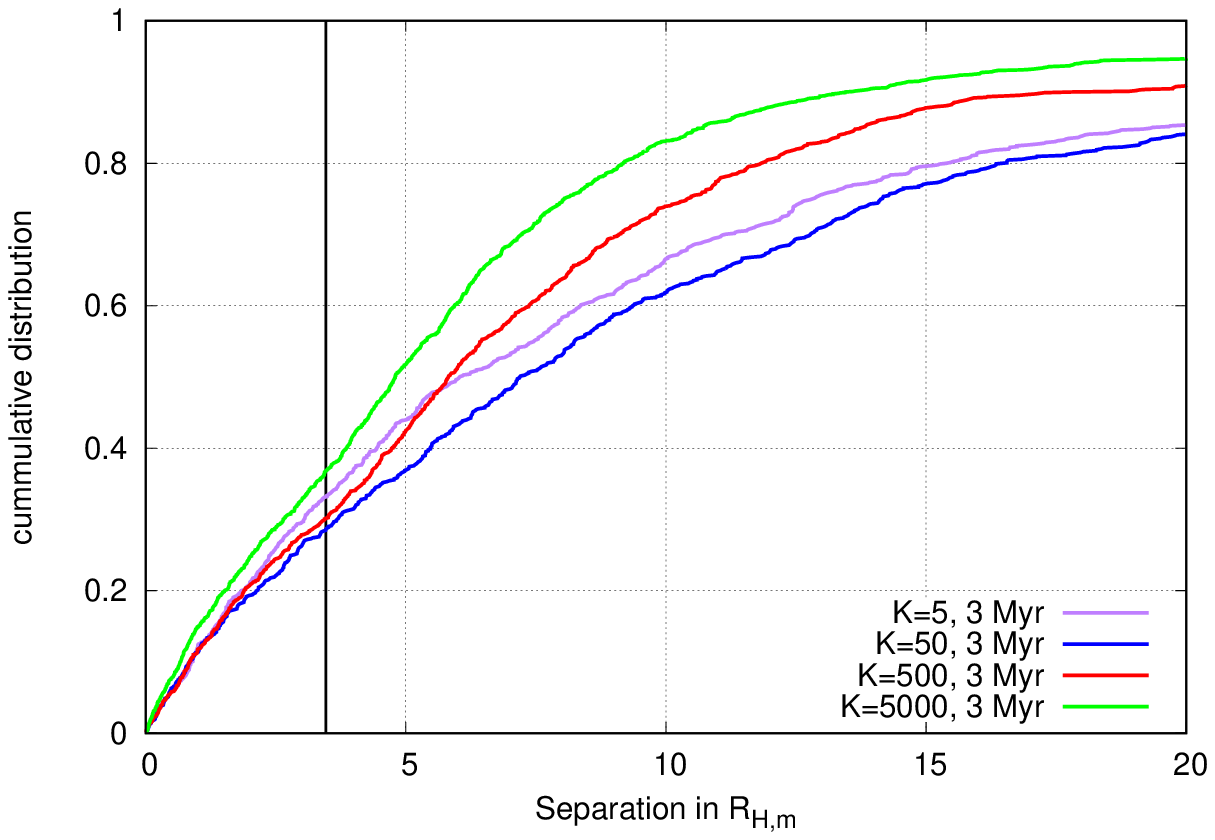}
 \includegraphics[scale=0.67]{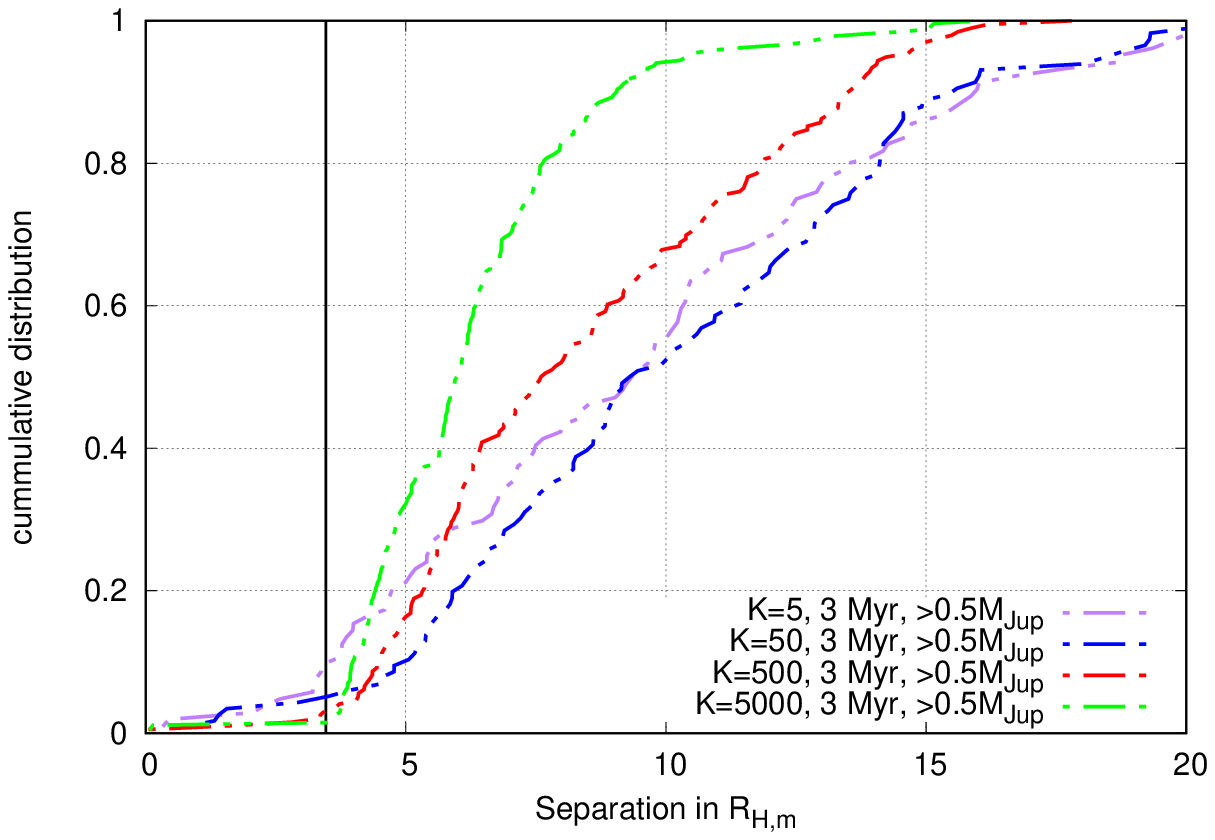}  
 \includegraphics[scale=0.67]{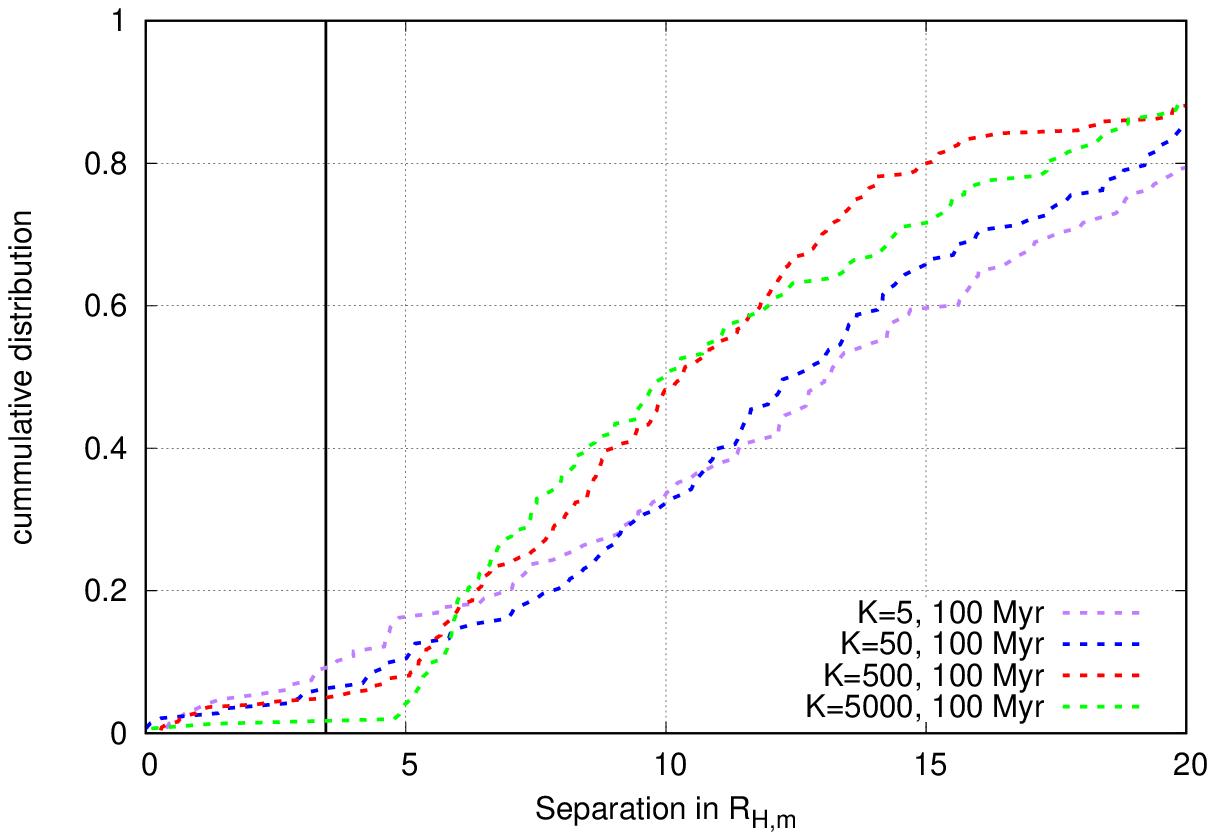}    
 \caption{Distances of adjacent planetary pairs of all planets in the simulations starting with 60 planetary embryos (including the small bodies) after 3 Myr (end of the gas disc phase, top) and 100 Myr (dashed, bottom). In addition, we show the distances after 3 Myr only for planets larger than 0.5 Jupiter masses (long-dashed, middle panel), to indicate how the stability of the system is influenced by the larger bodies. The black vertical line marks a separation of $2\sqrt{3}$ mutual Hill radii, at which two planets on circular orbits avoid mutal encounters for all time \citep{1993Icar..106..247G}. We only show the separation in units of mutual Hill radii for simulations with initially 60 bodies. The trends are very similar for simulations with 15 and 30 initial embryos.
   \label{fig:distance}
   }
\end{figure}

\begin{figure}
 \centering
 \includegraphics[scale=0.67]{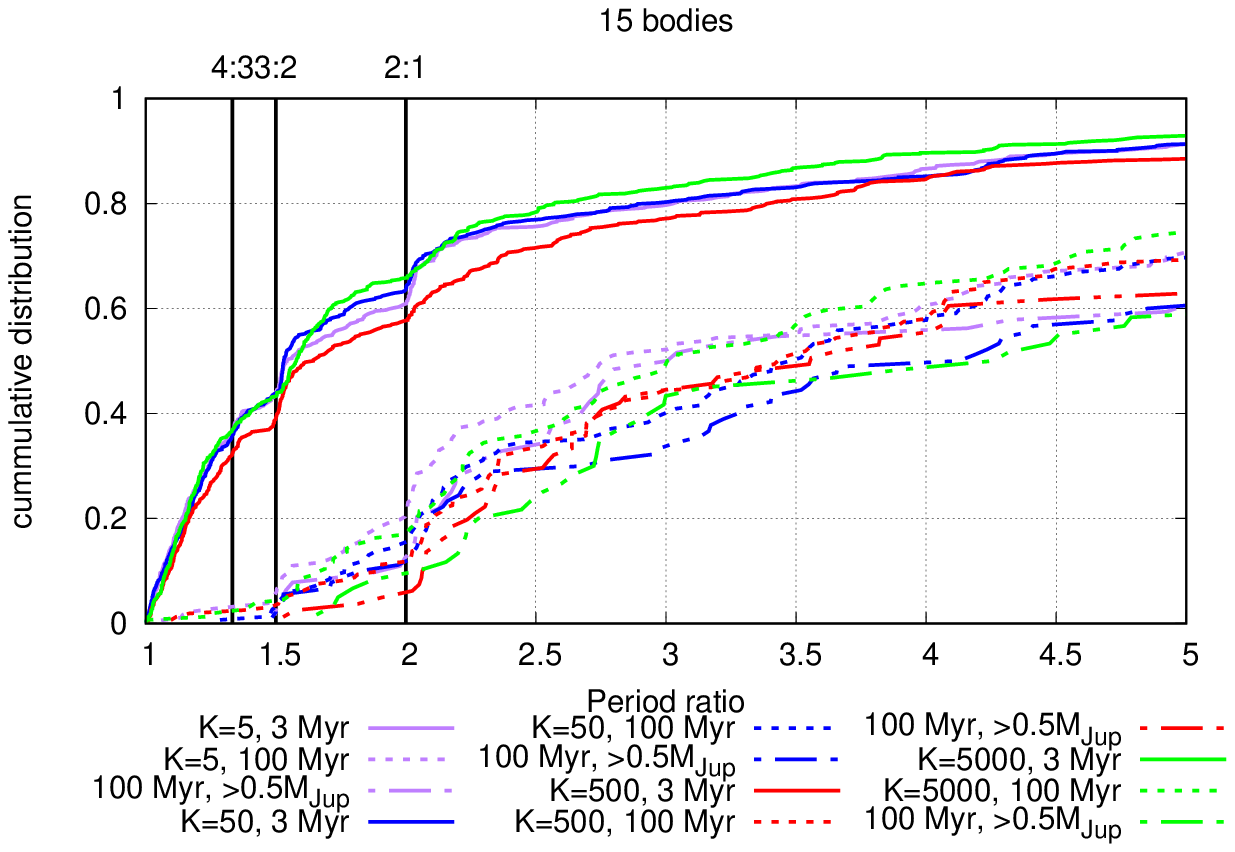} 
 \includegraphics[scale=0.67]{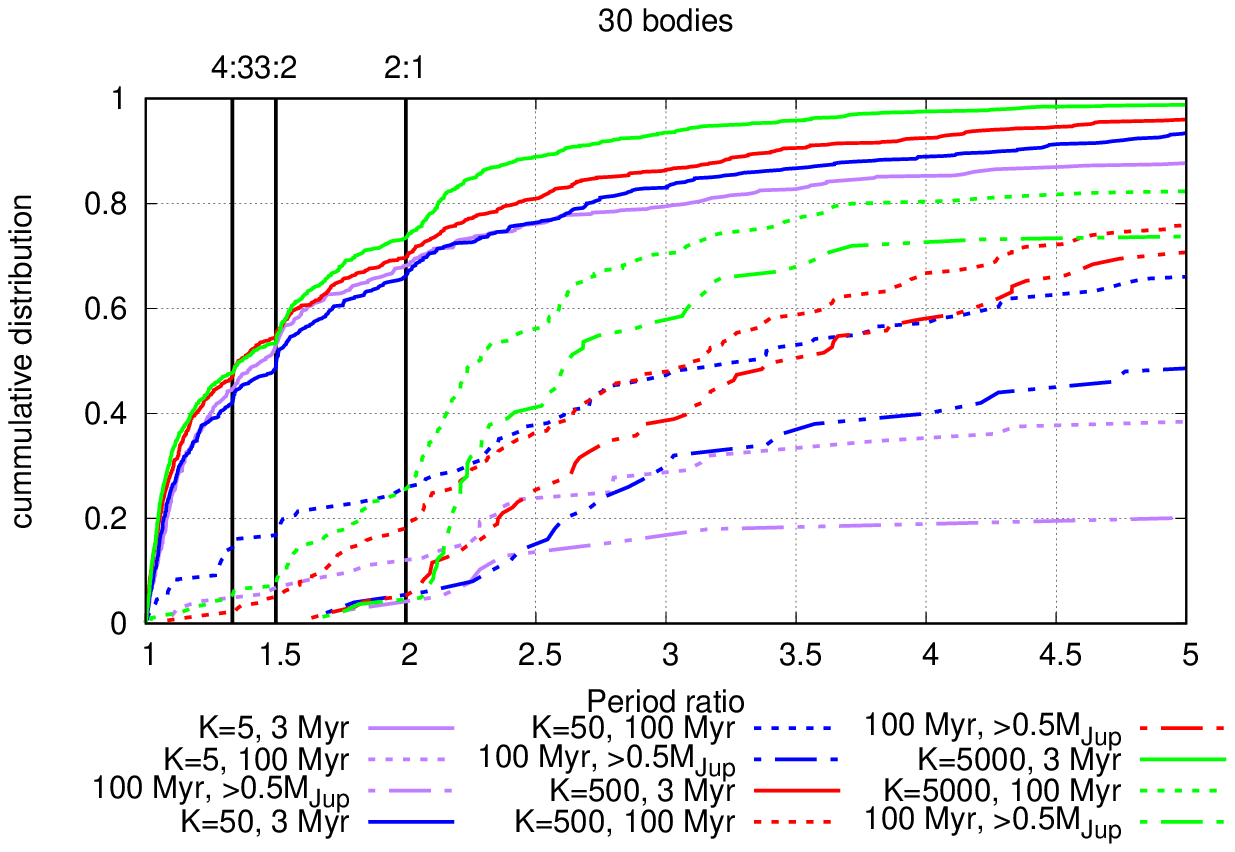} 
 \includegraphics[scale=0.67]{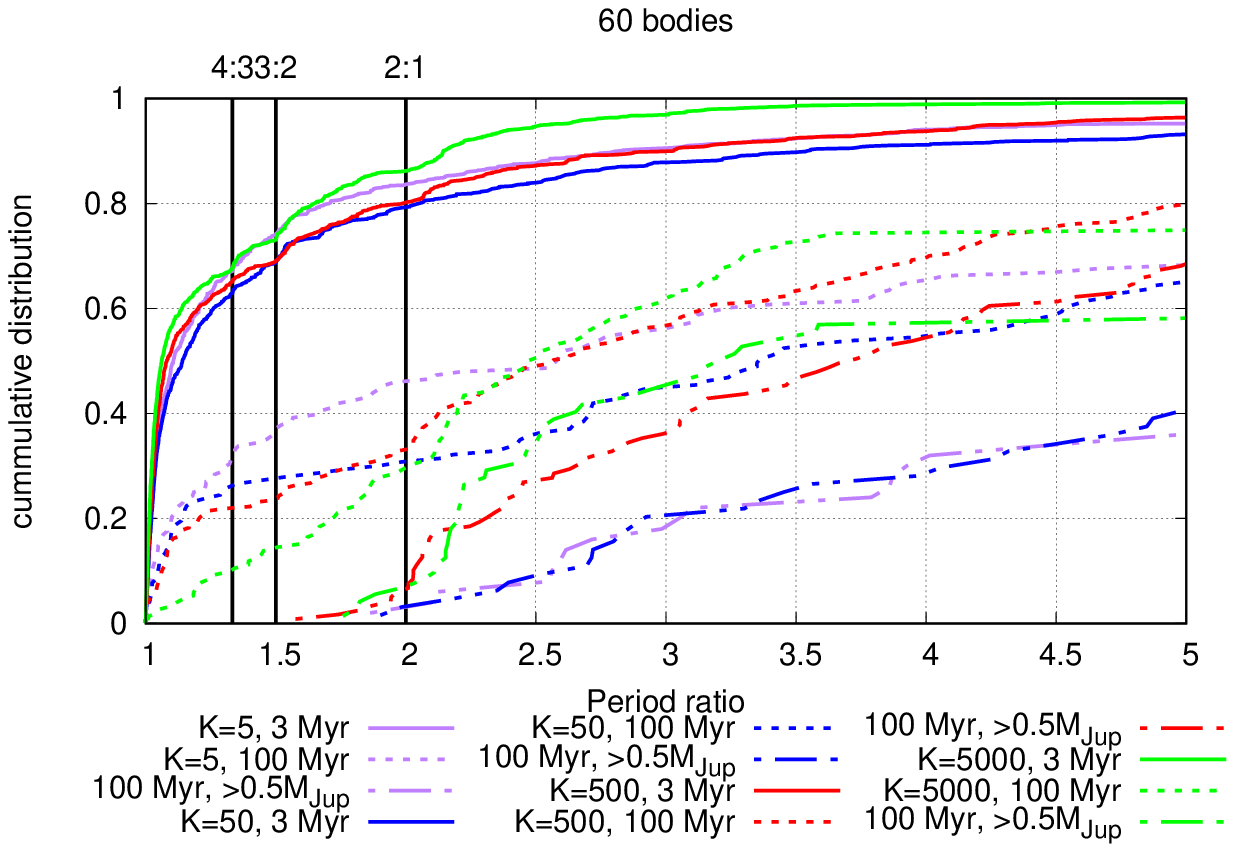}  
 \caption{Period ratios of adjacent planetary pairs of all planets in the simulations (including the small bodies) after 3 Myr (end of the gas disc phase) and 100 Myr (dashed). In addition, we show the period distribution including only planets with masses larger than 0.5 Jupiter masses after 100 Myr (long-dashed lines). We do not observe a significant pile-up of planets close to mean-motion resonances, in contrast to the super-Earth simulations of \citet{2019arXiv190208772I}, which we attribute to the slower migration speed in the here used simulations, which prevents the formation of resonant chains at the inner edge of the protoplanetary discs. Nevertheless, the systems become unstable and the instabilities result in a wide spacing of the planets after 100 Myr.
   \label{fig:periods}
   }
\end{figure}

In this section we discuss the different aspects of the dynamical evolution of the planetary systems formed in our simulations. We focus on the separation of the planets (Fig.~\ref{fig:distance}), the period ratio between adjacent planets (Fig.~\ref{fig:periods}) and how many giant planets, defined as planets with masses above 0.5 Jupiter masses, are in each system (table~\ref{tab:average} and table~\ref{tab:RVaverage}).

In Fig.~\ref{fig:distance} we show the distances between adjacent planetary pairs in the simulations at the end of the gas disc lifetime (3 Myr) and at 100 Myr for the simulations with 60 initial planetary embryos for the different eccentricity and inclination damping values. The separation between the planets is shown in mutual Hill radii $R_{\rm H,m}$ defined as
\begin{equation}
\label{eq:mHill}
R_{\rm H,m} = \frac{1}{2}\left(\frac{M_1 + M_2}{3 M_\odot}\right)^{1/3} (a_1 + a_2) \ .
\end{equation}
Here $M_1$ and $M_2$ are the masses of the two planets and $a_1$ and $a_2$ their semi-major axes. In the case of two planets on circular orbits, a separation larger than $2\sqrt{3}$ in units of the mutual Hill radii is sufficient to ensure that planets avoid mutual close en-
counters for all time \citep{1993Icar..106..247G}. The distances between planetary pairs for the cases of 15 and 30 initial planetary embryos follows the same trend as for 60 planetary embryos.

Comparing the distances of the planets for the different simulations at 3 Myr (end of the gas disc lifetime) shows that some planets are closer to each other in units of mutual Hill radii than $2\sqrt{3}$, indicating that these planets can not avoid mutual encounters for all time \citep{1993Icar..106..247G}. However, there are basically no planets with separation of less than 5 mutual Hill radii, if only planets with masses larger than 0.5 Jupiter masses are considered, at the end of the gas disc lifetime at 3 Myr. This indicates a certain stability of the planetary systems.

\citet{2009ApJ...699L..88R} studied the eccentricity evolution of giant planets by putting already fully formed giant planets at separations of 4-5 $R_{\rm H,m}$ to induce scattering events. In our simulations, only a small fraction of giant planets is within this separation at the end of the gas disc phase at 3 Myr and thus only a small fraction of our systems formed by pebble and gas accretion reflect the initial conditions used in \citet{2009ApJ...699L..88R}, which has important consequences for the eccentricity distribution of the giant planets (see below).

After 100 Myr of evolution, the systems have undergone scattering events, which increases the separation between adjacent planet pairs (dotted lines in Fig.~\ref{fig:distance}). In particular, for $K$=5 the planets are very widely spaced at 100 Myr, indicating more scattering events compared to the simulations with high $K$ (fast damping). The mutual separations at 3 and 100 Myr look very similar for all three sets of simulations with initially 15, 30, and 60 planetary embryos.

%For 15 initial planetary embryos, the distances in units of mutual Hill radii are nearly independent of the damping efficiency (top in Fig.~\ref{fig:distance}). However, this trend breaks for simulations starting with a larger number of embryos and low $K$ values. In the case of a larger number of embryos and low $K$ values, the distances in units of mutual Hill radii are larger compared to their counterparts with high $K$ at 100 Myr. In particular, for $K$=5 the planets are very widely spaced at 100 Myr, indicating more scattering events compared to the simulations with high $K$ (fast damping).

In Fig.~\ref{fig:periods} we show the period ratios of adjacent planetary pairs in our simulations with different initial number of embryos (top to bottom) as well as for the different damping values. At the end of the gas disc lifetime, about 50\% of adjacent planet pairs are interior of the 3:2 period ratio, with no substantial difference for the different damping values for all simulations. Our simulations show a slight pile up of planets around the 3:2 and 2:1 period ratios at this phase, but the clear majority of planets are clearly not locked in first order mean motion resonacnes.

In our simulations, mostly the inner super-Earths are trapped in resonant configuration at the end of the gas disc lifetime. The inner two pairs of super-Earths shown in Fig.~\ref{fig:30bodyK5000} are actually in a 3:2 resonance configuration. However, our simulations do not show very long chains of super-Earths as in previous simulations dedicated to the formation of super-Earths \citep{2017MNRAS.470.1750I, 2019arXiv190208772I, 2019arXiv190208694L}, where the chains can accomodate even up to 9 super-Earths. In our simulations we have a maximum of 4-5 super-Earths per system. This difference is caused by the slower migration rates used in the here presented work compared to \citet{2019arXiv190208772I} and \citet{2019arXiv190208694L}. In our simulations the planets that form in the inner regions of the disc and stay at super-Earth mass (e.g. Fig.~\ref{fig:30bodyK5000}) do not always migrate to the inner edge of the disc at 0.1 AU and can thus not always form a resonant chain by this mechanism. In addition, the super-Earth systems can be destroyed by gravitational interactions between the planets (see below). Furthermore, the most massive planetary cores in our simulations start to accrete gas and become gas giants.

During the next 100 Myr of evolution, the planetary systems undergo dynamical instabilities that increase the distances between adjacent planet pairs (Fig.~\ref{fig:distance}) and as a consequence also the period ratios between planetary pairs. In fact now only a very small fraction of planet pairs are closer than the 2:1 period ratio.

Doing the same cut as before, by only looking at planets with masses larger than 0.5 Jupiter masses, reveals that only a tiny fraction of systems feature giant planets with a period ratio smaller than 2:1. However, our simulations show a pile-up of planets just exterior to the 2:1 ratio. Nevertheless, there is no clear preferred period ratio between the giant planets in our simulations. In addition, our simulations suggest that pairs of giant planets in resonance should be rare. As already noted above, a lower $K$ damping value results in more violent instabilities, also pushing the period ratios between adjacent planet pairs to larger values.

%In Fig.~\ref{fig:number} we show the number of planets with masses larger than 0.5 Jupiter masses at the end of our simulation time of 100 Myr (solid lines) for the simulations with different initial number of embryos and damping prescriptions. In addition, we also show (in dashed lines) the planets that would be visible by RV detections of a sensitivity of 1 m/s with a distance of up to 5.2 AU (corresponding to 11.8 years) and masses above 0.5 Jupiter masses.

In table~\ref{tab:average} we show the average number of planets with masses larger than 0.5 Jupiter masses at the end of the simulation at 100 Myr. The average number of planets slightly increases from $K$=5 to $K$=500, but then drops off slightly again for $K$=5000.

In our simulations we observe a complex interplay between the efficency of eccentricity and inclination damping on the one hand and the number of planetary embryos on the other hand. The interplay between these two variables then influences how stable the formed planetary system is and thus how many giant planets survive. However, from within our simulations it is difficult to observe a very clear trend regarding the distances between the planets, their period ratios or the number of surviving planets.

%There are some overall trends visible that can be explained quite easily by the initial number of seeds. A larger initial number of seeds allows slightly more giant planets to form, where the maximum number of giant planets formed in one simulation rises from 5 to 7, when increasing the initial number of planetary embryos from 15 to 60 after 100 Myr of evolution. However, these large numbers of giant planets are only reached in simulations with $K \geq 500$. For $K$=50, the maximum number of giant planets after 100 Myr is 5 in the case of 60 initial embryos, but 4 otherwise.

%As the damping becomes more efficient, less scattering events and collisions between the planets happen, naturally increasing the number of giant planets present at the end of the simulations. The slight decrease for $K$=5000 could be explained by the fact that more giant planet survive the gas phase, which then lead to more scattering events after the gas disc phase, reducing the average number of giant planets in our simulations.

%As already mentioned before, the number of giant planets emerging from the gas disc phase is only weakly dependent on the initial number of embryos, resulting in only a slight variation in the final average number of planets for the same damping value.

\begin{table*}
\centering
\begin{tabular}{c|c|c|c|c}
\hline
K & 15 seeds & 30 seeds & 60 seeds & 30 seeds, $S_{\rm peb}=5.0$ \\ \hline \hline
5 & 2.42 $\pm$ 0.95 & 1.74 $\pm$ 0.85  & 1.96 $\pm$ 1.03 & 1.74 $\pm$ 0.83 \\
50 & 2.36 $\pm$ 1.01 & 2.02 $\pm$ 0.89 & 2.26 $\pm$ 1.11 & 2.82 $\pm$ 1.17 \\ 
500 & 2.64 $\pm$ 1.24 & 2.96 $\pm$ 1.41 & 3.44 $\pm$ 1.62 & 4.62 $\pm$ 1.31 \\ 
5000 & 2.26 $\pm$ 1.05 & 2.68 $\pm$ 1.44 & 2.46 $\pm$ 1.51 & 4.54 $\pm$ 2.01 \\ \hline
\end{tabular}
\caption[Average number of planets]{Average number of planets with masses larger than 0.5 Jupiter masses at the end of our simulations at 100 Myr with different $K$ damping values and initial number of planets (seeds). We also show the 1$\sigma$ deviation of the mean values.}
\label{tab:average}
\end{table*}

Table~\ref{tab:RVaverage} shows the average number of planets with masses larger than 0.5 Jupiter masses that could be detected by RV measurements with 1m/s and a distance up to 5.2 AU. There, the trends of the total average number of giants with masses larger than 0.5 Jupiter masses is reflected as well. But, of course, the total number of planets detected by the RV measurements is much lower. In fact the simulations with slow damping (low $K$) predict an average number of less than 1.5 giant planets per system. This implies that currently observed planetary systems should have one or two giant planets, but should host on average around 2-3 planets according to our simulations (tab.~\ref{tab:average}), where the maximal number of giant planets in our simulations is 7.

Most of the giant planets in our simulations are within 10 AU, they are unikely to be observed by direct imaging. However, our simulations predict that if systems harboring giant planets are observed for longer time by RV measurements, extending the orbital distance to which planets can be found, the average number of giant planets per system should increase by roughly 50\%.

\begin{table*}
\centering
\begin{tabular}{c|c|c|c|c}
\hline
K & 15 seeds & 30 seeds & 60 seeds & 30 seeds, $S_{\rm peb}=5.0$ \\ \hline \hline
5 & 1.64 $\pm$ 0.85 & 1.00 $\pm$ 0.90 & 1.16 $\pm$ 1.04 & 0.78 $\pm$ 0.84 \\
50 & 1.56 $\pm$ 1.01 & 1.30 $\pm$ 0.93 & 1.38 $\pm$ 1.11 & 1.60 $\pm$ 0.93 \\ 
500 & 1.62 $\pm$ 1.01 & 2.02 $\pm$ 1.20 & 2.30 $\pm$ 1.28 & 2.78 $\pm$ 0.95\\ 
5000 & 1.50 $\pm$ 1.09 & 1.74 $\pm$ 1.23 & 1.48 $\pm$ 1.33 & 2.42 $\pm$ 1.25 \\ \hline
\end{tabular}
\caption[Average number of planets]{Average number of planets with masses larger than 0.5 Jupiter masses at the end of our simulations at 100 Myr with different $K$ damping values and initial number of planets (seeds) that are detectable by RV measurements with 1 m/s up to a distance of 5.2 AU. We also show the 1$\sigma$ deviation of the mean values.}
\label{tab:RVaverage}
\end{table*}

\subsection{Eccentricity distribution}

\begin{figure}
 \centering
 \includegraphics[scale=0.67]{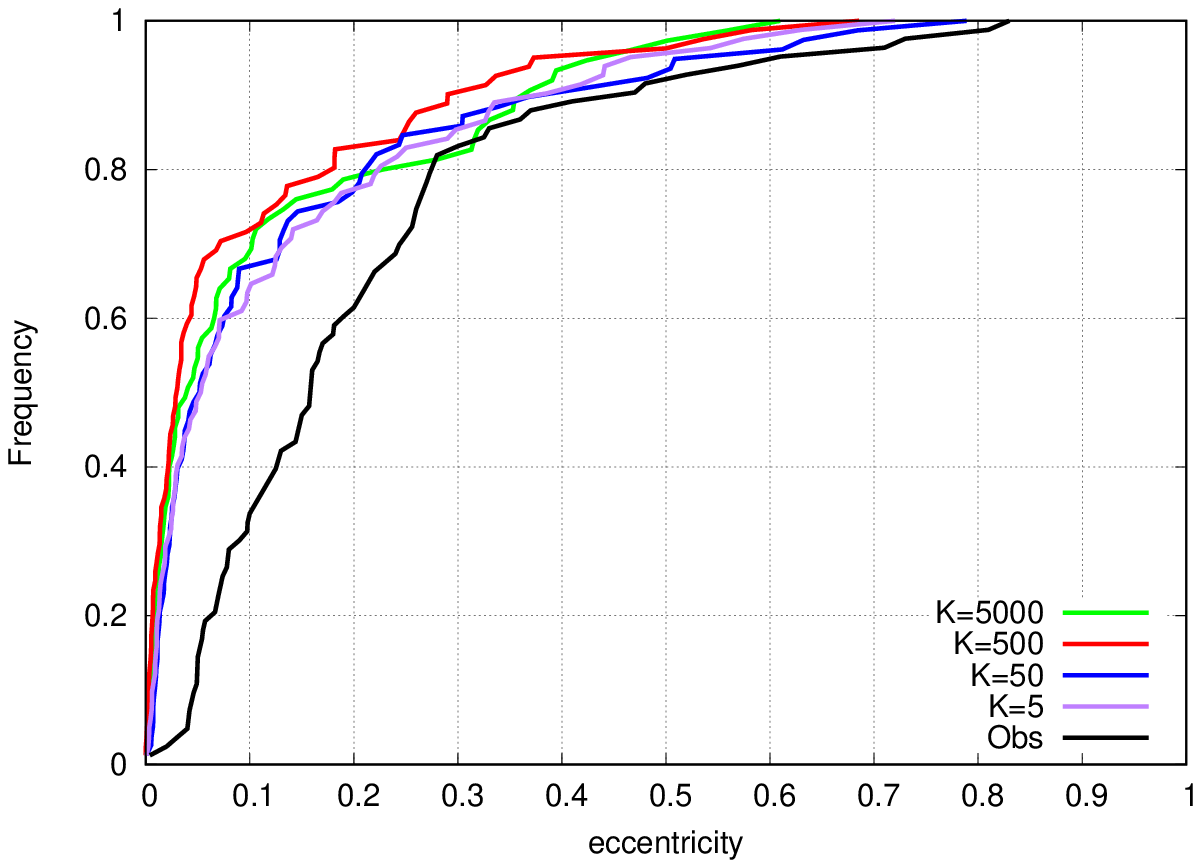} 
 \includegraphics[scale=0.67]{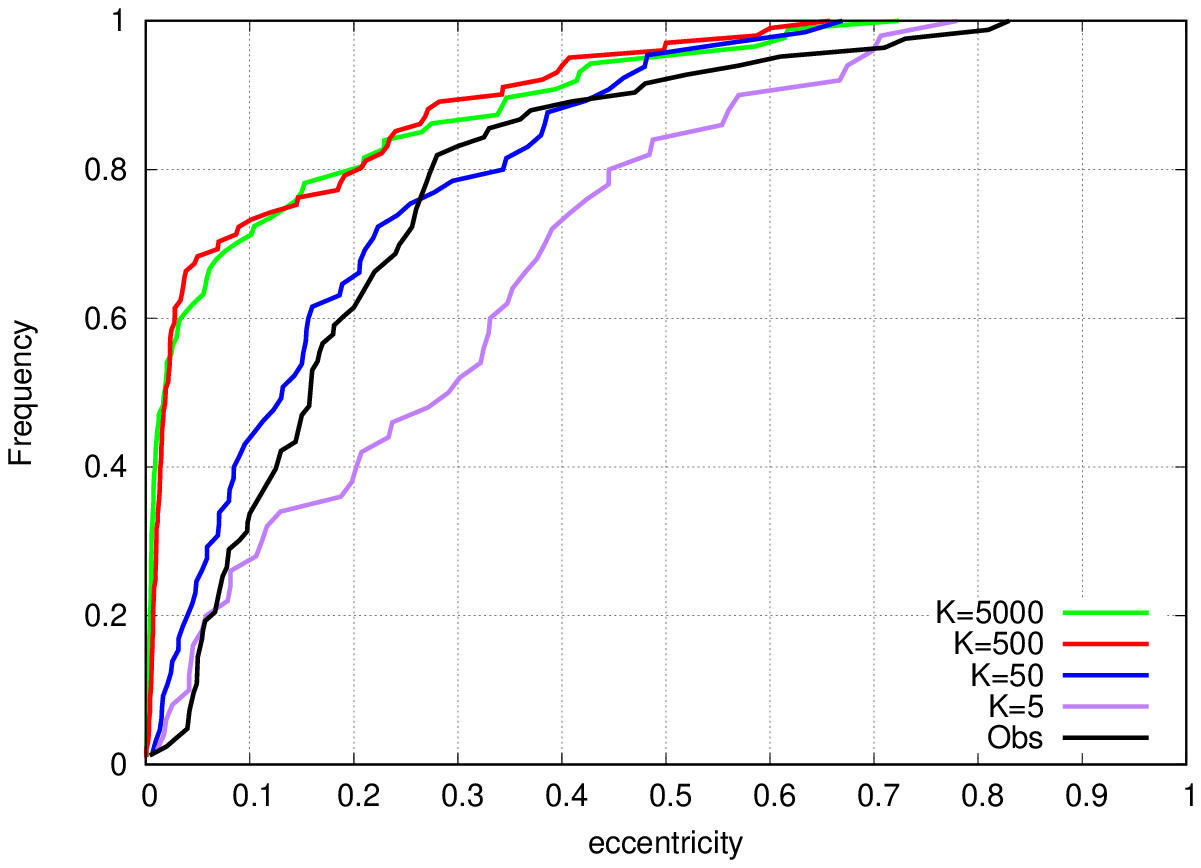} 
 \includegraphics[scale=0.67]{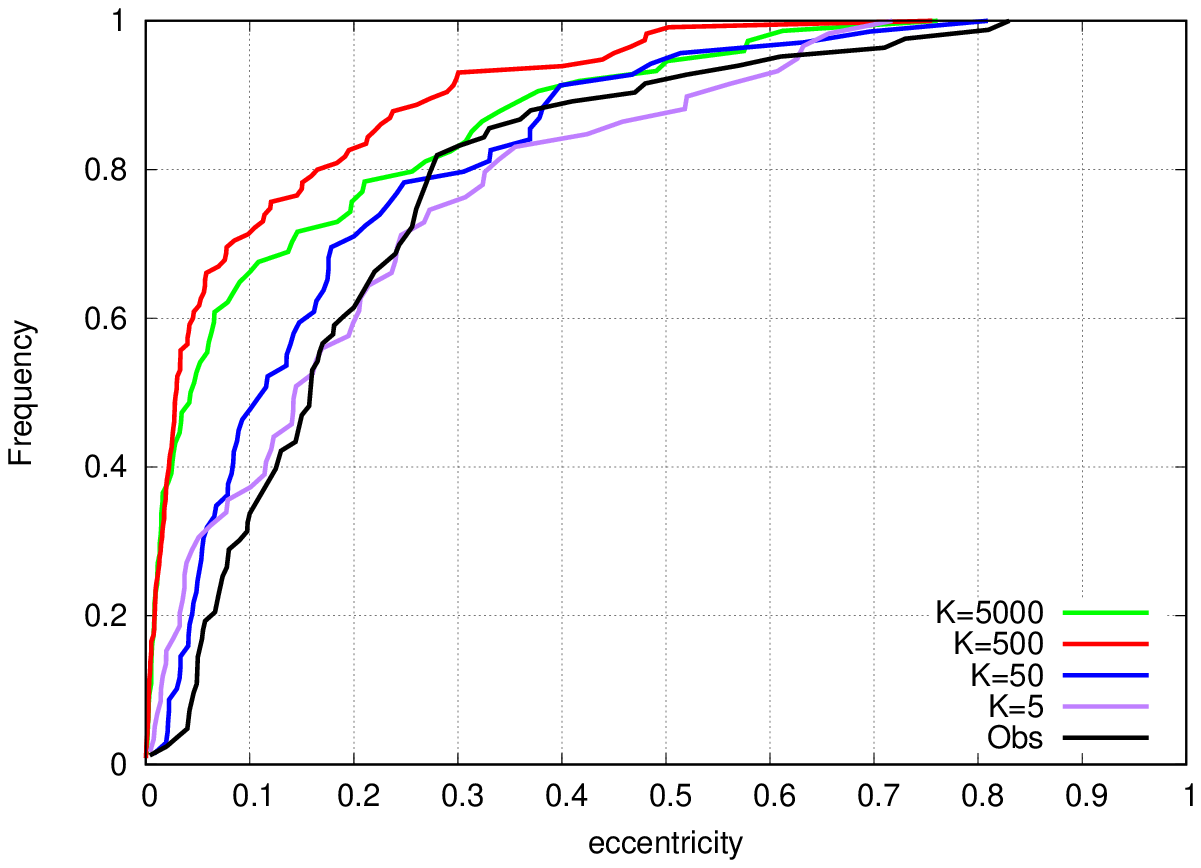}  
 \caption{Eccentricity distribution of the simulated planets within a distance of 5.2 AU from their host star with masses of at least 0.5 Jupiter masses as well as of RV observations of giant planets with masses between 0.5 and 1.25 Jupiter masses exterior to 0.1 AU. We show in colour the different damping rates and systems with initially 15, 30 and 60 embryos (from top to bottom).
   \label{fig:ecchist}
   }
\end{figure}

In Fig.~\ref{fig:ecchist} we show the eccentricity distribution for the planets formed in our simulations with different $K$ and different initial number of planetary embryos for planets that could be detected by RV measurements with a sensitivity of 1 m/s and up to 5.2 AU with masses above 0.5 Jupiter masses. In addition, we show the eccentricity distribution of the giant planets observed via RV from the exoplanet.eu database. We show here only planets with a minimum mass of 0.5 Jupiter masses and maximal $M\sin(i)$ of 1.25 Jupiter masses for the observations, but no upper mass limit of planets for our simulations, which barely reach masses larger than 1 Jupiter masses for this set of simulations. In addition, we limit ourselves to planets with orbital distance larger than 0.1 AU, in order to avoid a contamination by hot Jupiters, which are abundant in their observations due to their detectability, but are not very common in general \citep{2011arXiv1109.2497M, 2013Sci...340..572H, 2018ARA&A..56..175D}. In all our simulations we only produce about $1\%$ of hot Jupiters with semi-major axis smaller than 0.1 AU, which are also excluded from the following analysis, because we do not include effects of tides in our simulations.

Clearly, larger $K$ values result in an eccentricity distribution that is too steep to explain the observations. For $K\geq 500$, around $\sim$70\% of all giant planets with masses larger than 0.5 Jupiter masses have an eccentricity below 0.1. This clearly does not match the eccentricity distribution from the observations. In the case of 15 initial planetary embryos, this is also true for $K$=50 and $K$=5. However, when increasing the initial number of planetary embryos, the number of planets with eccentricities below 0.1 drops to around $\sim$45\% or below for $K$=50 and $K$=5, resulting in an eccentricity distribution closer to the observations for that particular eccentricity bin.

On the other hand, the simulations with $K$=50 under predict the eccentricity distribution of the observations for $0.1 \leq e \leq 0.3$, while they seem to slightly over predict the frequency of giant planets with $e>0.3$ in the case of 30 or 60 initial planetary embryos. In the case of $K$=5, this effect becomes even more prominent. Our simulations actually under produce planets with low eccentricities and predict more planets with larger eccentricities. This is caused by the slow damping during the gas disc phase which allows the planets to acquire already some small eccentricities, leading to instabilities at the end of the gas disc phase.

The comparisons in Fig.~\ref{fig:ecchist} already give a good clue that low $K$ values are needed in our simulations to reproduce the eccentricity distribution of the RV observations. It seems that a $K$ value between 5 and 50 seems to give the best results regarding the eccentricity distribution. We show in table~\ref{tab:pvalue} the results of a Kolmogorov-Smirnov-Test (KS-test) for the comparison between the eccentricity distribution in our simulations with the eccentricity distribution of the observed giant planets. The higher the value of the KS-test, the better the match of the simulations with the observations. Clearly, a $K$ value between 5 and 50 seems to match the observations best. However, this does not contain the radial distribution of the planets and if the giant planets with large eccentricities originate from close-in or far away orbits.

\begin{table*}
\centering
\begin{tabular}{c|c|c|c|c}
\hline
K & 15 seeds & 30 seeds & 60 seeds & 30 seeds, $S_{\rm peb}=5.0$ \\ \hline \hline
5 & 1.72e-6 & 1.01e-3 & 0.037 & 0.114 \\
50 & 1.18e-6& 0.369 & 0.249 & 7.37e-3 \\ 
500 & 5.55e-12 & 1.25e-12 & 1.86e-12 & 1.22e-57 \\ 
5000 & 6.15e-8 & 4.68e-13 & 3.37e-8 & 1.22e-52 \\ \hline
\end{tabular}
\caption[p-value]{p-value of a K-S test of the eccentricity distributions from our simulations compared to the observed eccentricities of exoplanets. We note here again, that the exoplanet sample is cut at 1.25 Jupiter masses for $S_{\rm peb}=2.5$, but continues to 5 Jupiter masses for $S_{\rm peb}=5.0$, because the planets grow more massive in that case. We presume that our simulations match observations if the KS-test returns a p-value larger than 0.05.}
\label{tab:pvalue}
\end{table*}

\begin{figure}
 \centering
 \includegraphics[scale=0.67]{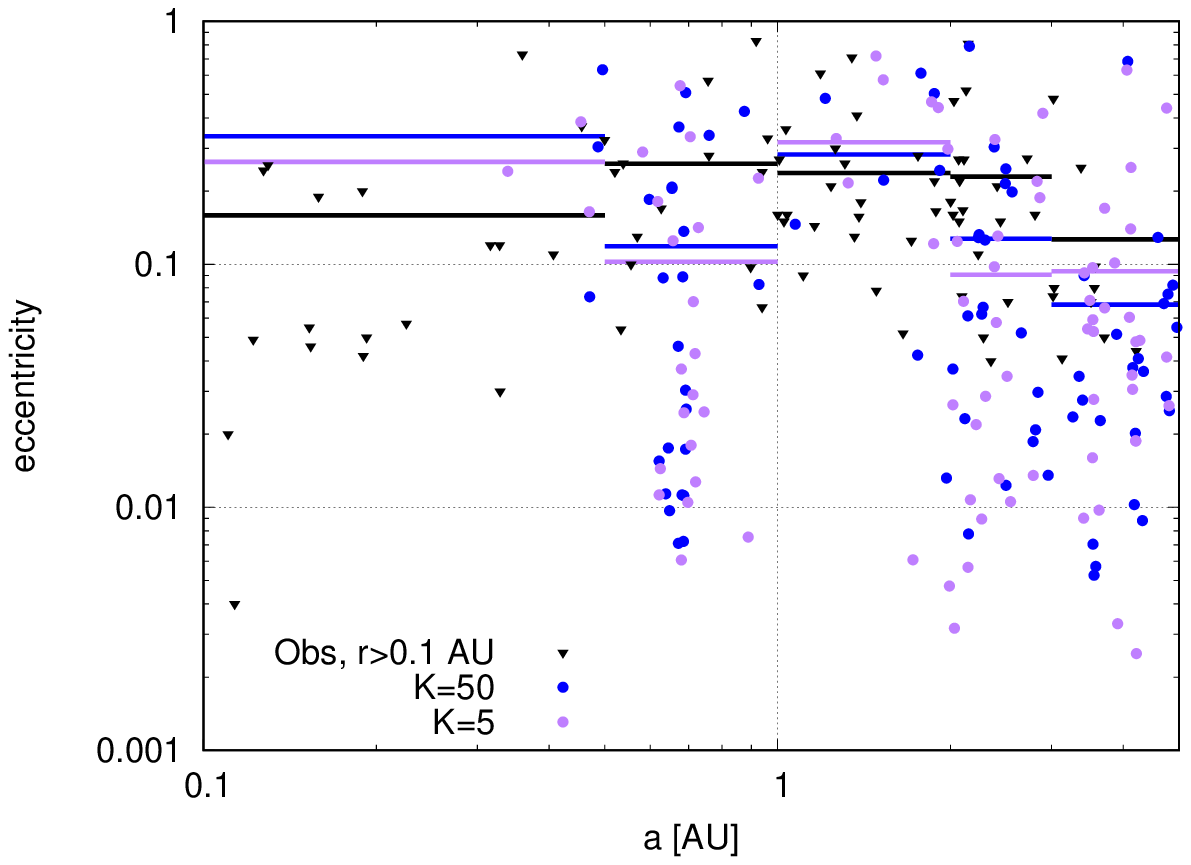} 
 \includegraphics[scale=0.67]{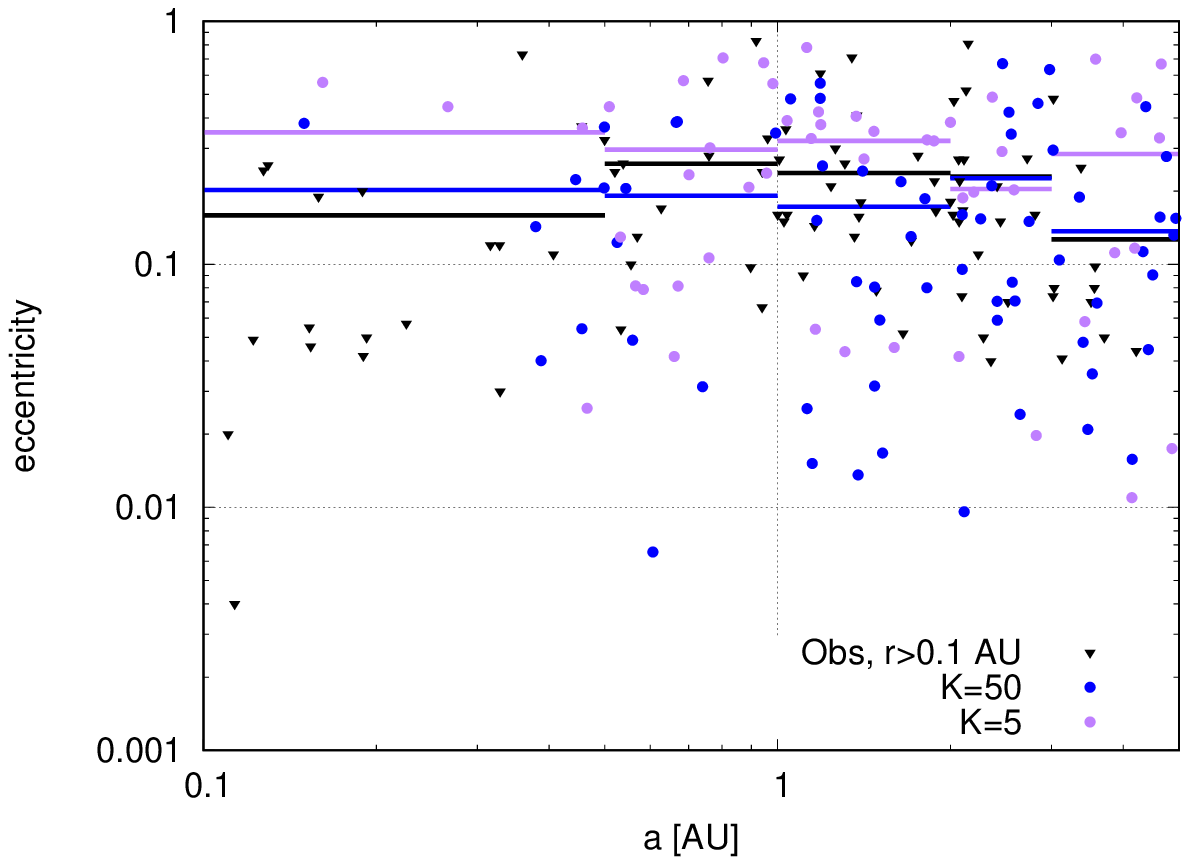} 
 \includegraphics[scale=0.67]{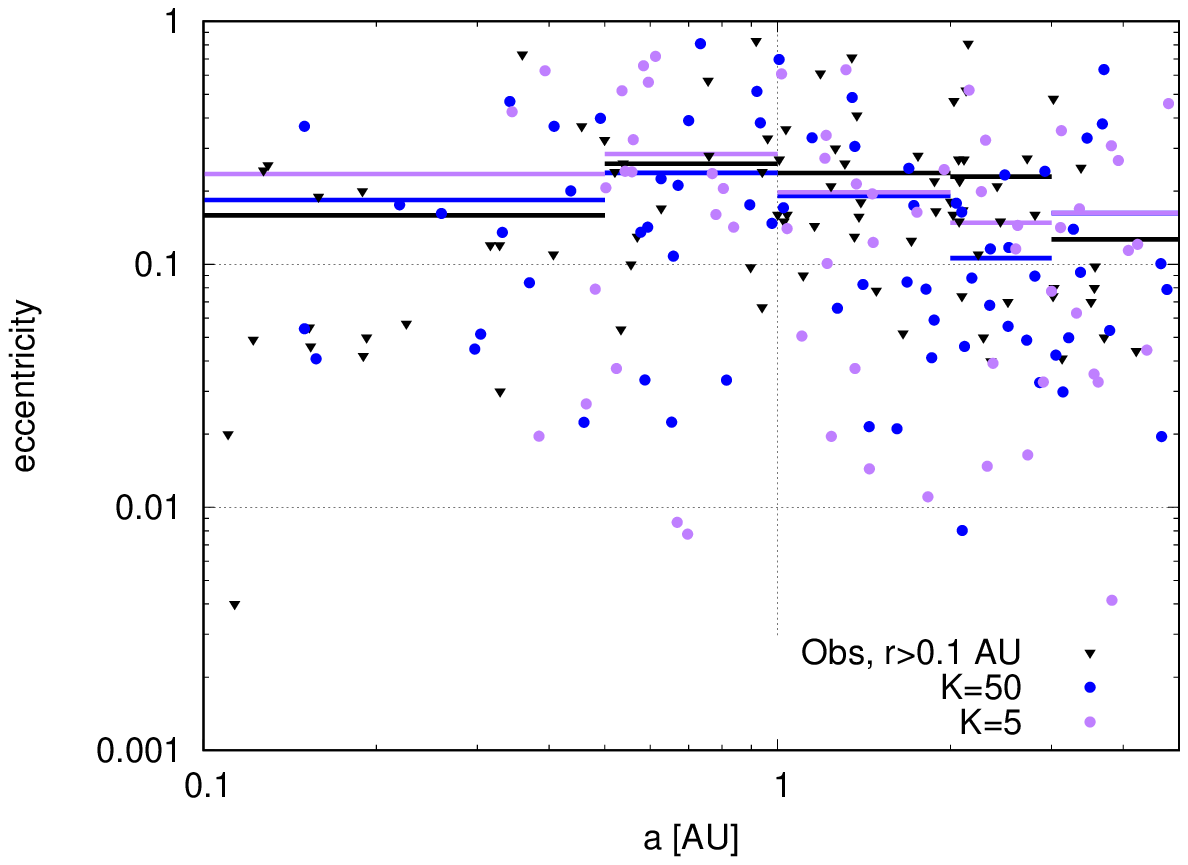}  
 \caption{Eccentricity as a function of semi-major axis of simulated planetary systems with different damping rates and different amount of initial embryos (15, 30, 60 from top to bottom). We only show the planets with masses of at least 0.5 Jupiter masses (for simulations and observations) and only for the $K$ values that match the observations of the eccentricity distribution best (Fig.~\ref{fig:ecchist}). The horizontal lines depict the mean eccentricity within each radial bin. The black symbols show the data from exoplanet.eu with planetary masses ranging from 0.5 to 1.25 Jupiter masses, which is roughly the masses the planets in our simulations reach.
   \label{fig:eccorg}
   }
\end{figure}

In Fig.~\ref{fig:eccorg} we show the eccentricity of the planets as function of their semi-major axis for the same set of simulations as in Fig.~\ref{fig:ecchist} and for the same observations. However, we limit ourselves to $K$=5 and $K$=50, because these simulations match the eccentricity distribution of the observations best. In addition, we show the average eccentricity for 5 semi-major axis bins.

%Curiously, our simulations show that for large damping values ($K$>500), the eccentricity of the giant planets from 0.1 to 0.5 AU can be larger than inferred from the observations. At larger orbital distances, the simulations with larger $K$ clearly fail to reproduce the eccentricity distribution. There also seems to be only minor differences in the mean eccentricity as function of orbital distance for simulations with initially a different number of planetary embryos.

Our simulations allow a quite nice match to the observed eccentricity distribution of the giant planets. For $K$=5, our simulations show a slightly too large frequency of giant planets with large eccentricity (see also Fig.~\ref{fig:ecchist}). From Fig.~\ref{fig:eccorg} it is clear that especially planets at large orbital distances ($r\geq$3 AU) have a larger mean eccentricity compared to the observations for 30 or 60 initial planetary embryos. In the case of 15 initial planetary embryos the mean eccentricity at these large distances is too low compared to the observations, which is related to the reduced number of scattering events for these simulations. 

Our simulations clearly show that either larger $K$ values are inappropriate (see also \citealt{2002ApJ...567..596L}) or the planetary masses in our simulations are not large enough to allow sufficient scattering to match the eccentricity distribution of the giant planets observed via RV detections. We relax the later assumption in the next section, where planets are allowed to grow faster and bigger. In addition, our simulations show that the initial number of planetary embryos does not play a significant role in the final number of giant planets in simulations using a slow damping rate (small $K$), but only seems to become slightly more important at fast damping rates (large $K$). We discuss the biases and limits of the observation of eccentricity giant planets in section~\ref{sec:discussion}.

\section{Fast growth}
\label{sec:growth}

\begin{figure*}
 \centering
 \includegraphics[scale=0.7]{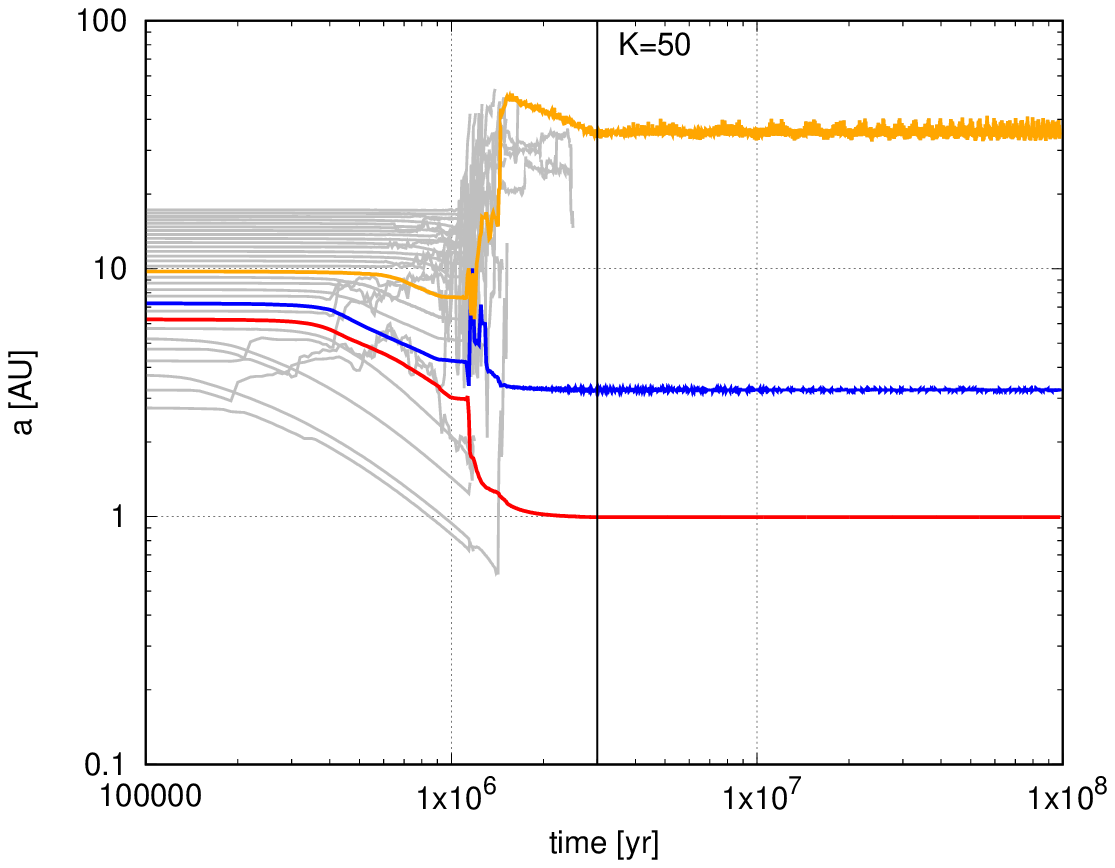}
 \includegraphics[scale=0.7]{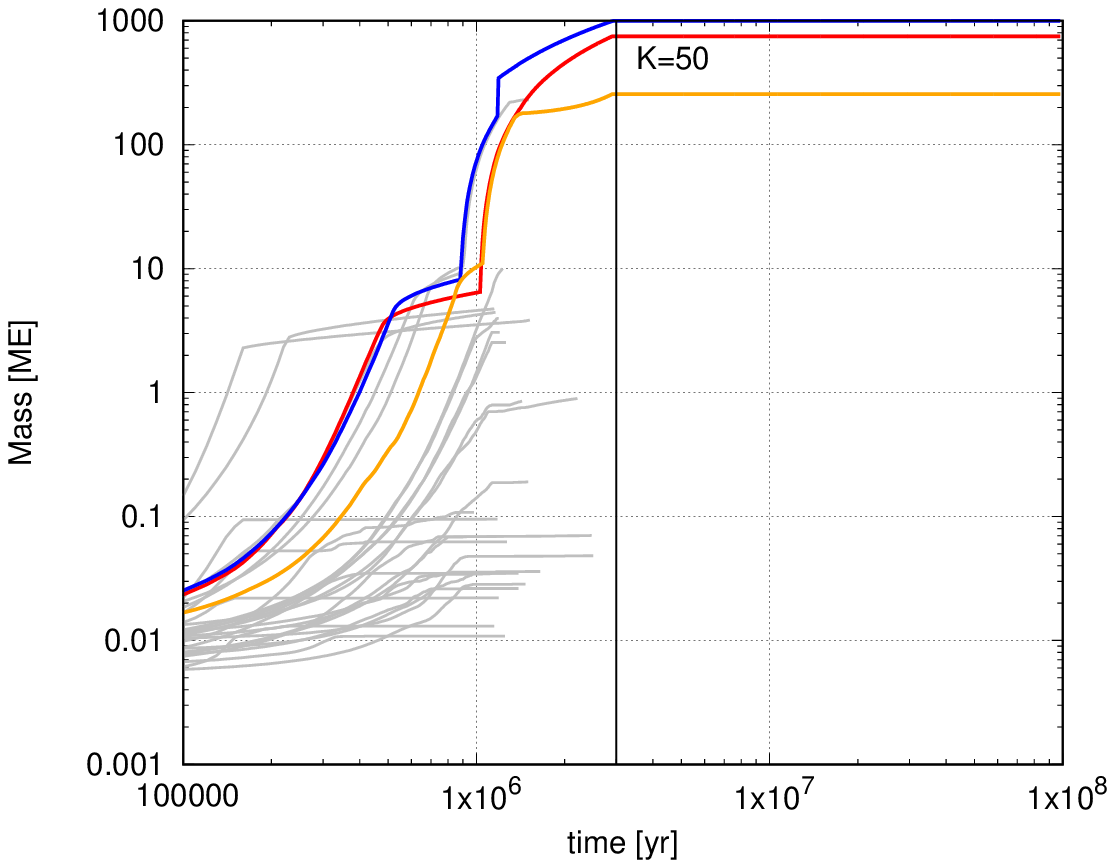}
 \includegraphics[scale=0.7]{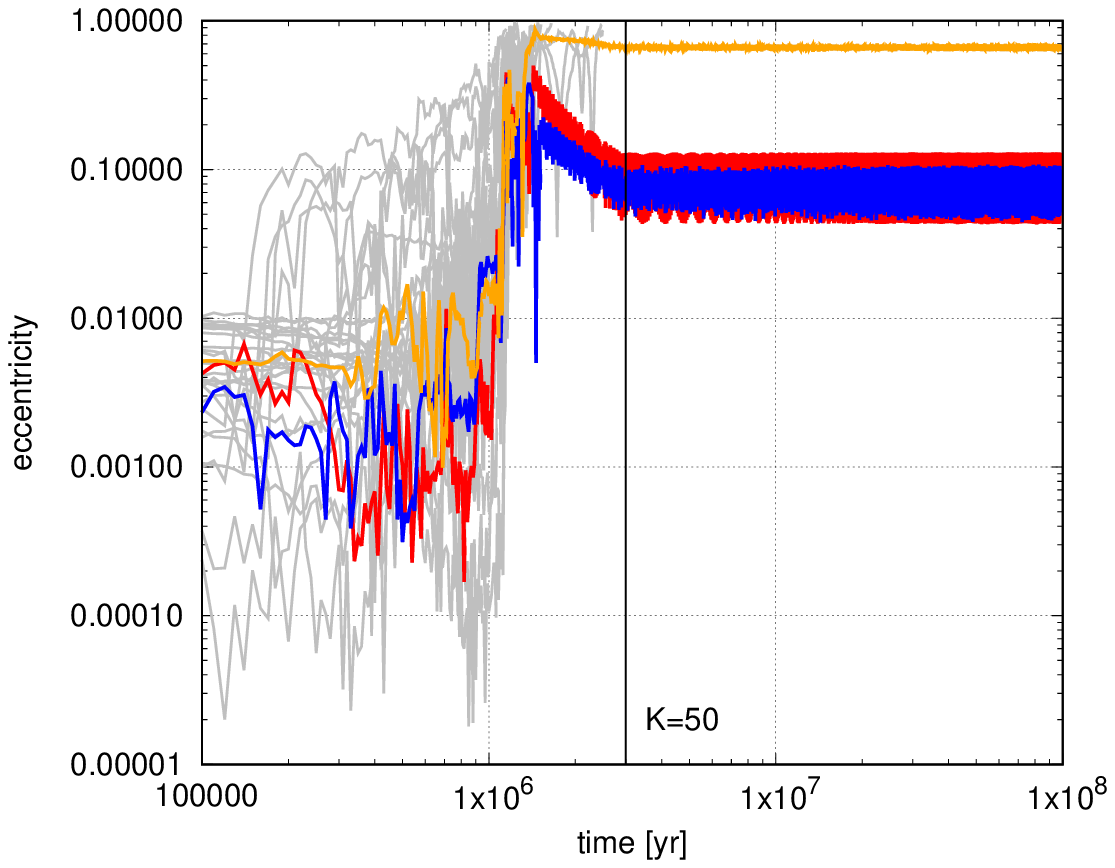}
 \includegraphics[scale=0.7]{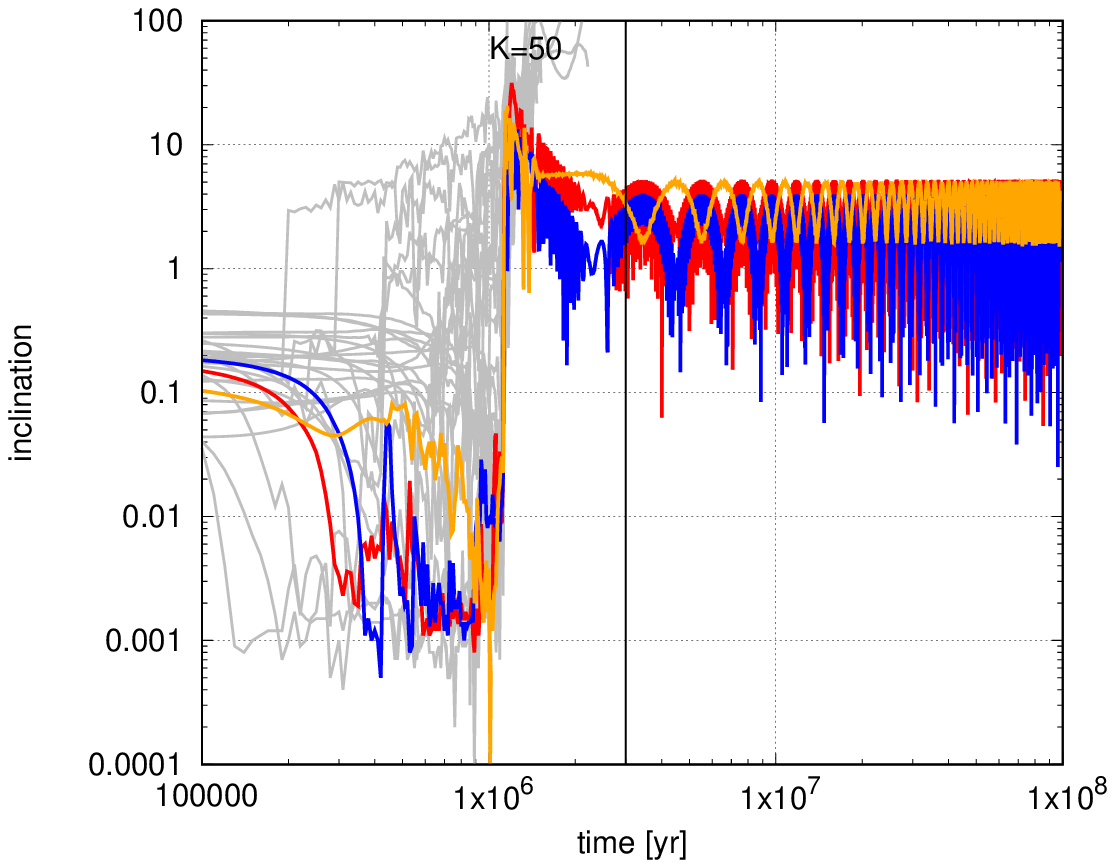}   
 \caption{Evolution of a system with a damping factor of K=50 where the pebble flux is twice as large as in Fig.~\ref{fig:30bodyK50}. The lines have the same meaning as in Fig.~\ref{fig:30bodyK50}. Due to the massive planets, an instability already occurs during the gas phase of the disc, increasing the eccentricity of the planets. However, during the remaining gas phase of 1.5 Myr after the instability, the eccentricities of the planets get damped efficiently and the two inner planets have eccentricities of around 0.1.
   \label{fig:30bodyK50growth}
   }
\end{figure*}

In this section we relax the previous restrictions to our simulations and allow planets to grow faster ($S_{\rm peb}$=5.0) and beyond 1 Jupiter mass by pure gas accretion. However, we limit ourselves here to simulations with $30$ initial planetary embryos, but vary the damping rates. We show the evolution of such a system in Fig.~\ref{fig:30bodyK50growth}.

By comparing the simulation shown in Fig.~\ref{fig:30bodyK50growth} with the simulation shown in Fig.~\ref{fig:30bodyK50}, a few differences become immediately visible. In the case of $S_{\rm peb}$=5.0, the planets accrete pebbles more efficiently and some planets reach pebble isolation mass well before 1 Myr. As the planets grow faster, they reach a larger pebble isolation mass \citep{2015A&A...582A.112B, 2019A&A...630A..51B}, which in turn allows them to contract their envelope in a shorter time. This result in the planets starting to reach runaway gas accretion at about 1 Myr. In contrast the planets growing in a disc with lower pebble flux ($S_{\rm peb}$=2.5) planets start their runaway gas accretion at $\approx$2 Myr (Fig.~\ref{fig:30bodyK50}).

The faster growth of the planets around 1 Myr of evolution results in stronger interactions between the planets, where the eccentricity is growing significantly. This effect is enhanced due to the slow damping ($K$=50) by the gas disc, which is too slow to keep the eccentricities small. As a consequence, the system undergoes an instability after about 1.3 Myr, where then only three giant planets survive and the remaining planetary embryos are ejected from the system already during the gas phase. 

After the scattering event, the eccentricities and inclinations of the remaining giant planets are damped by the interactions with the gas. This results in an eccentricity of around 0.05-0.1 for the inner two planets and of 0.65 for the outer giant planet. During the damping phase, the giant planets also continue to grow, where the inner planets reach around 3 Jupiter masses each and the outer planet is slightly below Jupiter mass. The outer planet grows very slowly, because of the low gas densities in the outer regions of the disc.

After the end of the gas disc phase, the eccentricity and inclinations of the inner planets oscillate, which has important consequences for the comparison with observations (see section~\ref{sec:discussion}). These oscillations are also very common in the simulations with $S_{\rm peb}$=2.5, independently of the initial number of embryos.

\begin{figure}
 \centering
 \includegraphics[scale=0.7]{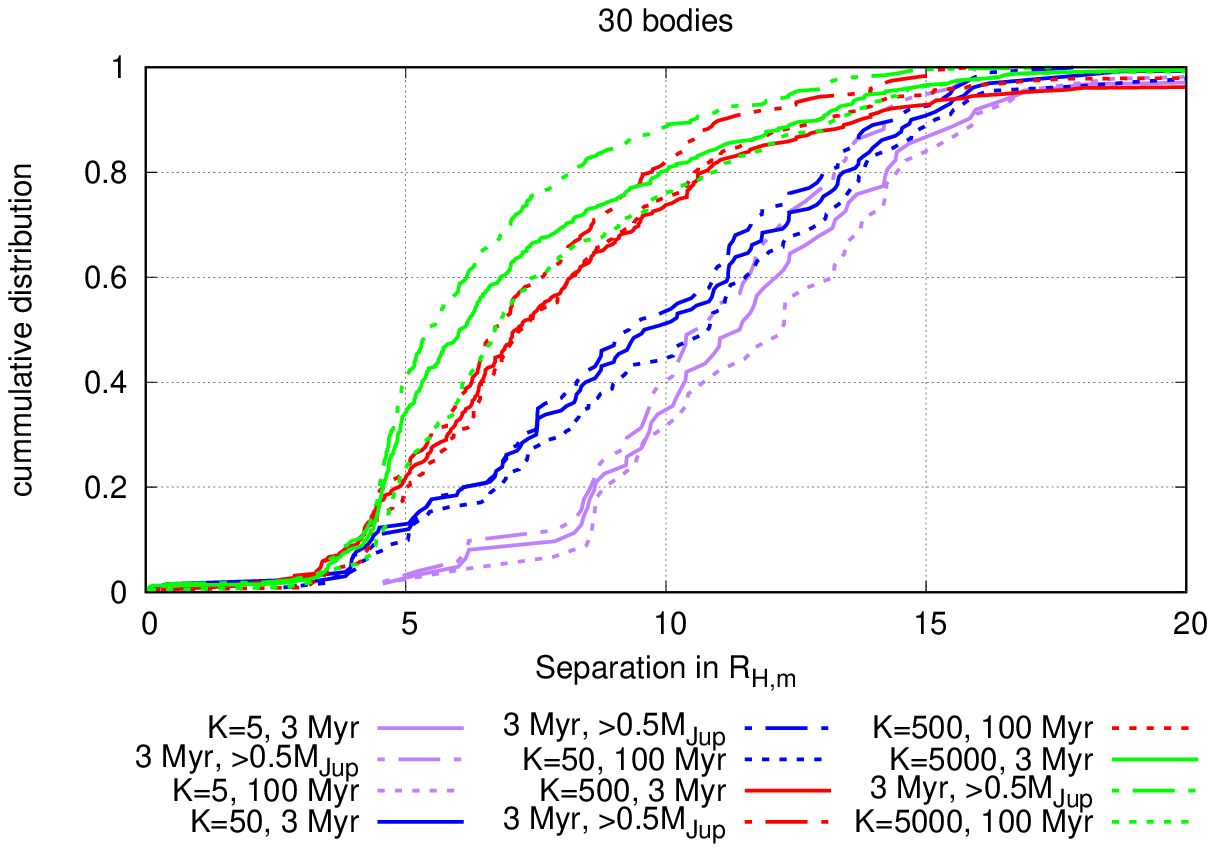}  
 \includegraphics[scale=0.7]{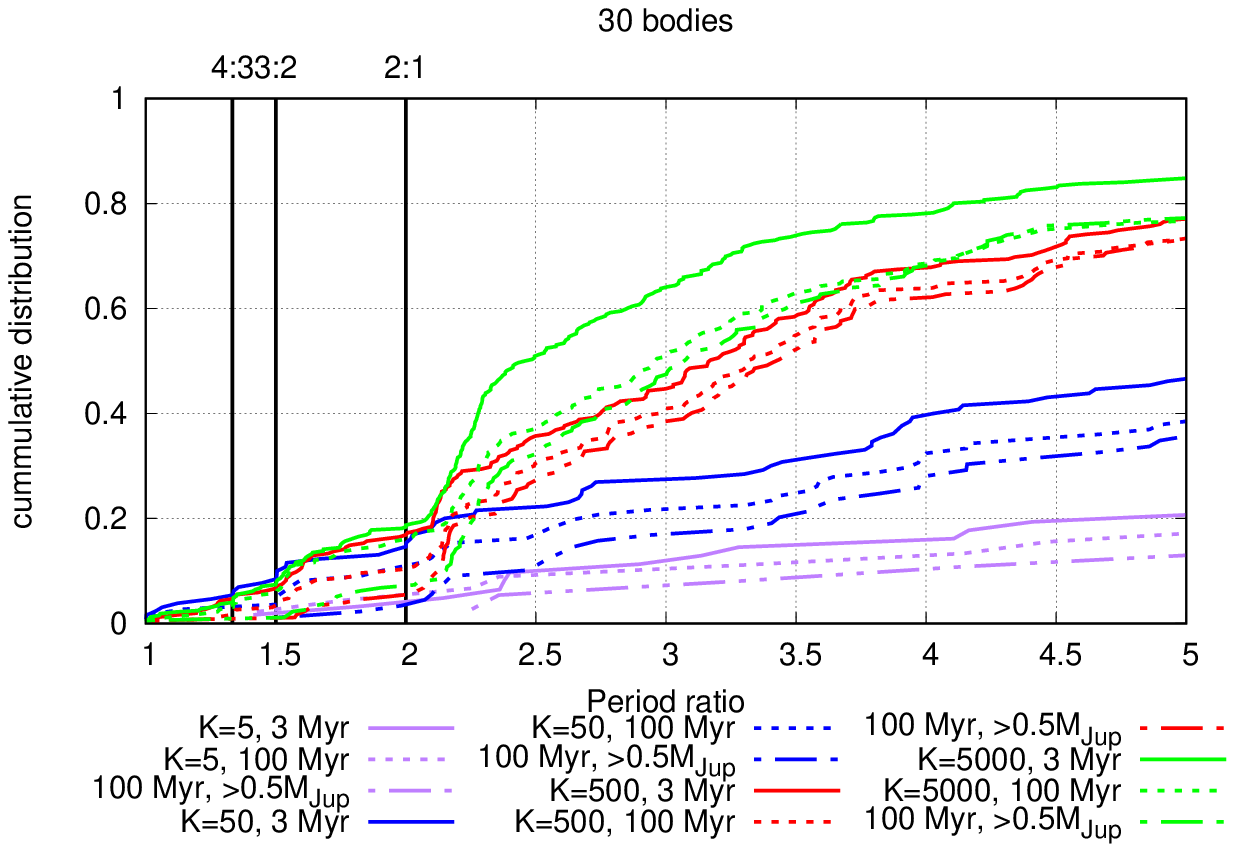} 
 \caption{Distances (top) and period ratios (bottom) of planets formed in our simulations with $S_{\rm peb}$=5.0 and different damping values. For this plot we cut the period ratio at 5.
   \label{fig:pdgrowth}
   }
\end{figure}

In Fig.~\ref{fig:pdgrowth} we show the orbital separation between adjacent planets (top) and their period ratios (bottom) for all our simulations with $S_{\rm peb}$=5.0. A clear difference with respect to the simulations with $S_{\rm peb}=2.5$ is visible (Fig.~\ref{fig:distance} and Fig.~\ref{fig:periods}). In the simulations with $S_{\rm peb}=5.0$ only a very tiny fraction of planets is closer than 5 mutual Hill radii at the end of the gas disc phase at 3 Myr, independently of the damping rate of eccentricity and inclination. As mentioned before, a distance of 4-5 mutual Hill radii was used in \citet{2009ApJ...699L..88R} to start out their simulations in order to achieve scattering events in short times. In addition, there is no large difference if only the separations between planets with masses larger than 0.5 Jupiter masses is taken into account.

Due to the more rapid growth, the planets grow quickly and become very massive, allowing instabilities to happen already during the gas disc phase. As a result the orbital separations between the planets are increased. The surviving giant planets migrate in the slow type-II migration regime (eq.~\ref{eq:migII}) preventing them to migrate close to each other. In particular, in the case of $K$=5, the instabilities during the gas disc phase are very common and 90\% of the planets have separations larger than 7 mutual Hill radii already towards the end of the gas disc phase. On the other hand, as already seen in the simulations with $S_{\rm peb}=2.5$, fast damping results in more tightly packed systems at the end of the gas disc phase, because the efficient damping reduces the eccentricities of the planets and thus allows planets to be close to each other\footnote{In the case of two planets, \citet{1996Icar..119..261C} showed that planetary systems with eccentric orbits need larger separations in units of the mutual Hill radius in order to remain stable.}. Here, the faster growth combined with the efficient damping ($K\geq$500) leads to very stable systems at wide separations.

As the majority of the instabilities for simulations with $K\leq$50 already happened during the gas disc phase, the mutual separation between the planets does not change significantly after 100 Myr of evolution. For the simulations with $K\geq$500, we also do not observe a large change in the separations between the planets after 100 Myr. This is probably caused by the fact that the planets are already further away that 5 mutual Hill radii from each other at the end of the gas disc phase, preventing efficient scattering on timescales of 100 Myr \citep{1996Icar..119..261C}.

This is also reflected in the period ratios between the planets (bottom in Fig.~\ref{fig:pdgrowth}), where the number of planetary systems that host planets interior to the 2:1 period ratio is very small, especially when compared to the simulations with $S_{\rm peb}$=2.5 (Fig.~\ref{fig:periods}). This is also caused by the fact that the planets grow faster and bigger, which reduces their migration speed due to the deeper planetary gap (eq.~\ref{eq:Kgapopen}), and thus prevents them to come close to each other.

In agreement with the separation in mutual Hill radii, the period ratios of the systems do not change significantly when including only planets with masses larger than 0.5 Jupiter masses or after 100 Myr. In Fig.~\ref{fig:pdgrowth} we plot only up to period ratios of 5:1 for visibility reasons. The period ratios between the planets are largest for the simulations featuring $K$=5. This is caused by the stronger scattering interactions between the planets, which separates the planets further and thus increases their period ratios.

In table~\ref{tab:average} we show the total number of giant planets with masses larger than 0.5 Jupiter masses in our simulations and the number of these giant planets that are detectable with RV measurements of 1m/s up to 5.2 AU is shown in table~\ref{tab:RVaverage}. In comparison to the simulations with $S_{\rm peb}$=2.5, the simulations with $S_{\rm peb}$=5.0 and large $K$ can harbor systems with more giant planets (see also table~\ref{tab:average}). This can be explained through a larger pebble flux through the disc, which causes the planets to grow faster. This implies that more planets start initially to grow and are thus more resistant to the instabilities that will later follow at the late stages of the gas disc phase where the small bodies are ejected from the system and the larger ones remain in the disc due to the efficient damping.

In the case of slow damping (small $K$), the average number of giant planets is quite similar for both pebble fluxes. This could imply that the growth does not necessarily play the major role in determining the final structure of the planetary system, but the damping rates of eccentricity and inclination once the planets become massive and open gaps. In the case of slow damping, we thus only observe small variations in the number of planets that could be found via RV detections for the different pebble fluxes (table~\ref{tab:RVaverage}).

%Regarding the planets in our simulations that could be found via RV detections (dashed lines in Fig.~\ref{fig:numbergrowth}), we observe a small difference between the simulations with different pebble fluxes (table~\ref{tab:RVaverage}). In the case of the high pebble flux, fewer planets are observeable with RV simulations compared to the planets formed in systems with low pebble flux and small $K$. This is related to the slow damping rates, which allow more interactions already during the gas disc phase, cleaning out the systems more efficiently. 

% \begin{figure}
%  \centering
%  \includegraphics[scale=0.7]{Distribution/Growth/plots/Number30growthbodies.eps}  
%  \caption{Number of giant planets with masses larger than 0.5 Jupiter masses in our simulations with $S_{\rm peb}$=5.0 after 100 Myr as well as those that could be detected by RV measurements with a distance up to 5.2 AU.
%    \label{fig:numbergrowth}
%    }
% \end{figure}

\begin{figure}
 \centering
 \includegraphics[scale=0.7]{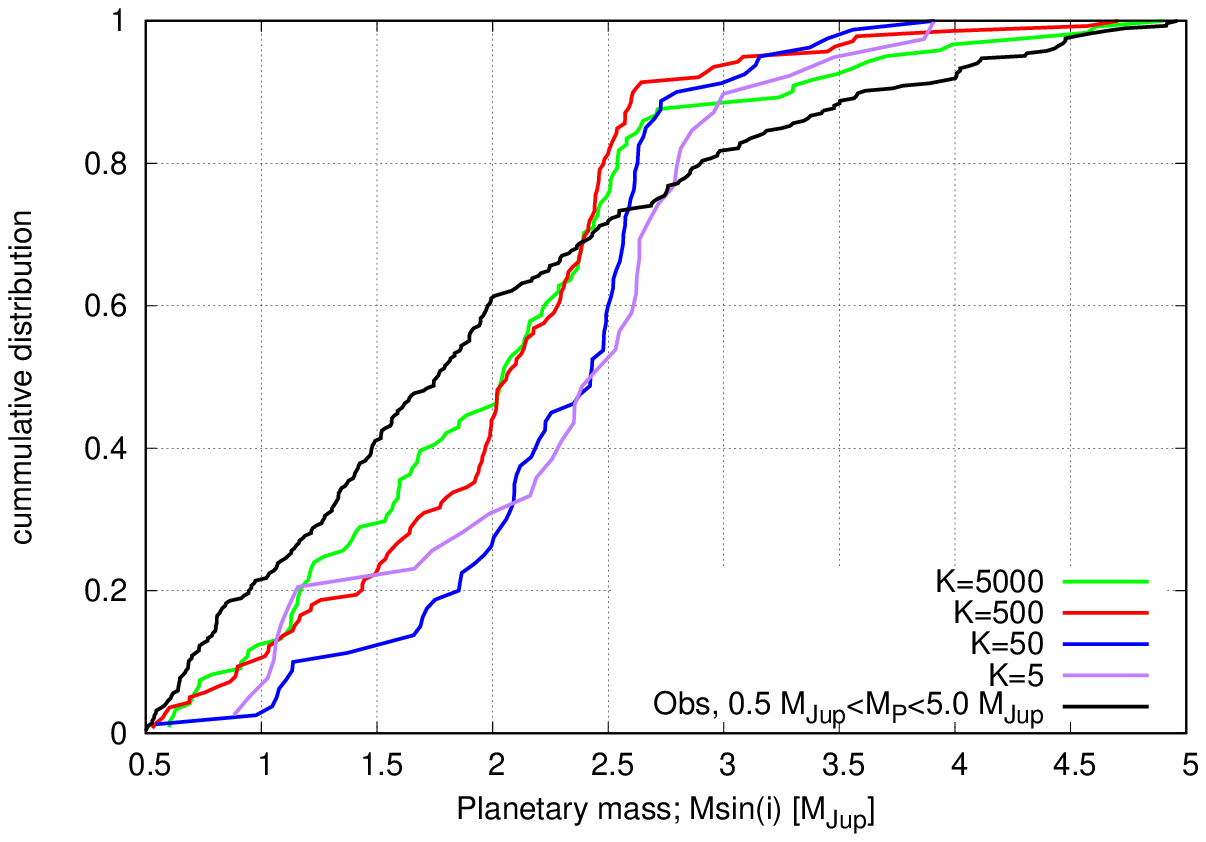}  
 \caption{Mass distribution of the planets formed with masses larger than 0.5 Jupiter masses in our simulations and from RV observations, where we limit the observations to 5.0 Jupiter masses, as this is roughly the largest masses we reach in our simulations.
   \label{fig:masses}
   }
\end{figure}

In Fig.~\ref{fig:masses} we show the mass distribution of the giant planets formed in our simulations using $S_{\rm peb}$=5.0 and those of the RV observations with masses larger than 0.5 Jupiter masses, but limited to 5 Jupiter masses. We chose this upper limit, because planets with larger masses could predominantly be formed via gravitational instabilities \citep{2018ApJ...853...37S}.

The giant planets in our simulations have an average mass larger than infered from the observations, indicating that the growth in our simulations might be too efficient compared to reality. In addition, the planets formed in the simulations with slow damping (low $K$) seem to be heavier than their counterparts formed in the simulations with fast damping (high $K$). This difference could originate from planet-planet collisions, which are more frequent in the simulations with slow damping. However, even though the mean mass is higher in our simulations compared to the observations, it seems that our simulations under produce planets with masses larger than 2.5-3 Jupiter masses. This could be related to the limited disc lifetime in our simulations, where the planets only grow 3 Myr in a gas disc environment, however, discs in reality can live longer \citep{2009AIPC.1158....3M}. Planets growing in discs that live longer have more time to accrete gas and could thus grow bigger. In addition, disc masses are quite widely spread \citep{2013ApJ...771..129A}, which would provide more material for the planets to grow. In addition, planets froming by gravitational instabilities could populate this area of parameter space \citep{2018ApJ...853...37S}.

\begin{figure}
 \centering
 \includegraphics[scale=0.7]{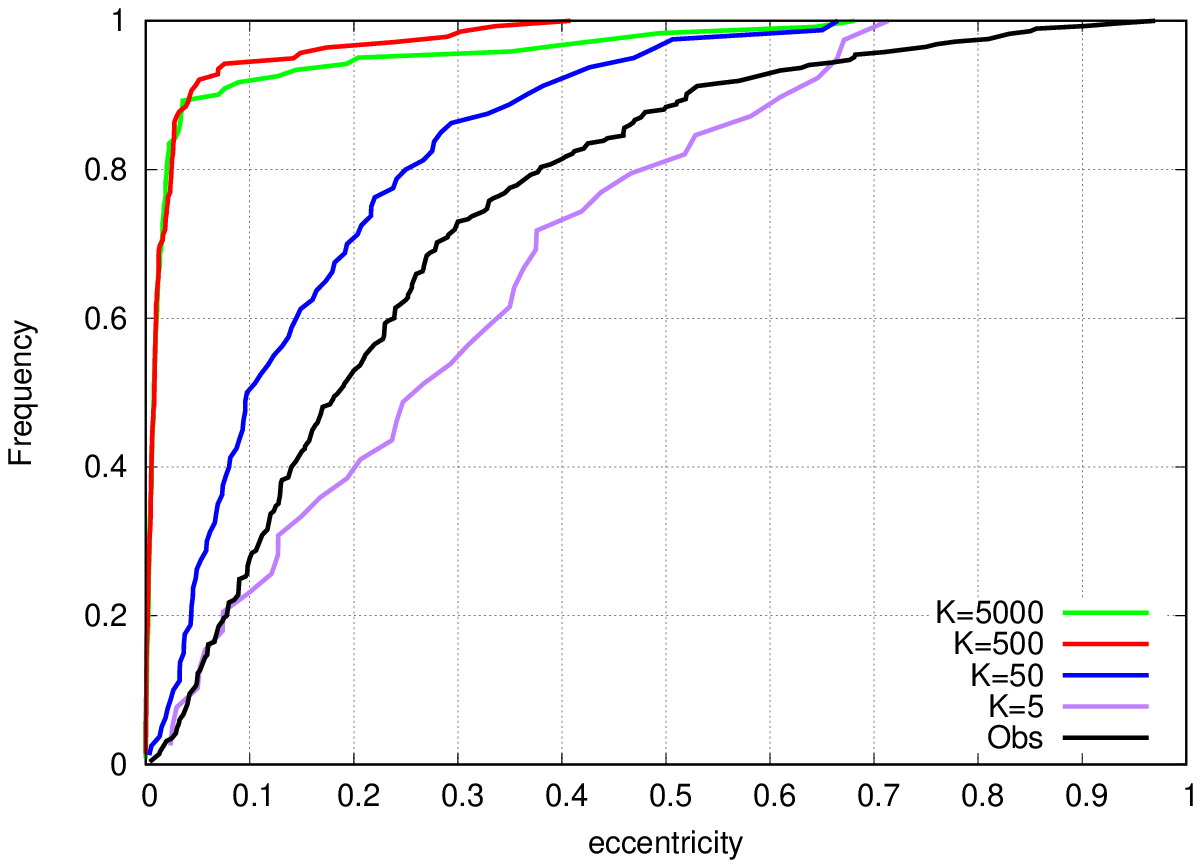} 
 \caption{Eccentricity distribution of the planets formed in our simulations with faster growth ($S_{\rm peb}$=5.0) and the observations with planetary masses from 0.5 to 5 Jupiter masses.
   \label{fig:eccgrowth}
   }
\end{figure}

In Fig.~\ref{fig:eccgrowth} we show the eccentricity distribution of the giant planets with masses larger than 0.5 Jupiter masses in our simulations with $S_{\rm peb}$=5.0 as well as of RV observations taken from exoplanet.eu with masses between 0.5 and 5.0 Jupiter masses. Compared to fig.~\ref{fig:ecchist} we now include planets with larger masses for the RV observations.

Clearly, a fast damping of eccentricity and inclination (large $K$) results in a very small number of instabilities and thus in a very steep eccentricity distribution, similar to the simulations with $S_{\rm peb}$=2.5. This clearly indicates that also planets with larger masses can not compensate for the efficient damping in this case. In particular, due to the larger masses, the planets migrate slower and have larger mutual distances at the end of the gas disc lifetime (Fig.~\ref{fig:pdgrowth}) compared to the planets formed in the simulations with $S_{\rm peb}$=2.5, which prevents most scattering events after the gas disc phase in the 100 Myr of system evolution, resulting in systems with basically no significant eccentricity.

On the other hand, the planets formed in simulations with slow damping (small $K$) show a significant eccentricity distribution. For $K$=50, our simulations slightly over predict the number of planets with $e<0.1$, match quite nicely for $0.1<e<0.3$, but under predict the number of planets at larger eccentricities in agreement with the simulations using $S_{\rm peb}$=2.5. However, the number of planets with $e<0.1$ is larger for $S_{\rm peb}$=5.0. We again attribute this to the faster growth, resulting in slower migration and thus in wider separation of the giant planets at the end of the gas disc phase (Fig.~\ref{fig:pdgrowth} and Fig.~\ref{fig:distance}), resulting in more stable systems.

But for $K$=5, our simulations show an under prediction of planets with eccentricities smaller than 0.3, but over predict planets with $e>0.3$ compared to the observations in line with the simulations using $S_{\rm peb}$=2.5. Even though the planets have wide separations at the end of the gas disc lifetime (Fig.~\ref{fig:pdgrowth}), they feature a significant eccentricity distribution, which is caused by the very inefficient damping, resulting in instabilities during the gas disc phase.

\begin{figure*}
 \centering
 \includegraphics[scale=1.4]{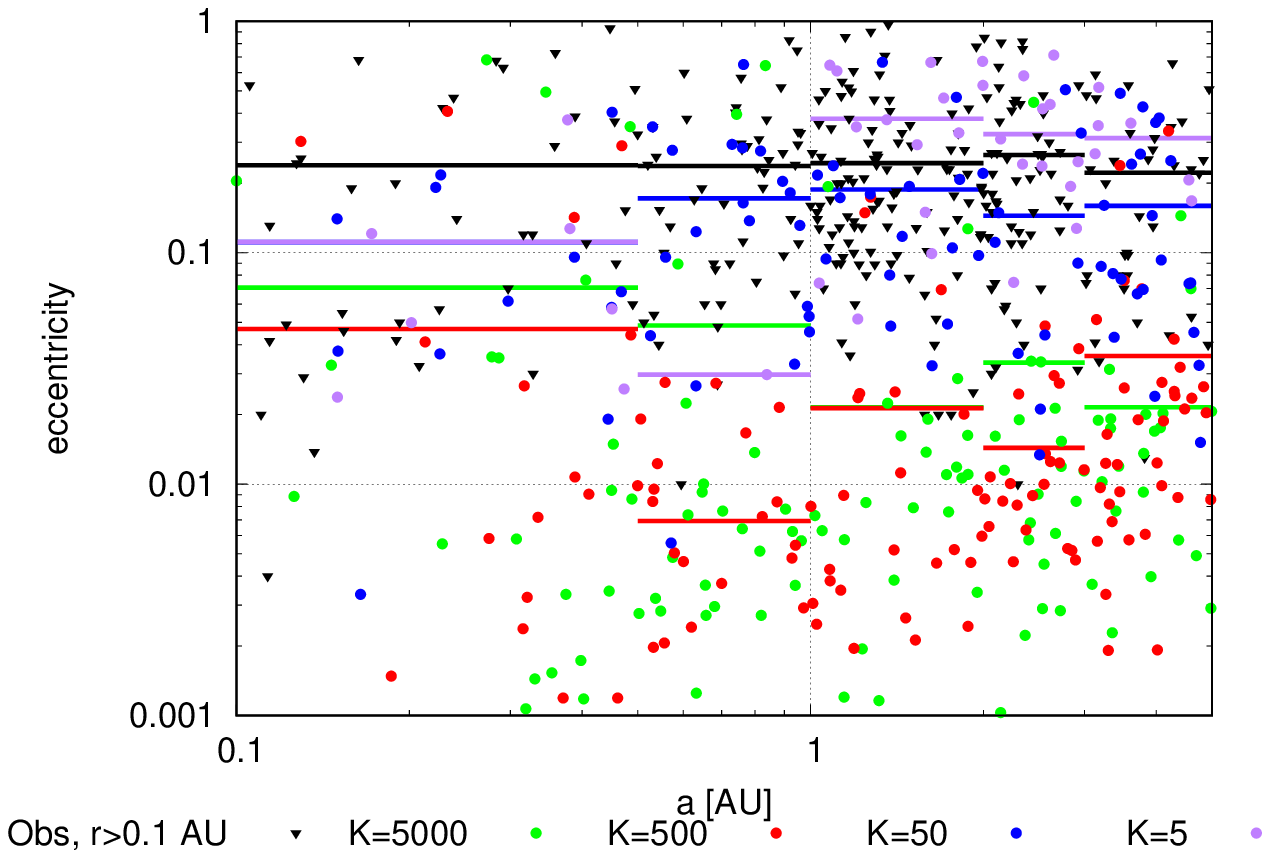} 
 \caption{Eccentricity - semi-major axis distribution of the observed giant planets (black triangles) with masses of 0.5 < M$_{\rm P}$<5.0 M$_{\rm J}$ and our simulations with different K-damping values (coloured circles) after 100 Myr of evolution. The horizontal lines show the mean eccentricity within each orbital distance bin. While our simulations with low K reproduce the e-a distribution exterior to 1 AU, the simulations show too small eccentricities for the planets interior to 1 AU. On the other hand, this might also be related to the limited number of giant planets within this region in our simulations. For planets in the order of Jupiter mass, the low K values match quite nicely (Fig.~\ref{fig:eccorg}), due to the slightly larger amount of planets within that semi-major axis bin in our simulations.
   \label{fig:eccagrowth}
   }
\end{figure*}

We show the eccentricities of the planets with masses larger than 0.5 Jupiter masses formed in our simulations as function of their orbital distance in Fig.~\ref{fig:eccagrowth}. We show also the eccentricities of giant planets of masses between 0.5 and 5.0 Jupiter masses detected via RV. In addition, we show the mean eccentricity for five orbital distance bins. As we include now more massive planets from the observations, the mean orbital eccentricity in the outermost bin increases significantly compared to Fig.~\ref{fig:eccorg}. This change in the mean eccentricities is caused by the larger sample of planets, which now include more massive planets. Our sample now includes 287 planets, in contrast to the 85 planets in the mass range of 0.5 to 1.25 Jupiter masses used for Fig.~\ref{fig:ecchist} and Fig.~\ref{fig:eccorg}. We discuss this more in section~\ref{sec:discussion}.

The differences compared to the simulations with $S_{\rm peb}$=2.5 can already be infered from our previous discussion. For the fast damping rates (large $K$), the eccentricities of the giant planets are lower compared to the simulations where planets grow with $S_{\rm peb}$=2.5 (Fig.~\ref{fig:eccorg}). It is very clear that these fast damping rates fail to reproduce the eccentricity distribution of giant planets infered from observations.

Our simulations suggest that a $K$ value between 5 and 50 is needed to match the eccentricity distribution of the observations exterior to 1 AU. This also becomes apparent from a K-S test (see table~\ref{tab:pvalue}), where only the simulations with low $K$ give a reasonable answer to match the observed eccentricity distribution. When mixing the simulations of K=5 and K=50 in a 3:1 fashion, we get a p-value of 0.39, indicating that the real $K$ for these kind of simulations should be in between 5 and 50. 

On the other hand, our simulations with $S_{\rm peb}$=5.0 clearly under predict the eccentricities interior of 1 AU. We attribute this to the low number of planets within 1 AU in our simulations. In fact the 50 simulations using $K$=5 only show one planet within 0.5 and 1 AU. On the other hand, the simulations with $S_{\rm peb}$=2.5 and $K$=5 show a nice match to the observations within 1 AU (Fig.~\ref{fig:eccorg}). 

In addition, the damping of eccentricity and inclination depends on the gap that giant planets open \citep{2001A&A...366..263P, 2006A&A...447..369K, 2013A&A...555A.124B}. The gap opening in turn depends on the disc's viscosity and aspect ratio \citep{2006Icar..181..587C, 2018arXiv180511101K}, which is smaller in the inner disc for flared disc profiles as we use here. This indicates that the faster growing planets will open their gap earlier, especially in the inner regions. As the type-II damping rate is slower than the type-I damping rate, this enhances the eccentricities of the growing planets, leading to more scattering events eventually depleting the number of giant planets in the inner regions for our simulations. On the other hand, the disc's viscosity is set by the disc's turbulence, which varies with orbital distance due to the operation of different instabilties (e.g. \citealt{2019ApJ...871..150P}), indicating that taking radially varrying profiles of viscosity into account could alter the damping rates and thus the interactions of growing planets.

It seems that the faster growth rate and thus more massive planets have only a significant impact on the eccentricity distribution if the damping of eccentricity and inclination is slow (low $K$). However, the overall evolution of the planetary systems is very different. In the case of slow growth, the instabilities happen towards the end of the disc lifetime, when the damping forces reduced due to the decay of the disc. On the other hand, in the case of fast growth, the instabilities can already happen in the middle of the gas disc lifetime, with important implications for the evolution of the system and how these systems could be observed, as we discuss in detail in Section~\ref{sec:discussion}.

%\begin{itemize}
 %\item Instabilities during the gas disc phase are common
 %\item leads to wider spacing and eccentricity distribution at the end of the gas disc lifetime compared to the other simulations
 %\item as the instabilities happen during the gas phase, the planetary orbits can be damped after the instability, if the damping is fast. Only for low damping values can some eccentricities remain.
 %\item include plot with mass distribution of planets
%\end{itemize}

\section{Super-Earths - cold Jupiter relation}
\label{sec:super}

In inner regions of planetary systems (within 1 AU), super-Earths seem to be the most common type of planets, where 30-50\% of all systems appear to harbour a super-Earth \citep{2018AJ....156...24M}. On the other hand, only 10-20\% of all planetary systems should host a cold gas giant beyond 1 AU. In classical formation scenarios, the super-Earths might be the failed cores that did not make it to become giant planets \citep{2014arXiv1407.6011C}.

RV Observations by \citet{2018arXiv180408329B} investigated if systems hosting cold gas giants also host inner super-Earths. However, within their sample of 20 systems they did not find any super-Earth, resulting in the conclusion that at maximum 10\% of cold Jupiter systems should harbour also inner super-Earths. However, \citet{2018arXiv180502660Z} and \citet{2018arXiv180608799B} concluded opposite, namely that cold Jupiters should have inner super-Earth in up to 90\% of the cases. The difference to \citet{2018arXiv180408329B} is explained in \citet{2018arXiv180502660Z} through the detection limits by the RV survey that can only observe planets down to 15 Earth masses, while the Kepler satellite, whose data was used in \citet{2018arXiv180502660Z} and \citet{2018arXiv180608799B}, can find much smaller objects.

The simulations by \citet{2015ApJ...800L..22I} showed that gas giants that are formed interior to super-Earth can act as barriers to the faster migrating super-Earths. This scenario is built on the assumption that gas giants interior to super-Earths exist. In the solar system context this could explain the formation of Uranus and Neptune via collisions between the outer super-Earth cores \citep{2015A&A...582A..99I}.

Here, our model shows the opposite formation pathway, namely that inner super-Earths form before the giant planets emerge from gas accretion (e.g. Fig~\ref{fig:30bodyK5000}) and interior to them. We now discuss within our simulations how many systems harbor outer giant planets and inner super-Earths.

We show in table~\ref{tab:superEarths} the fraction of systems from our simulations that harbour outer gas giants with inner super-Earths. We define here planets that are dominated in mass by their core as super-Earths in our simulations. Our simulations show that for the fast damping cases a significant fraction of giant planet systems should host inner super-Earths or if the initial number of planetary embryos is small. The number of systems hosting inner super-Earths and outer gas giants clearly decreases when the initial number of planetary embryos is larger. This is related to the dynamical history of the system. As discussed in the previous section, a faster damping rate (larger $K$) and a lower number of initial planetary embryos results in planetary systems that are less likely to undergo dynamical instabilities, which then keep the inner systems intact.

On the other hand, the simulations that show a 30-40\% occurrence of inner super-Earths and outer gas giants, feature 1.5-2.8 gas giants per system that could be detected via RV measurements (table~\ref{tab:RVaverage}). Nevertheless, most observed systems with inner super-Earths feature only one gas giant, indicating a bit of discrepancy between the observations and our models. Of course, some systems like HD 160691 and HD 34445 actually feature two gas giants with orbits larger than 1 AU and with inner super-Earths, but these seem to be the exception.

\begin{table*}
\centering
\begin{tabular}{c|c|c|c|c}
\hline
K & 15 seeds & 30 seeds & 60 seeds & 30 seeds, $S_{\rm peb}=5.0$ \\ \hline \hline
5 & 44\% & 8\% & 10\%  & 6\% \\
50 & 30\% & 18\% & 10\% & 10\% \\ 
500 & 36\% & 32\% & 18\% & 34\% \\ 
5000 & 34\% & 36\% & 16\% & 40\% \\ \hline
\end{tabular}
\caption[Super-Earths]{Fraction of systems that harbour inner super-Earths with outer giant planets. The first three sets of simulations feature $S_{\rm peb}=2.5$, while the last set of simulations features $S_{\rm peb}=5.0$. We define a super-Earth in this context as a planet that did not enter runaway gas accretion, so a planet that is dominated by solids.}
\label{tab:superEarths}
\end{table*}

Planetary systems formed in our simulations with slow damping (small $K$) only form a small fraction of systems with inner super-Earths and outer gas giants for the case of 30 or 60 initial planetary embryos, contradicting the super-Earth cold Jupiter relationship from observations. This is related to the dynamical instabilities that happen in these systems, where giant planets become eccentric and thus destroy the inner planetary systems, as has been shown also by pure N-body simulations \citep{2015ApJ...808...14M}. On the other hand, the simulations with slow damping reproduce the eccentricity distribution of the giant planets very well. It seemed that fast damping of eccentricity and inclination is favourable to explain the super-Earth cold Jupiter relation in our simulations, but these same simulations clearly fail to reproduce the eccentricity distribution of the giant planets.

We propose a few ways to solve this apparent mystery. In the planetary systems formed from simulations with $S_{\rm peb}$=5.0 and $K$=5 and $K$=50 (Fig.~\ref{fig:strucK5} and Fig.~\ref{fig:strucK50}), it is evident that systems of inner super-Earths mostly exist in systems where the outer gas giants have a very low eccentricity, implying that these systems have not undergone major scattering events. Our simulations feature also a significant fraction of warm Jupiters (r$<$1 AU), which could have a significant influence on the stability of the inner systems once scattering events take place.

The study by \citet{2018arXiv180502660Z} suggested that up to 90\% of the cold Jupiter population should have inner super Earths. However, our simulations show that violent scattering events are needed to explain the eccentricity distribution of the observed cold Jupiter population, but these scattering events destroy the inner planetary systems, so that giant planets on eccentric orbits should not harbour inner super-Earth systems. One exception in our simulations is run three in Fig.~\ref{fig:strucK5} which features an inner super-Earths and a giant planet with an eccentricity of around $\sim$0.6 at the end of the 100 Myr evolution.

\begin{figure}
 \centering
 \includegraphics[scale=0.7]{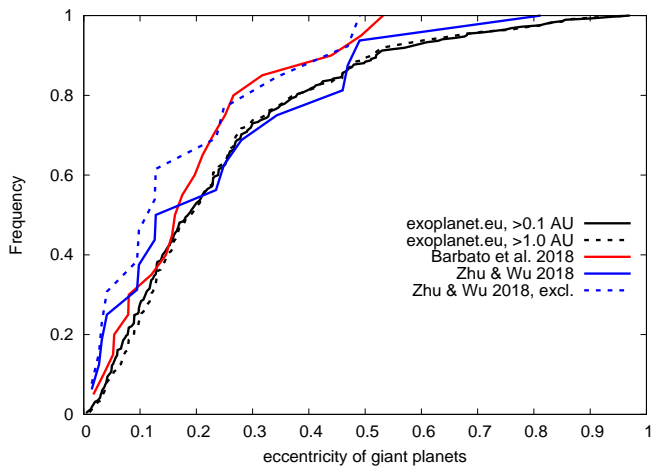} 
 \caption{Eccentricity distribution of the sample of exoplanet.eu with masses larger than 0.5 Jupiter masses where the planets have a minimal distance to their host star of 0.1 AU (solid black line) or 1.0 AU (dashed black line). In red we show the eccentricity distribution of the planets observed in \citet{2018arXiv180408329B}, while we show eccentricity of the giant planets used in the study of \citet{2018arXiv180502660Z} in blue. In the dashed blue solid line we also show the sample of \citet{2018arXiv180502660Z}, but exclude two planetary systems where the companions are more massive than 5 Jupiter masses, which could inherit their eccentricity from planet-disc interactions.
   \label{fig:SEhist}
   }
\end{figure}

We show the eccentricity distribution of the observed giant planets used in our study and those used by \citet{2018arXiv180408329B} and \citet{2018arXiv180502660Z} in Fig.~\ref{fig:SEhist}. It is clear that the eccentricity distribution of the giant planets in the sample of \citet{2018arXiv180502660Z} features eccentricities that are on average lower than the eccentricities of the giant planets extracted from exoplanet.eu, when taking their selection criteria (the planets should have larger distances than 1 AU to their central star) into account (dashed black line).

In addition, two systems, HD 219828 and HD 125612 (both within the sample of \citet{2018arXiv180502660Z}) feature highly eccentric giant planets with inner super Earths. The giant planets in these systems have several Jupiter masses. Massive planets like this can actually require their eccentricities through interactions with the protoplanetary disc itself \citep{2001A&A...366..263P, 2006A&A...447..369K, 2013A&A...555A.124B}, potentially explaining their eccentricities without scattering events and thus leaving their inner super-Earth systems undisturbed. Excluding these two systems from their sample leads to a much steeper eccentricity distribution in the \citet{2018arXiv180502660Z} sample. In this case (blue dashed line vs. black dashed line), there seems to be a factor of $\sim$2 difference in the frequency of giant planets with eccentricities below 0.1 compared to the observations, which could significantly influence the survival rate of inner super-Earth systems as our simulations show.

The data used by \citet{2018arXiv180502660Z} suggests that giant planets on eccentric orbits ($e>0.1$) do not harbour any super-Earth exterior to a 10 day orbit ($\approx 0.1$ AU), while giant planets on close to circular orbits can harbour super-Earths with much longer periods. Planets that are closer to the central star are deeper anchored within the stellar potential well and are thus harder to eject by exterior giant planets, making these planets prone to survive instabilities easier than their wider orbit companions. However, in our simulations the type-I migration speed is quite slow (due to the low surface density and low viscosity), so that super-Earths barely reach 0.1 AU, which could result in basically no survivors in the inner systems once the outer giant planet system becomes unstable. We plan to test this hypothesis in future work.

Of the 20 planetary systems observed in the work by \citet{2018arXiv180408329B}, only 6 systems feature gas giants with an eccentricity below 0.1, while 7 planets feature eccentricities larger than 0.2. Our formation simulations suggest that the search for inner super-Earth planets in the majority of the systems observed by \citet{2018arXiv180408329B} was doomed from the start for most of their systems, potentially explaining why they did not find any inner super-Earths or mini Neptunes.

Alternatively, the systems of inner super-Earths could form late and after the scattering events. Our simulations with $S_{\rm peb}$=5.0 and slow damping (low $K$) suggest that the scattering events happen early and can actually take place during the gas disc phase. If planetesimals survive in the inner disc\footnote{Pebbles would be blocked by the already existing giant planets \citep{2015Icar..258..418M, 2016Icar..267..368M, 2019arXiv190208694L}.}, these could form systems of inner super-Earths after the scattering events similar to the terrestrial planets in our solar system that finished their formation after the gas disc phase \citep{2009Icar..203..644R}. However, this scenario requires that enough material is present in the inner disc to allow efficient planet formation.

Nevertheless, the results of our simulations suggest that systems with giant planets on very eccentric orbits should harbor only very close-in super-Earths, if at all. In addition, the more massive the outer giant planet is\footnote{As long as the planet is not more massive than about 5 Jupiter masses, where planet-disc interactions can drive the planets eccentricity \citep{2001A&A...366..263P, 2013A&A...555A.124B}.}, the less likely it is that the inner super-Earth survives the scattering event that gives the outer planet its eccentricity. Reversely, the less massive and less eccentric an outer giant planet is, the more likely it is that the inner super-Earth system stayed intact. In addition, the close a giant planet is to the inner super-Earth region, the less like it is for super-Earths to exist within the same system (Fig.~\ref{fig:strucK5} and Fig.~\ref{fig:strucK50}). If the formation channel presented in this work (pebble accretion, gas accretion and planet migration of planetary embryos forming in the outer disc) is correct, it implies that searching for terrestrial planets in systems with outer gas giants will only be of success if the outer gas giants have very low eccentricities and that an analysis based on stability limits alone without taking the formation chanels into account \citep{2015ApJ...808...14M, 2018MNRAS.477.3646A, 2020MNRAS.492..352K} might be misleading.

Our simulations with surviving inner super-Earths and outer gas giants harbour a maximum of three inner super-Earths in the case of slow damping, but can have up to 4-5 super-Earth in case of faster damping of eccentricity and inclination for giant planets. Chains of more inner super-Earths did not form within our simulations. However, in the simulations of \citet{2019arXiv190208772I} chains with up to 9 super-Earths can form in the inner disc, indicating that systems with many super-Earths should not harbor outer giant planets, if both (super-Earths and gas giants) form in the outer disc and migrate inwards.

\section{Discussion}
\label{sec:discussion}

In this section we discus the shortcomings of the eccentricity observations of giant planets as well as of our model. In addition, we relate our results to previous works and discuss about the implications of our results for protoplanetary disc and exoplanet observations.

\subsection{Eccentricity distribution of observations}

Determining the true eccentricity of giant planets from RV observations is quite difficult. From the orbital analysis of RV measurements one can determine the minimum planetary mass, the semi-major axis and the eccentricity. The success of this approach depends on the RV precision, temporal baseline of the data, their number, and also on the magnitude of the Doppler signal induced by the planet.

In particular, if only a few RV measurements have been taken, the orbital fit might result in a too large eccentricity of the planet. By taking more RV measurements of already known planetary systems, the orbital fit can be improved. It has been shown that the RV signal of single eccentric planets with only a few measurements could actually be caused by two giant planets in resonance \citep{2019MNRAS.484.4230W, 2019MNRAS.484.5859W}. This could apply to up to $30\%$ of the giant planet systems detected via radial velocities, where the single planet has an eccentricity below 0.5. However, our simulations do not show a significant pile-up of giant planets close to first order mean motion resonances (Fig.~\ref{fig:periods} and Fig.~\ref{fig:pdgrowth}). Nevertheless, the impact on the overall eccentricity distribution of exoplanets might be small, but would imply that slightly faster damping rates (slightly larger $K$ values) might be needed to model the eccentricity distribution of giant planets.

In our simulations, the eccentricity of the giant planets are caused by scattering events either during or after the gas disc phase, which was also observed in hydrodynamical simulations \citep{2013MNRAS.431.3494L}. However, if more than one giant planet survives, the remaining giant planets can still interact, resulting in a variation of the planetary eccentricity and inclination in time (e.g. Fig.~\ref{fig:30bodyK50}). The oscillations happen normally on times of several thousand (or much more) orbits. On the other hand, the observations of seemingly single giant exoplanets span a maximum of a few orbits, with usually no observational evidence of additional longer-period planets that may be part of the system. The orbital parameter uncertainties, together with the unknown physical and orbital properties of the hypothetical companions, make it impossible to estimate the magnitude of the mutual orbital perturbations, and thereafter, impossible to determine if the measured eccentricity of an the observed giant planet is currently on its minimal, maximal, or on any other phase of the eccentricity evolution.

\subsection{Model assumptions}

Our model of planet formation is based on pebble accretion, gas accretion, protoplanetary disc evolution as well as on planet migration. In addition, our model assumes that the planetary embryos are already fully formed at the beginning of the simulations. All these ingredients have many assumptions flowing into it, which we briefly discuss here. We do not attempt to vary all parameters in our model, because our work is a proof of concept and the here presented simulations are designed to form giant planets.

In our simulations we have used the disc model of \citet{2015A&A...575A..28B}, where the thermal structure is derived from just micrometer sized grains. In reality, these grains can grow through coagulation \citep{2008A&A...480..859B} and condensation \citep{2013A&A...552A.137R, 2019A&A...629A..65R}, while their maximal size is limited by fragmentation, leading to a size distribution of these grains. The first models of hydrodynamical simulations including a full grain size distribution to calculate the heating/cooling effects have just become available \citep{2020arXiv200514097S} and will be used in future simulations. We start our N-body simulations in a disc that is already two Myr old \citet{2015A&A...575A..28B}. We chose this approach to be consistent with our previous work \citep{2019A&A...623A..88B}. The disc's aspect ratio in the inner disc reduces quite fast due to the reduction in viscous heating, resulting in a small pebble isolation mass in the inner disc in line with the masses of super Earths \citep{2019ApJ...874...91W, 2019A&A...630A..51B}.

Recent disc evolution models including the effects of disc winds, show different radial profiles, where the gas surface density can generate pile-ups around 1 AU \citep{2016arXiv160900437S, 2016ApJ...821...80B, 2019ApJ...879...98C}. This could prevent the inward migration of planetary cores, generating a pile-up of planetary cores that then undergo runaway gas accretion in the outer regions of the disc. As these gas giants are in the outer disc, their scattering events could disturb the inner super-Earth population less compared to gas giants forming in the inner disc. A similar effect could be achieved by photoevaporation, which can carve a hole in the disc around 1 AU, preventing the inward migration of giant planets \citep{2012MNRAS.422L..82A}. These effects should be investigated in future simulations.

In the disc model of \citet{2015A&A...575A..28B}, the $\alpha$ viscosity parameter is set to 0.0054 to calculate the thermal structure. Here we keep the same disc structure, but use a reduced $\alpha$ value for the planet migration, $\alpha_{\rm mig}=10^{-4}$ as in our previous simulations. We motivate this choice by the theory that the accretion of discs is driven by winds on the surface, which transport the angular momentum. The gas falls in and is accreted efficiently onto the star, while the bulk viscosity in the disc midplane remains low \citep{2016arXiv160900437S}. The value of $\alpha_{\rm mig}$ is in line with observations of dust settling towards the midplane \citep{2016ApJ...816...25P, 2018ApJ...869L..46D}.

The low viscosity used for planet migration, results in a slow inward migration of the growing planets. Larger values of viscosity, on the other hand, would result in faster inward migration, because the growing planets transition later to the slow type-II migration resulting in more inward migration, which is probably not in line with the structure of the solar system \citep{2019A&A...623A..88B}.

In the classical $K$-damping prescriptions \citep{2002ApJ...567..596L}, $K \propto 1/h$. This indicates that our simulations that match the eccentricity distribution of the giant planets imply in this simplistic way a different disc structure compared to the simulations that show a strong correlation between inner super-Earths and outer gas giants. This aspect deserves further investigation especially under the aspect that disc masses and thus disc structures can vary a lot from star-to-star \citep{2013ApJ...771..129A} and could even invoke differences in the super-Earth systems found by Kepler \citep{2020arXiv200308431K}.

In our simulations, the planets grow by pebble accretion. We use here a model where the planets only accrete the dominating grain size (in the drift limit), without modelling a full size distribution. However, the majority of the mass of a size distribution of pebbles is in the large grains, which is the sizes the planets accrete. In addition, our pebble flux model is based on an equilibrium between growth and radial drift, where the pebble flux reduces as the accretion rate of the disc reduces in time as well. Reducing the accretion rate all over the disc, as implied by $\dot{M}$ disc models, results in an decrease in the gas surface density and with it automatically in a reduction of the solid density (which is bound to the gas surface density through the dust-to-gas ratio). This reduction, however, results in pebble densities in the outer disc that are too low compared to observations \citep{2018A&A...609C...2B}, which is why we increase the pebble flux $S_{\rm peb}$ by either 2.5 or 5.0 in our model, as in our previous works \citep{2019A&A...623A..88B, 2019arXiv190208772I}. This increase in the pebble flux is artificial, but the resulting pebble densities are in agreement with the observations. In addition, we do not aim here to study planet population synthesis model, where simulation parameters have to be as self consistent as possible, but investigate a simpler question regarding the origin of the eccentricities of giant planets.

In our simulations we start with an initial distribution of planetary embryos between 3 and 18 AU, where the planetary embryos have initially all around 0.01 Earth masses. Planetesimals formed by the streaming instability, however, are only around 100 km in size \citep{Johansen2015, 2016ApJ...822...55S, 2018ApJ...861...47S}, much smaller than 0.01 Earth masses. At these sizes pebble accretion is also very inefficient \citep{2016A&A...586A..66V, Johansen2017}, so these planetesimals need to collide first to form bigger objects until pebble accretion can take over and becomes more efficient than planetesimal accretion, especially exterior to 1 AU \citep{2019A&A...631A..70J, 2020arXiv200403492V}. This growth phase by planetesimals depends crucially on the amount of available planetesimals, so the planetesimal surface density. How and where planetesimals in discs form is still under debate, where some ideas suggest that the water ice line is a favourable position \citep{2016ApJ...828L...2A, 2017A&A...608A..92D, 2016A&A...596L...3I}, while other simulations are based on the concept that planetesimals form in vorticies that can exist all over the disc \citep{2019ApJ...874...36L}. These different ideas lead to a different initial distribution of planetesimals and thus planetary embryos \citep{2020arXiv200403492V}. Future models of planet formation should include this step self consistently.

In our model, the gas accretion process is modelled in two steps. First we follow a contraction phase of the planetary envelope, following the approach of \citep{2014ApJ...786...21P}, which is based on \citet{2000ApJ...537.1013I}, where the planet then enters the runaway accretion phase once the planetary envelope becomes as massive as the planetary core. The contraction phase then basically decides the fate of the planet. Planets with slow contracting envelopes remain small, like super-Earths or mini-Neptunes, while planets with fast contracting envelopes can grow to gas giants. This contraction phase depends mostly on the planetary mass itself and the opacity in its envelope. We use here a constant opacity for the envelope following \citet{2008Icar..194..368M}. We do not investigate here different envelope opacities, as the goal of our simulations is to generate giant planets and to study their dynamical interactions. However, our simulations show a very large abundance of warm Jupiters (r$<$1 AU), which could interfere with the survival of super-Earths. Future simulations with less efficient gas accretion might shine light on this relation.

\subsection{Previous simulations}

Since the first giant exoplanets on eccentric orbits have been discovered, scattering among a few of these objects was proposed as an explanation for their orbital properties \citep{2001Icar..150..303F, 2005Natur.434..873F, 2008ApJ...686..603J}. Many simulations since then have been undertaken to study how the eccentricity distribution could be originating from scattering events \citep{2009ApJ...699L..88R, 2017A&A...598A..70S}.

The simulations by \citet{2009ApJ...699L..88R} start with three giant planets with an initial separation of 4-5 mutual Hill radii. These small distances between the planets lead very soon to interactions, resulting in scattering events. The resulting eccentricity distribution matched the observations of exoplanets quite nicely. In addition, \citet{2009ApJ...699L..88R} include the effects of an outer planetesimal disc, which can induce scattering, similar to what is proposed to have happened in the solar system \citep{2005Natur.435..459T}.

\citet{2017A&A...598A..70S} improved on this concept a step further, by including the type-II migration phase of the giant planets during the gas disc. In their simulations, the giant planet were also already fully formed. In addition, they kept the innermost of their three giant planet fixed, which allowed the outer planets to migrate closer, resulting in recreating the initial conditions of \citet{2009ApJ...699L..88R} and thus also the eccentricity distribution of the giant planets.

In our simulations, we include also the growth phase of the planet from a planetary embryo all the way to a gas giant. In addition, our simulations include also self consistent the formation of inner super-Earths, not included in the previous simulations.

\subsection{Resonances of giant planets}

The growth of planets by pebble accretion happens inside-out in our simulations (e.g. Fig.~\ref{fig:30bodyK50}), where planets in the inner disc reach the pebble isolation mass first. These planets can then start to contract their gaseous envelope and become gas giants. Once these planets reach masses of several 10 Earth masses, they start to open gaps in the disc and eventually migrate in the slower type-II migration regime. Planets starting further out, grow slower and are thus longer in the faster type-I migration regime. This faster type-I migration allows them to catch up to the inner, slower migrating giant planets.
This behaviour is a typical outcome of planet migration simulations \citep{2001MNRAS.320L..55M}, if the outer planet is smaller than the inner planet and can lead to planets getting trapped into a resonance configuration. This has also been applied to the solar system either as starting conditions for the Nice-model \citep{2005Natur.435..459T} or also as an ingredient of the Grand Tack model\footnote{There it is actually assumed that Saturn migrates in type-III migration \citep{2003ApJ...588..494M} instead of type-I migration.} \citep{2011Natur.475..206W}, where the giant planets then migrate outwards.

However, this migration behaviour does not necessarily need to result in capture in resonance \citep{2014ApJ...795L..11P}, but can also lead to instabilities during the end of the gas phase (e.g. Fig.~\ref{fig:30bodyK5000} and \citealt{2013MNRAS.431.3494L}). In most of our simulations, if the outer planets are in resonance configuration in the outer disc, these resonances of giant planets are broken when the system becomes dynamically unstable. The instability could also be aided by the growth of the planets in resonance, which can destabilise the system \citep{2020ApJ...893...43M}. In systems that do not undergo any instability, these resonant configurations are maintained (e.g. system 5 in Fig.~\ref{fig:strucK50}, where the outer three planets are in 3:2 and 2:1 resonance configuration). However, the majority of the systems in our simulations do not show resonant configuration of the outer giant planets.

\subsection{Metallicity - eccentricity correlation}

Recent observations have revealed that the eccentricity of cold gas giants is related to the host star metallicity \citep{2013ApJ...767L..24D, 2018arXiv180206794B}. This can be explained by the fact that giant planets are more abundant around metal rich stars\footnote{In the core accretion scenario this is explained by the fact that more metal rich stars should have more planetary building blocks which makes the formation of giant planets easier.} \citep{2004A&A...415.1153S, 2005ApJ...622.1102F, J2010, 2018AJ....156..221N}, which enhances the probability that these giant planets scatter after the gas disc dissipated, resulting in eccentric giant planets. On the other hand, if the host star is less metal rich, giant planet formation is hindered and maybe only one gas giant might form, which will then remain probably on a very circular orbit, due to the lack of partners to scatter with.

In this work we have tested the outcome of our simulations with two different pebble fluxes and three different number of initial planetary embryos, which can function as a proxy of the host star metallicity. The final eccentricities of the giant planets formed in the low pebble flux simulations and for the same initial number of embryos are generally a bit lower (Fig.~\ref{fig:ecchist} and Fig.~\ref{fig:eccorg}) than in the simulations with larger pebble flux (Fig.~\ref{fig:eccgrowth} and Fig.~\ref{fig:eccagrowth}). In addition, the simulations with less initial embryos show a steeper eccentricity distribution, implying that not enough planets are available to lead to massive scattering events (Fig.~\ref{fig:ecchist}). However the largest influence on the eccentricities of the giant planets originates from the different damping rates of eccentricity and inclination. In general the trend in our simulations thus seem to confirm the idea that scattering is mostly responsible to explain the different eccentricities of giant planets around stars with different metallicities.

\subsection{Hot Jupiters}

The first exoplanet found around a main sequence star was a Jupiter type planet orbiting its host star in just a few days \citep{1995Natur.378..355M}. These planets were named hot Jupiters and even though they are very easy to detect, they should only exist around 1-2\% of stars \citep{2011arXiv1109.2497M}. In addition, the host star metallicity, which is a proxy for the amount of building blocks available to form planets, is similar to the host star metallicity of giant planets on eccentric orbits \citep{2018arXiv180206794B}, hinting potentially at a common origin of these planets. 

In our simulations, the inner edge of the disc is at 0.1 AU, corresponding roughly to a 10 day orbit. In our simulations only less than 1\% of the giant planets end up closer than 0.1 AU. Considering the limitations of our model, this result is actually quite encouraging, as future models that can also include more physics (e.g. tides).

\subsection{Heavy element content of giant planets}

The large heavy element content of giant planets presents a puzzle for planet formation theories, where a Jupiter mass planet on average should harbor around 60 Earth masses of heavy elements \citep{2016ApJ...831...64T}. This result is based on the match between observed planetary masses and their radii with interior models. With pure core accretion this is hard to achieve, because the pebble isolation mass in the inner disc is much lower than that and of the order of 10-20 Earth masses \citep{2018arXiv180102341B}.

In a recent work by \citet{2020arXiv200612500G}, it was proposed that collisions between the giant planets could increase the heavy element fraction of the surviving planet. In their model, the planets need to merge multiple times in order to achieve the large heavy element contents predicted by the observations. However, \citet{2020arXiv200612500G} did not model the growth, migration or scattering of these giant planets directly. 

In our simulations we observe an average of 2.5 collisions per system in the case of slow damping, which matches the eccentricity distribution of the giant planets best. In about $35\%$ of our systems do planet experience more than one merger event and only about $30\%$ of those experience more than two merger events\footnote{In our model, we consider perfect mergers, which is why the heavy element content of the planets does not increase substantially due to collisions.}. However, in most of the cases these events happen for super-Earths planets with small cores. In addition, not all the planets that underwent merger events survive the dynamical instabilities in the system. This implies that collisions could only account for a very tiny fraction of planets with large heavy element content and we thus deem that the scenario of \citet{2020arXiv200612500G} is probably not the main reason for the heavy element content of the giant planets.

Another mechanism that could explain the heavy element content of giant planets is the evaporation of drifting pebbles, which locally enriches the gas by their volatile contributions released into the case \citep{2017MNRAS.469.3994B}. This could account for large heavy element contents in the giant planet atmospheres, but needs to be tested within a framework of multi planet formation.

\subsection{Implications for protoplanetary disc observations}

Recent observations of protoplanetary discs with ALMA have revealed an amazing level of substructures in these discs \citep{2018ApJ...869L..41A}, where basically all of these large discs have rings and gaps in the mm emissions. There are many ideas what could cause these rings and gaps, for instance, ice lines \citep{2015ApJ...806L...7Z}, MHD effects \citep{2015A&A...574A..68F}, but also massive planets that generate pressure perturbations in the protoplanetary discs where pebbles accumulate \citep{2006A&A...453.1129P, 2012A&A...545A..81P, 2012A&A...538A.114P, 2018ApJ...869L..47Z}. 

Recent observations of the gas velocity dispersions in protoplanetary discs have revealed that some discs could indeed harbour giant planets \citep{2018ApJ...860L..12T, 2018ApJ...860L..13P}, which cause rings and gaps. However, the real cause of these rings and gaps is still under debate.

In our simulations, the growing planets generate pressure perturbations, where dust could accumulate and form the rings we observe with ALMA. However, in the simulations with slow damping (low $K$) and $S_{\rm peb}$=5.0, the planets already interact gravitationally and scatter during the gas disc phase. This would destroy the nearly axisymmetric rings and gaps observed by ALMA. In particular multiple gas giants on eccentric orbits can result in a very chaotic gas distribution \citep{2013MNRAS.431.3494L}, which might not be in line with the observations of clean rings and gaps in protoplanetary discs. On the other hand, the structures generated by multiple giant planets are in line with the observation of the PDS70 system \citep{2018A&A...617A..44K, 2019ApJ...884L..41B}.

These recent ALMA observations now give another constraint on planet formation theories. Not only do the models have to match the exoplanet observations, but they should also be in line with the observations of their birth phase, namely their natal protoplanetary discs. We think that this avenue needs to be explored in much more detail in the future to constrain planet formation models.

\section{Summary}
\label{sec:summary}

In this work we have combined an N-body framework with pebble and gas accretion as well as planet migration in an evolving protoplanetary discs following our previous works \citep{2019A&A...623A..88B}. We investigated the influence of different gas damping rates of eccentricity and inclination of giant planets as well as the influence of a different number of initial planetary embryos to study the eccentricity distribution of giant planets. In addition, our simulations formed self consistently systems with inner super-Earths and outer gas giants, which was not included in past simulations aimed to study the eccentricity distribution of giant planets \citep{2008ApJ...686..603J, 2009ApJ...699L..88R, 2017A&A...598A..70S}. We show some of the systems formed in our simulations in appendix~\ref{ap:structure}.

Our simulations show that fast damping (large $K$) results in planetary systems that do not undergo large scattering events during the gas disc phase and afterwards. The resulting eccentricities are too low to be close to the observations of giant planets. This result seems to be independent of how fast the planets actually grow in the disc.

Slow damping (low $K$) of eccentricity and inclination, on the other hand, results in scattering events already during the gas disc phase. In the end, these simulations reproduce the eccentricity distribution of the giant exoplanets, where $K$ needs to be between 5 and 50. These results are independent of how fast the planets grow in the disc as well, implying that the damping rates are more important in determining the fate of the planetary system. Future hydrodynamical simulations with multiple planets are needed to determine conclusively how damping of eccentricity and inclination evolves for systems of multiple giant planets.

In addition, we find only a small dependency on the number of giant planets formed in our simulation on the initial number of planetary embryos. If many embryos are implanted within the disc, the gravitational interactions are already initially quite strong, scattering the majority of objects and thus preventing their growth. In addition, our simulations suggest that the giant planets on average should not be single, but should come in multiples.

The simulations that match the eccentricity distribution of the giant planets best, undergo large scattering events. As a consequence of these scattering events, the inner super-Earth systems are destroyed and only a very small fraction of the systems formed in our simulations harbour inner super-Earths and outer gas giants. \citet{2020arXiv200705563S} studied the super-Earth - cold Jupiter relation as well, but used instead 300m planetesimals for the growth of their planetary embryos, not in agreement with the evidence in the solar system \citep{2005Icar..175..111B, 2009Icar..204..558M, 2019Sci...363..955S, 2019Sci...364.9771S}. However, they recover a similar result regarding the eccentricity of outer gas giants and surviving inner super-Earths.

On the first look this seems to contradict the observations using the Kepler data \citep{2018arXiv180502660Z, 2018arXiv180608799B}, while RV observations seem to agree with our simulation results \citep{2018arXiv180408329B}. However, these studies did not take the eccentricity of the giant planets into account. Our simulations predict that systems with lower mass giant planets on nearly circular orbits are more likely to harbour systems of inner super Earths. On the other hand, systems harboring massive giant planets on eccentric orbits should not harbor any inner super-Earths. More detailed statistics of the super-Earth - cold Jupiter relation including the eccentricity distribution of the giant planets could verify if our planet formation approach is correct and constrain future theories.

\begin{acknowledgements}

B.B., thanks the European Research Council (ERC Starting Grant 757448-PAMDORA) for their financial support. Andre Izidoro acknowledges NASA grant 80NSSC18K0828 to Rajdeep Dasgupta, during preparation and submission of the work. A. I. also thanks finnancial support from FAPESP via grants 16/12686-2 and 16/19556-7, and CNPq via grant  313998/2018-3. We thank the referee John Chambers, whose comments helped a lot to improve the manuscript.

\end{acknowledgements}

\appendix

\section{Evolution of individual planetary systems}
\label{ap:indi}

We show here additional evolution of systems with $K$=500 and 30 initial planetary embryos as well as simulations with 15 or 60 initial planetary embryos and $K$=50. All the here presented simulations feature $S_{\rm peb}=2.5$.

\begin{figure*}
 \centering
 \includegraphics[scale=0.7]{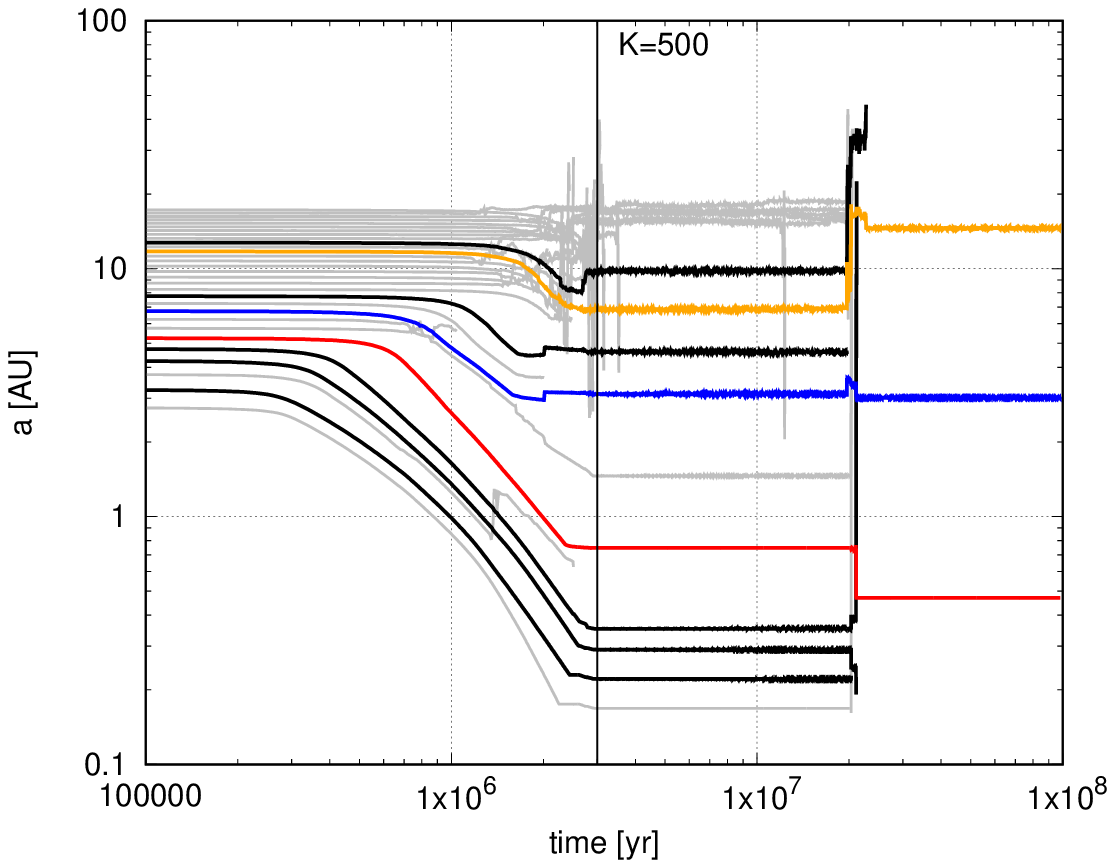}
 \includegraphics[scale=0.7]{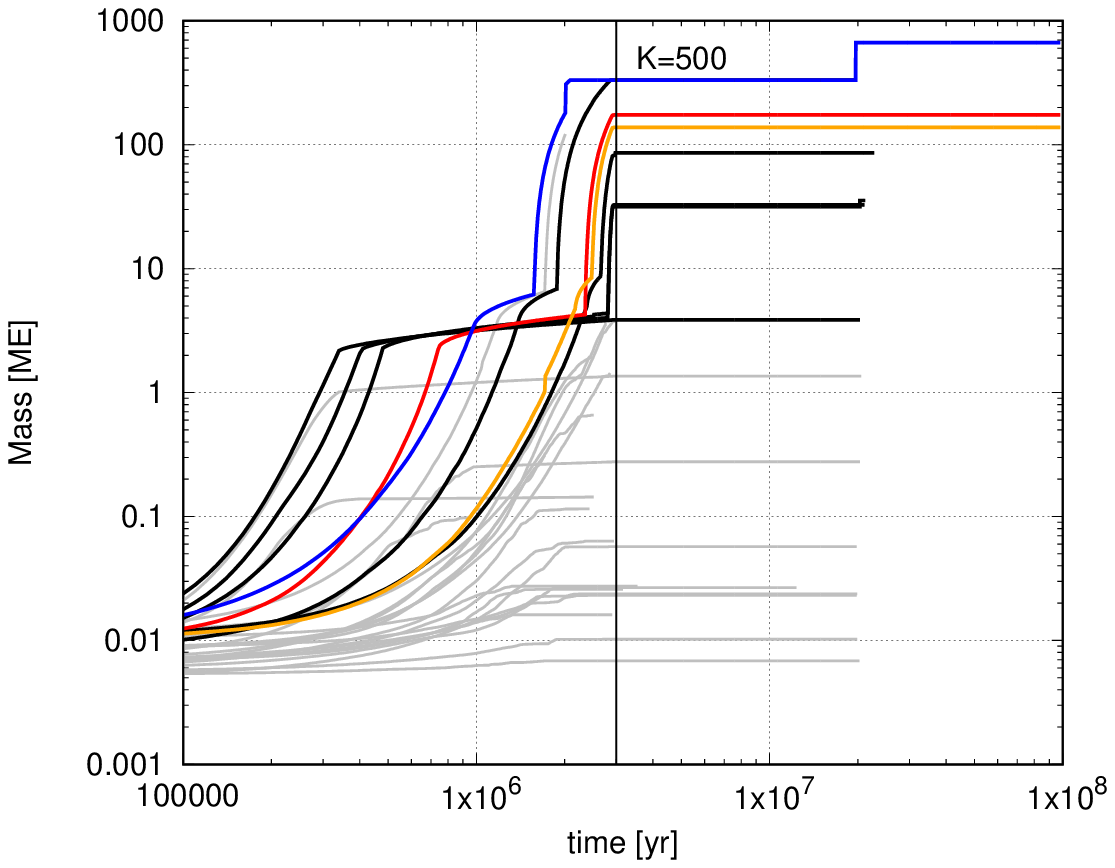}
 \includegraphics[scale=0.7]{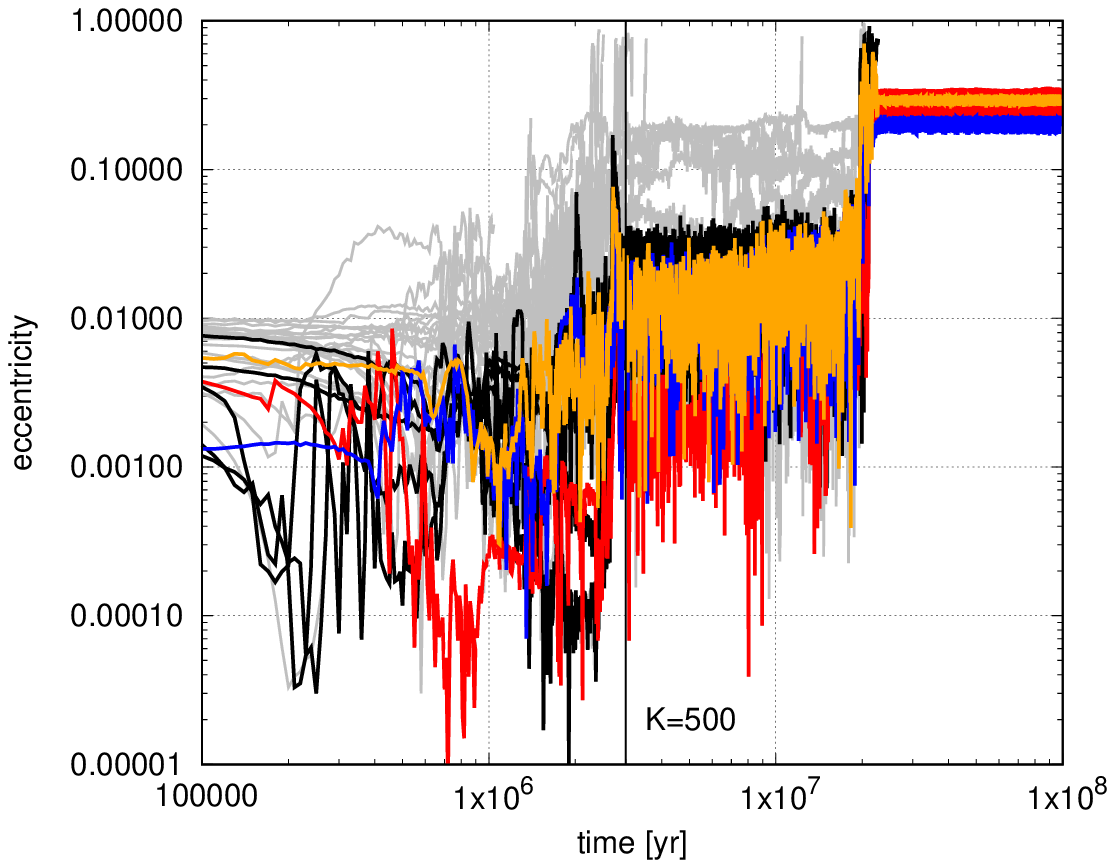}
 \includegraphics[scale=0.7]{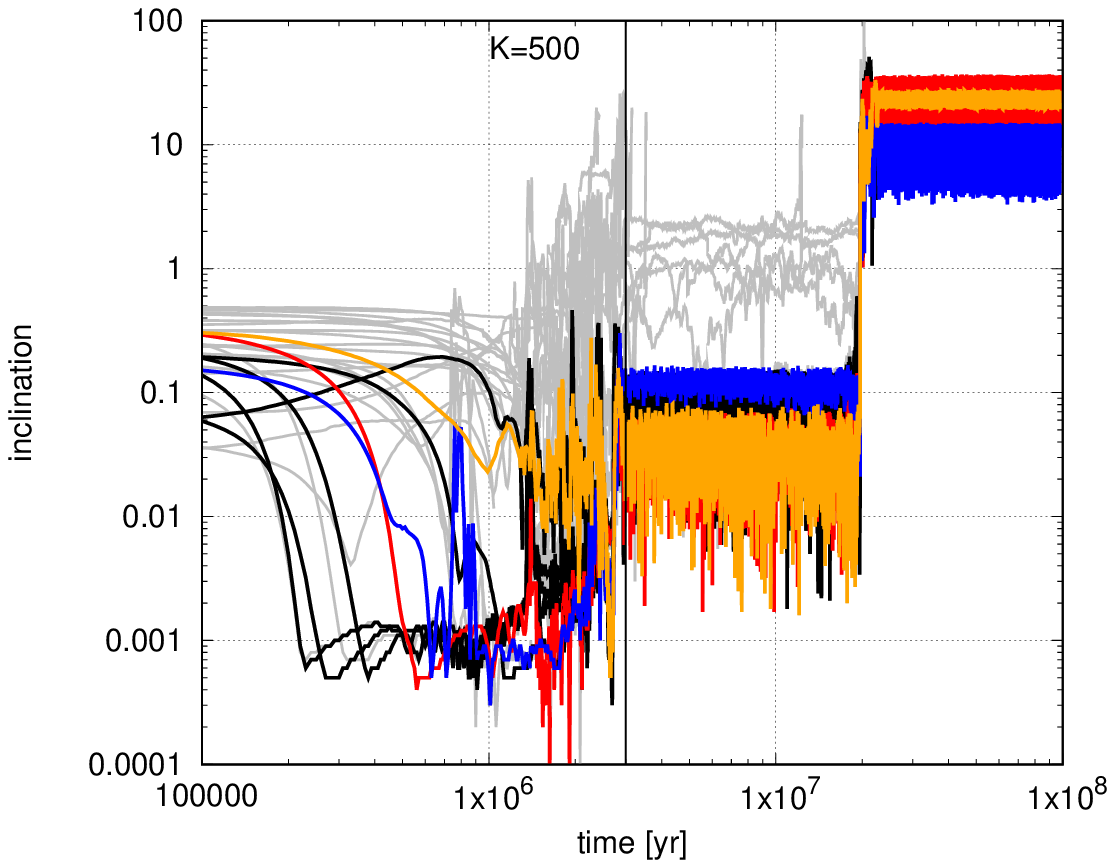}   
 \caption{Evolution of a system with a damping factor of $K=$500. The plots and lines have the same meaning as in Fig.~\ref{fig:30bodyK50}. In the end three gas giants survive with a significant eccentricity, originating from a scattering event about 15 Myr after gas disc dispersal.
   \label{fig:30bodyK500}
   }
\end{figure*}

We show the evolution of a planetary system using $K=500$ in Fig.~\ref{fig:30bodyK500}. The evolution of the planetary system is very similar as in the $K=50$ case during the early stages of the gas disc lifetime. However, due to the larger $K$ value, the damping is still efficient at the end of the gas disc lifetime, resulting in a multi planetary system with inner super-Earths and outer gas giants emerging from the gas disc. The eccentricities of these planets is of the order of a few percent.

But at about 20 Myr, the planetary system undergoes an instability event, including a collision between two planets, resulting in a planet of about 2 Jupiter masses. The smaller planets of the system have been ejected and only three gas giants on eccentric orbits (with an average eccentricity of 0.2-0.3) survive.

\begin{figure*}
 \centering
 \includegraphics[scale=0.7]{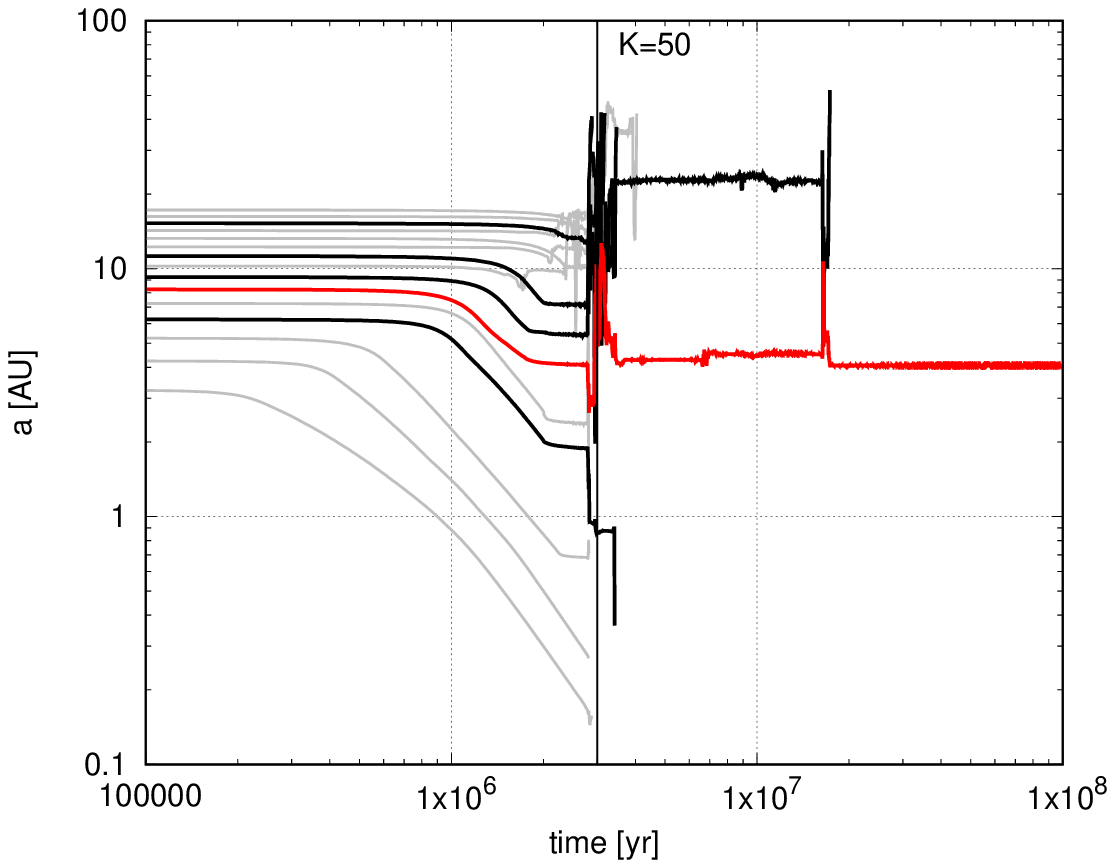}
 \includegraphics[scale=0.7]{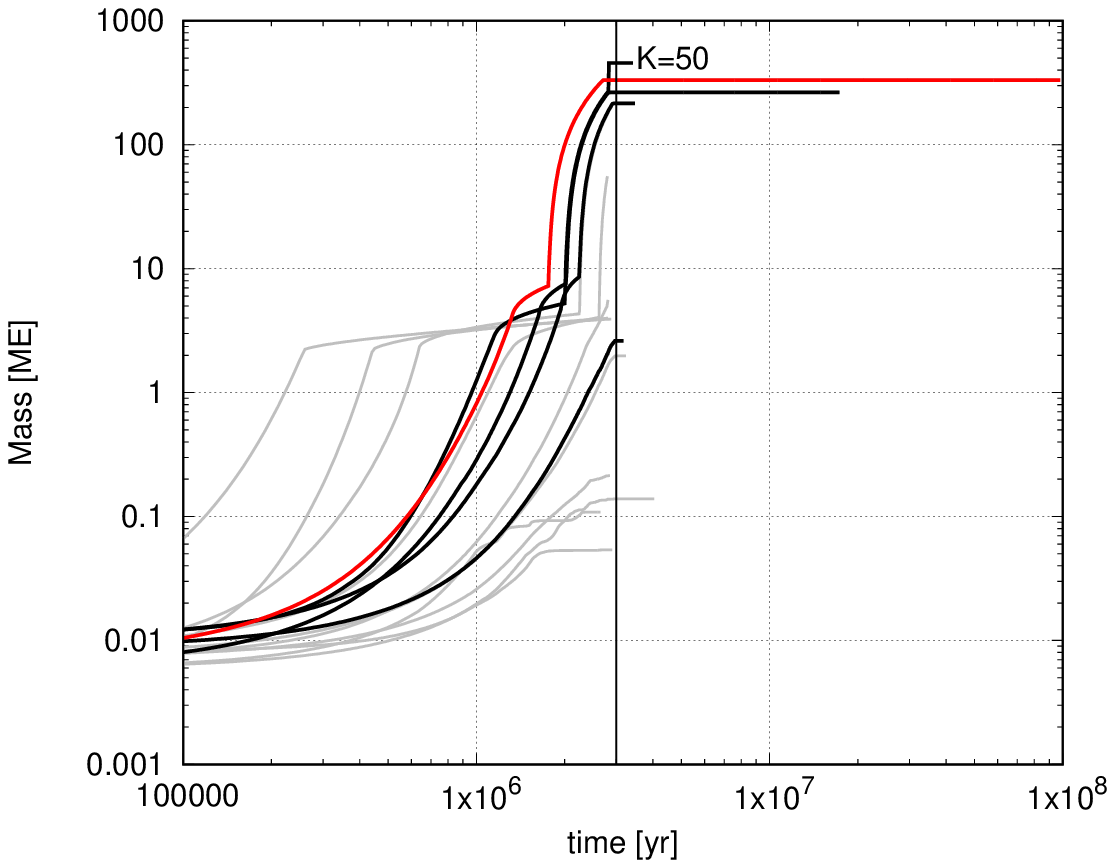}
 \includegraphics[scale=0.7]{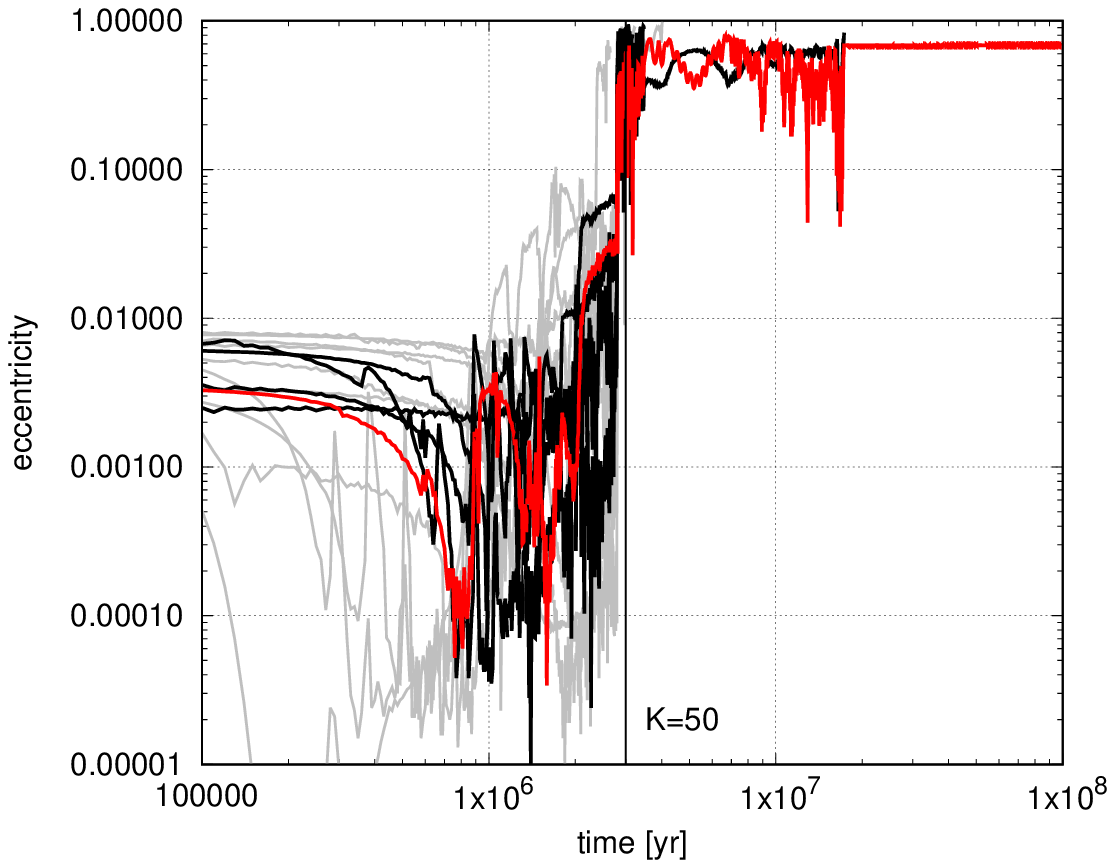}
 \includegraphics[scale=0.7]{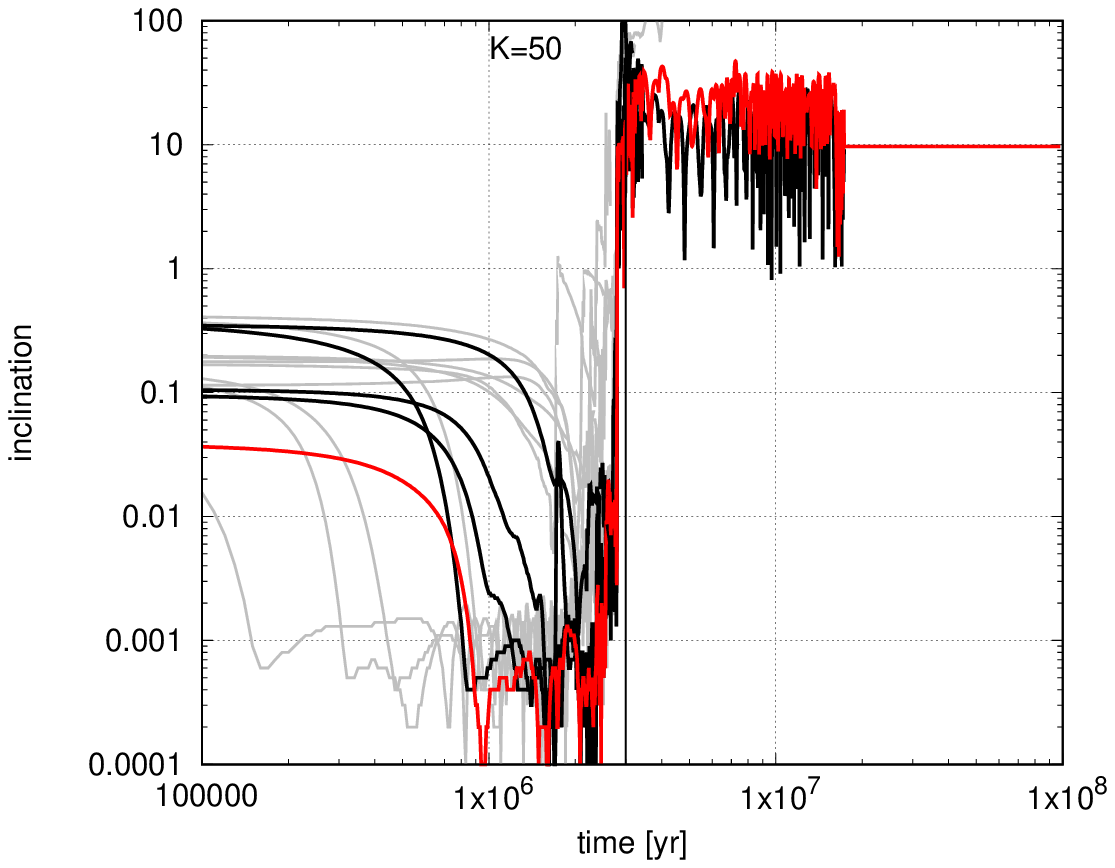}   
 \caption{Evolution of a system with a damping factor of K=50 and 15 initial embryos. The plots and lines have the same meaning as in Fig.~\ref{fig:30bodyK50}. After the scattering event shortly about 15 Myr after the gas disc phase, only one gas giant on a highly eccentric orbit survives.
   \label{fig:15bodyK50}
   }
\end{figure*}

In Fig.~\ref{fig:15bodyK50} we show a simulation using $K$=50, but in contrast to Fig.~\ref{fig:30bodyK50} with only 15 initial planetary embryos. The overall evolution during the gas disc phase is again very similar to the simulation with 30 embryos, except that even the small planetary embryos start to grow to at least $\approx$0.1 Earth masses. This is caused by the initially larger separation between the bodies, which prevents initially planet-planet interactions and thus keeps the eccentricities and inclinations low during this phase, allowing all planetary embryos to accrete some pebbles. However, as a few dominating bodies emerge, the eccentricities and inclinations start to increase and the small bodies stop growing.

Similar to the simulation with 30 bodies (Fig.~\ref{fig:30bodyK50}), the damping of the gas disc reduces towards the end of the disc lifetime and the planetary systems undergoes an instability event, which removes the smaller bodies from the disc, where only two gas giants on highly eccentric orbits remain. However, these two gas giants scatter again after 20 Myr and only one gas giant on a highly eccentric orbit survives.

\begin{figure*}
 \centering
 \includegraphics[scale=0.7]{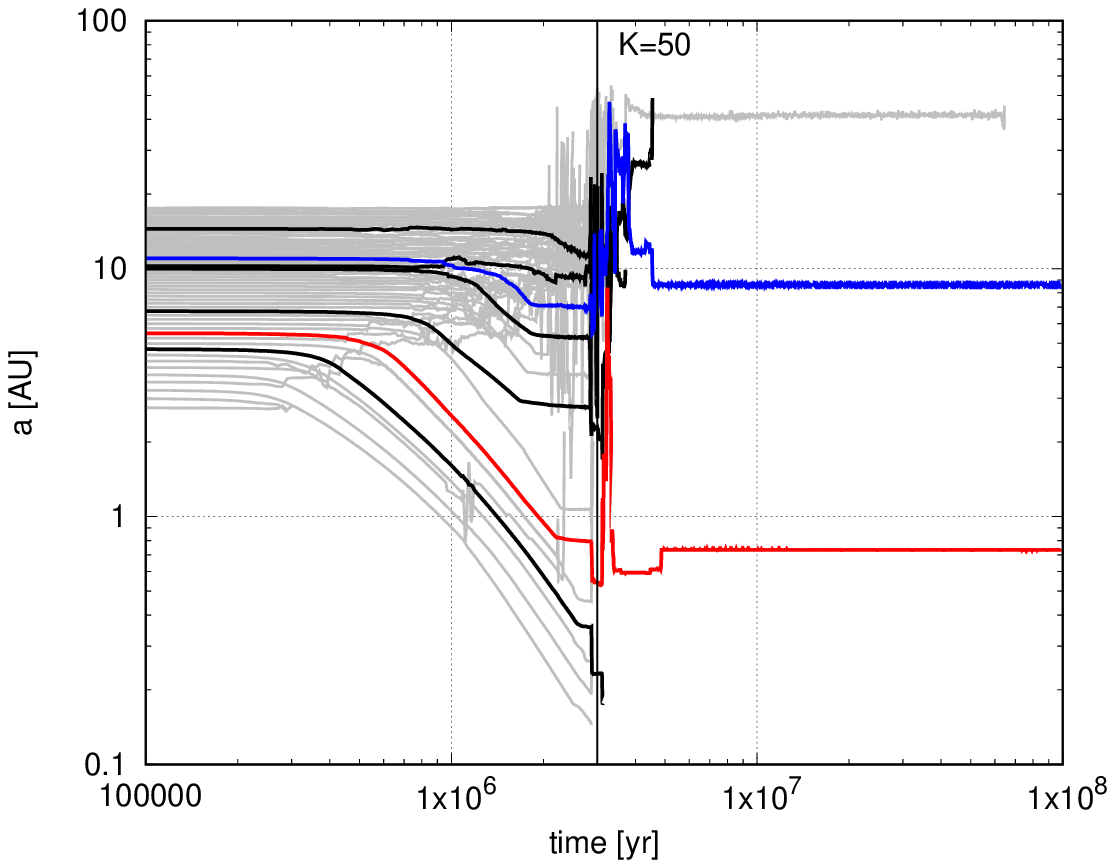}
 \includegraphics[scale=0.7]{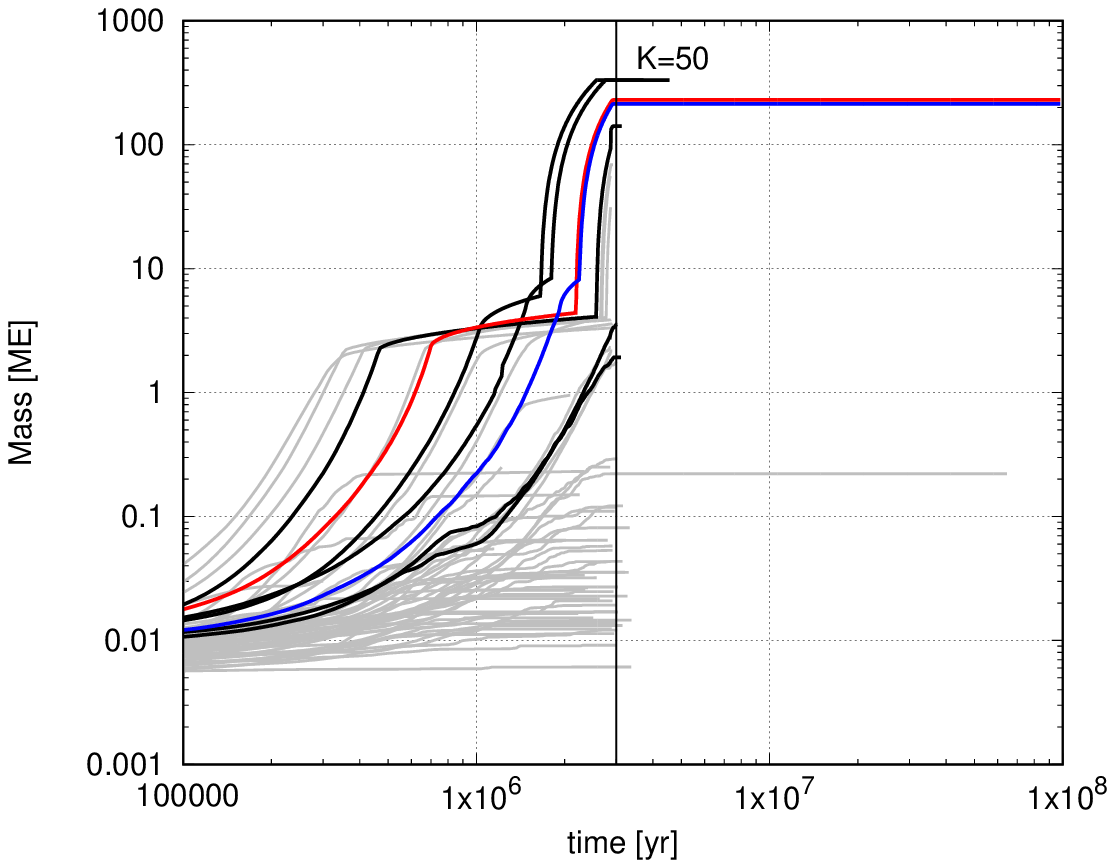}
 \includegraphics[scale=0.7]{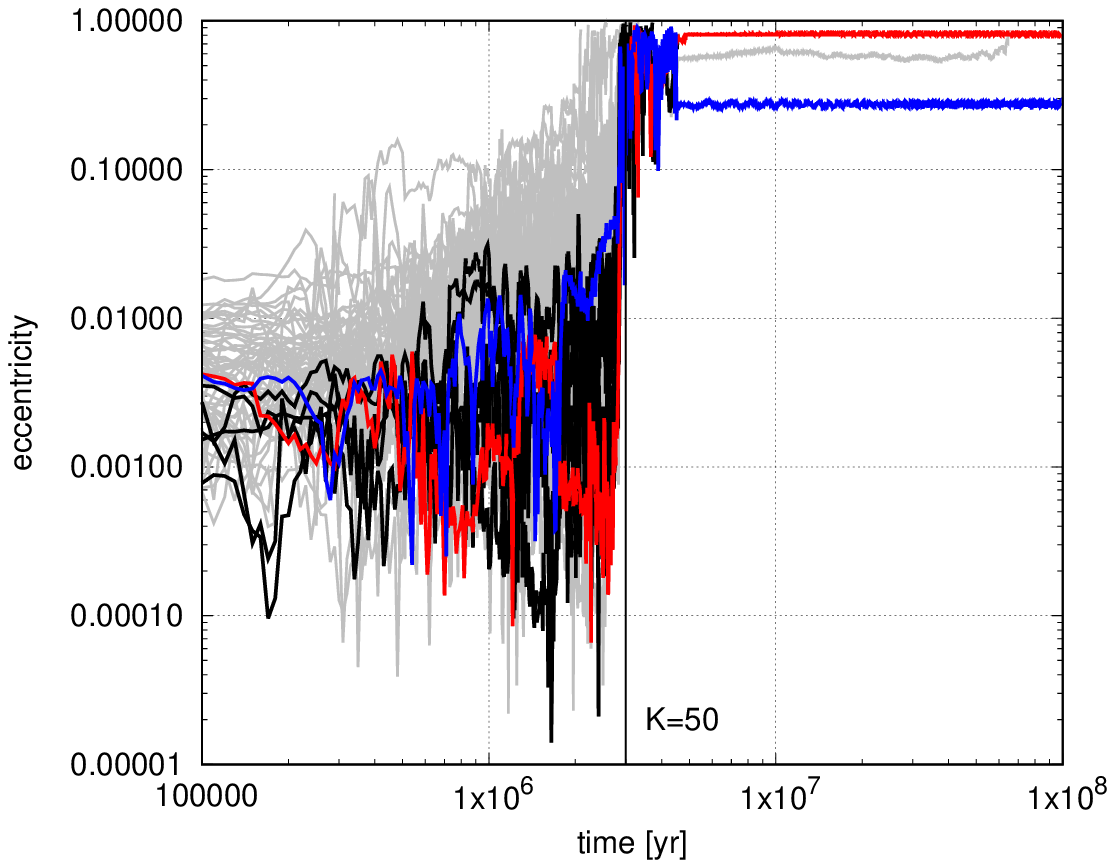}
 \includegraphics[scale=0.7]{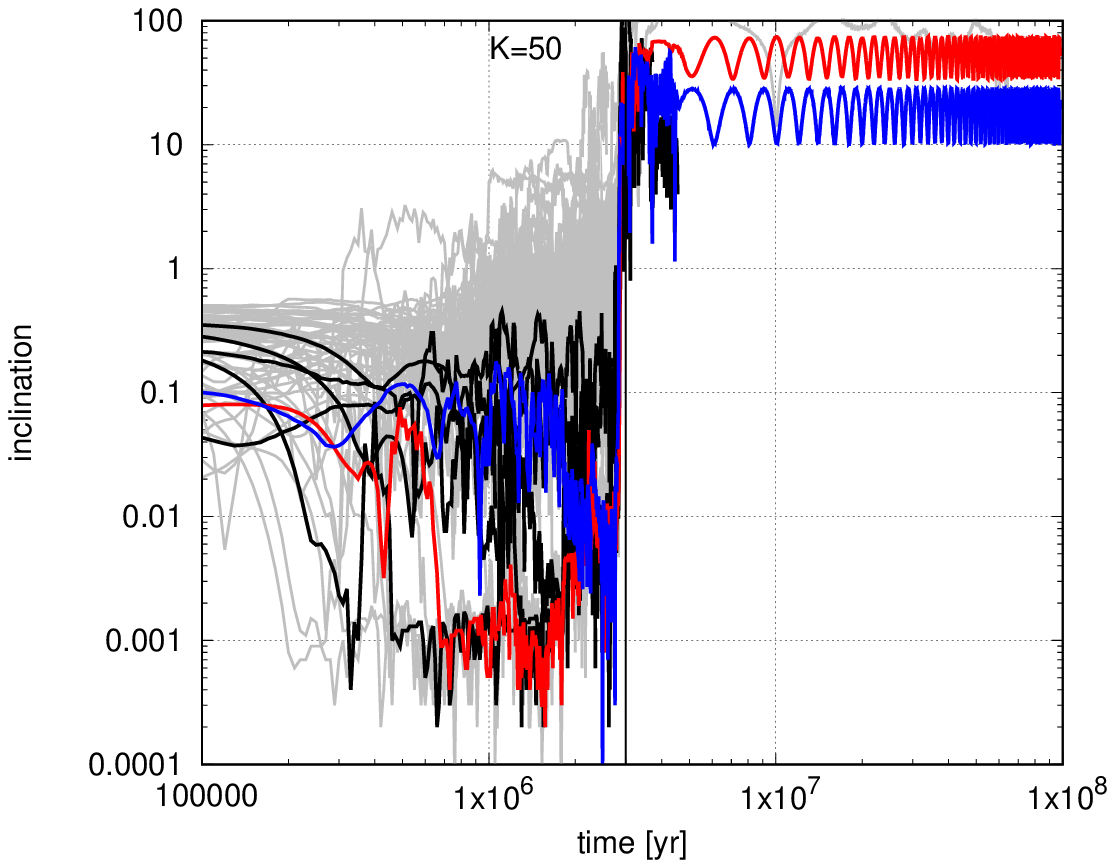}   
 \caption{Evolution of a system with a damping factor of $K$=50 and 60 initial embryos. The plots and lines have the same meaning as in Fig.~\ref{fig:30bodyK50}. After the scattering event shortly after the end of the gas disc phase, two gas giants on eccentric orbits survive.
   \label{fig:60bodyK50}
   }
\end{figure*}

Finally, we show in Fig.~\ref{fig:60bodyK50} a simulation with initially 60 planetary embryos and again with $K$=50. The initially large number of planetary embryos increases the interactions between them compared to the previous simulations. This results in an increase of eccentricity and inclination of the small bodies as soon as the first bodies start to reach masses above Earth mass. As a consequence, as before, only a few dominating bodies emerge. As in the previous simulation, towards the end of the disc lifetime the system becomes dynamically unstable and only two gas giants on highly eccentric orbits survive.

Even though the simulations start with a different number of initial planetary embryos, the resulting planetary systems are very similar for simulations with the same damping values. The reason for that is that only a few dominating bodies emerge in each simulation, similar to the N-body simulations with pebble accretion from \citet{2015Natur.524..322L}, who, however, did not include planet migration. If many bodies are initially present, many of them acquire an eccentricity of a few percent that prevents the planets to accrete pebbles efficiently and thus stopping their growth \citep{Johansen2015}. As a result only a few bodies start to grow efficiently. On the other hand, if only initially a small number of embryos is present, then obviously only a few bodies can grow. This is not only evident in the here presented simulations (Fig.~\ref{fig:15bodyK50}, Fig.~\ref{fig:30bodyK50} and Fig.~\ref{fig:60bodyK50}), but is true for the full sets of our simulations. 

However, if $K$ is large and thus damping of eccentricity and inclination is efficient, the initially larger number of planetary embryos can lead to a larger number of surviving planets, because the planets remain on nearly circular orbits during the gas disc phase (Fig.~\ref{fig:30bodyK500} and Fig.~\ref{fig:30bodyK5000}), increasing the stability of the systems also after the gas disc phase. In this case, accretion is very efficient and systems with many planets can emerge.

\section{Structures of planetary systems}
\label{ap:structure}

\begin{figure}
 \centering
 \includegraphics[scale=0.7]{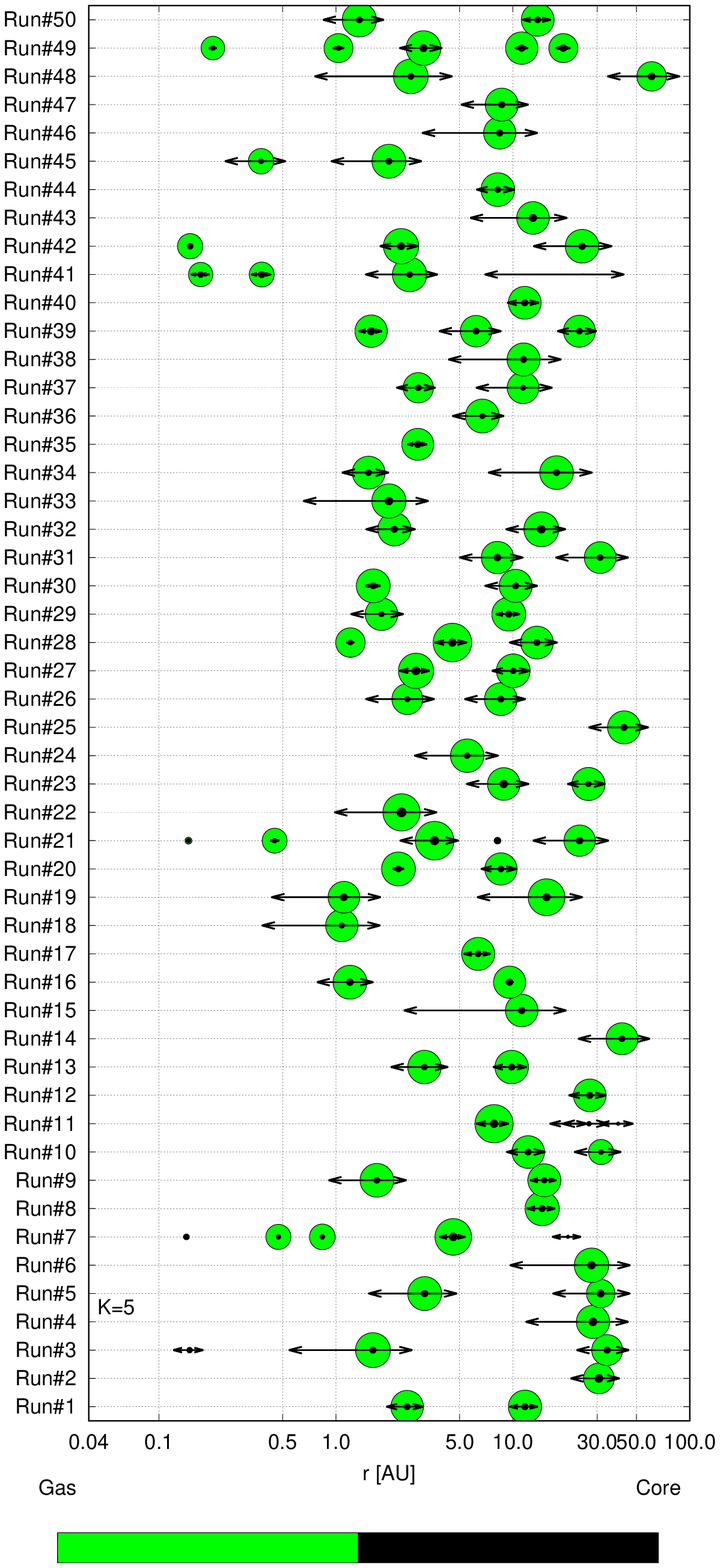}
 \caption{Final configurations after 100 Myr of integration of all our simulations with $S_{\rm peb}=5.0$ and $K$=5. The size of the circle is proportional to the total planetary mass (green) by the 3rd root and to the mass of the planetary core (black) also by the 3rd root. The black arrows indicate the aphelion and perihelion positions of the planet calculated through $r_{\rm P} \pm e\times r_{\rm P}$.
   \label{fig:strucK5}
   }
\end{figure}

\begin{figure}
 \centering
 \includegraphics[scale=0.7]{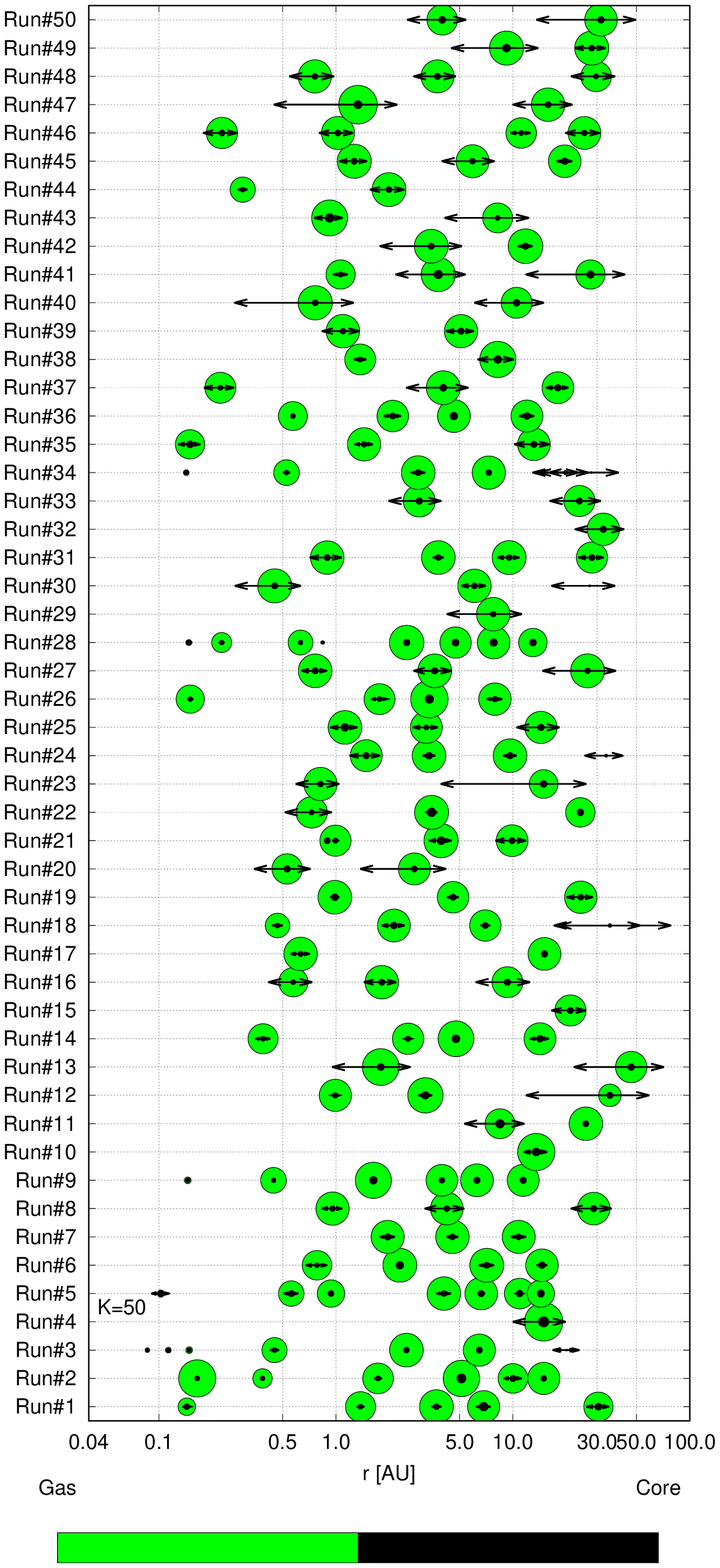}
 \caption{Same as Fig.~\ref{fig:strucK5}, except that $K$=50 was used.
   \label{fig:strucK50}
   }
\end{figure}

We show here the structure of planetary systems formed in discs with $K$=5 and $K$=50 in the case of 30 initial embryos and $S_{\rm peb}$=5.0 at 100 Myr in Fig.~\ref{fig:strucK5} and Fig.~\ref{fig:strucK50}. The size of the circle is proportional to the total planetary mass (green) by the 3rd root and to the mass of the planetary core (black) also by the 3rd root. The black arrows indicate the aphelion and perihelion positions of the planet calculated through $r_{\rm P} \pm e\times r_{\rm P}$.

Clearly, a larger $K$ value, which implies more efficient damping of eccentricity and inclination during the gas disc phase, results in planetary systems with more giant planets. In addition, it is also clear that the eccentricities of the giant planets are larger for $K$=5 compared to $K$=50.

\bibliographystyle{aa}
\bibliography{Stellar}

\begin{thebibliography}{149}
\expandafter\ifx\csname natexlab\endcsname\relax\def\natexlab#1{#1}\fi

\bibitem[{{Agnew} {et~al.}(2018){Agnew}, {Maddison}, \&
  {Horner}}]{2018MNRAS.477.3646A}
{Agnew}, M.~T., {Maddison}, S.~T., \& {Horner}, J. 2018, \mnras, 477, 3646

\bibitem[{{Alexander} \& {Pascucci}(2012)}]{2012MNRAS.422L..82A}
{Alexander}, R.~D. \& {Pascucci}, I. 2012, MNRAS, 442, pp.82

\bibitem[{{Andrews} {et~al.}(2018){Andrews}, {Huang}, {P{\'e}rez}, {Isella},
  {Dullemond}, {Kurtovic}, {Guzm{\'a}n}, {Carpenter}, {Wilner}, {Zhang}, {Zhu},
  {Birnstiel}, {Bai}, {Benisty}, {Hughes}, {{\"O}berg}, \&
  {Ricci}}]{2018ApJ...869L..41A}
{Andrews}, S.~M., {Huang}, J., {P{\'e}rez}, L.~M., {et~al.} 2018, \apjl, 869,
  L41

\bibitem[{{Andrews} {et~al.}(2013){Andrews}, {Rosenfeld}, {Kraus}, \&
  {Wilner}}]{2013ApJ...771..129A}
{Andrews}, S.~M., {Rosenfeld}, K.~A., {Kraus}, A.~L., \& {Wilner}, D.~J. 2013,
  \apj, 771, 129

\bibitem[{{Anglada-Escud{\'e}} {et~al.}(2010){Anglada-Escud{\'e}},
  {L{\'o}pez-Morales}, \& {Chambers}}]{2010ApJ...709..168A}
{Anglada-Escud{\'e}}, G., {L{\'o}pez-Morales}, M., \& {Chambers}, J.~E. 2010,
  \apj, 709, 168

\bibitem[{{Armitage} {et~al.}(2016){Armitage}, {Eisner}, \&
  {Simon}}]{2016ApJ...828L...2A}
{Armitage}, P.~J., {Eisner}, J.~A., \& {Simon}, J.~B. 2016, \apjl, 828, L2

\bibitem[{{Ataiee} {et~al.}(2018){Ataiee}, {Baruteau}, {Alibert}, \&
  {Benz}}]{2018A&A...615A.110A}
{Ataiee}, S., {Baruteau}, C., {Alibert}, Y., \& {Benz}, W. 2018, \aap, 615,
  A110

\bibitem[{{Ayliffe} \& {Bate}(2009)}]{2009MNRAS.393...49A}
{Ayliffe}, B.~A. \& {Bate}, M.~R. 2009, MNRAS, 393, 49

\bibitem[{{Bae} {et~al.}(2019){Bae}, {Zhu}, {Baruteau}, {Benisty}, {Dullemond},
  {Facchini}, {Isella}, {Keppler}, {P{\'e}rez}, \&
  {Teague}}]{2019ApJ...884L..41B}
{Bae}, J., {Zhu}, Z., {Baruteau}, C., {et~al.} 2019, \apjl, 884, L41

\bibitem[{Bai(2016)}]{2016ApJ...821...80B}
Bai, X.~N. 2016, ApJ, 821, id.80

\bibitem[{{Bai} \& {Stone}(2010)}]{2010ApJ...722L.220B}
{Bai}, X.~N. \& {Stone}, J.~M. 2010, ApJ, 722, L220

\bibitem[{{Barbato} {et~al.}(2018){Barbato}, {Sozzetti}, {Desidera}, {Damasso},
  {Bonomo}, {Giacobbe}, {Colombo}, {Lazzoni}, {Claudi}, {Gratton}, {LoCurto},
  {Marzari}, \& {Mordasini}}]{2018arXiv180408329B}
{Barbato}, D., {Sozzetti}, A., {Desidera}, S., {et~al.} 2018, A\&A, 615,
  id.A175

\bibitem[{{Baruteau} {et~al.}(2014){Baruteau}, {Crida}, {Paardekooper},
  {Masset}, {Guilet}, {Bitsch}, Nelson, {Kley}, \&
  {Papaloizou}}]{2013arXiv1312.4293B}
{Baruteau}, C., {Crida}, A., {Paardekooper}, S.~J., {et~al.} 2014, in
  {Protostars and Planets VI}, arXiv:1312.4293

\bibitem[{{Baruteau} \& {Masset}(2008)}]{2008ApJ...672.1054B}
{Baruteau}, C. \& {Masset}, F. 2008, \apj, 672, 1054

\bibitem[{{Baumann} \& {Bitsch}(2020)}]{2020arXiv200400874B}
{Baumann}, T. \& {Bitsch}, B. 2020, A\&A, 637, id. A11

\bibitem[{Ben\'{\i}tez-Llambay {et~al.}(2015)Ben\'{\i}tez-Llambay, Masset,
  Koenigsberger, \& Szul{\'a}gyi}]{2015Natur.520...63B}
Ben\'{\i}tez-Llambay, P., Masset, F., Koenigsberger, G., \& Szul{\'a}gyi, J.
  2015, Nature, 520, pp. 63

\bibitem[{Birnstiel {et~al.}(2012)Birnstiel, Klahr, \&
  Ercolano}]{2012A&A...539A.148B}
Birnstiel, T., Klahr, H., \& Ercolano, B. 2012, A\&A, 539, id.A148

\bibitem[{{Bitsch}(2019)}]{2019A&A...630A..51B}
{Bitsch}, B. 2019, \aap, 630, A51

\bibitem[{Bitsch {et~al.}(2013)Bitsch, Crida, Libert, \&
  Lega}]{2013A&A...555A.124B}
Bitsch, B., Crida, A., Libert, A.-S., \& Lega, E. 2013, A\&A, 555, id.A124

\bibitem[{{Bitsch} {et~al.}(2019){Bitsch}, {Izidoro}, {Johansen}, {Raymond},
  {Morbidelli}, {Lambrechts}, \& {Jacobson}}]{2019A&A...623A..88B}
{Bitsch}, B., {Izidoro}, A., {Johansen}, A., {et~al.} 2019, \aap, 623, A88

\bibitem[{Bitsch {et~al.}(2015{\natexlab{a}})Bitsch, Johansen, Lambrechts, \&
  Morbidelli}]{2015A&A...575A..28B}
Bitsch, B., Johansen, A., Lambrechts, M., \& Morbidelli, A. 2015{\natexlab{a}},
  A\&A, 575, id.A28

\bibitem[{Bitsch {et~al.}(2015{\natexlab{b}})Bitsch, Lambrechts, \&
  Johansen}]{2015A&A...582A.112B}
Bitsch, B., Lambrechts, M., \& Johansen, A. 2015{\natexlab{b}}, A\&A, 582,
  id.A112

\bibitem[{{Bitsch} {et~al.}(2018{\natexlab{a}}){Bitsch}, {Lambrechts}, \&
  {Johansen}}]{2018A&A...609C...2B}
{Bitsch}, B., {Lambrechts}, M., \& {Johansen}, A. 2018{\natexlab{a}}, \aap,
  609, C2

\bibitem[{{Bitsch} {et~al.}(2018{\natexlab{b}}){Bitsch}, {Morbidelli},
  {Johansen}, {Lega}, {Lambrechts}, \& {Crida}}]{2018arXiv180102341B}
{Bitsch}, B., {Morbidelli}, A., {Johansen}, A., {et~al.} 2018{\natexlab{b}},
  A\&A, 612, id.A30

\bibitem[{{Boisvert} {et~al.}(2018){Boisvert}, {Nelson}, \&
  {Steffen}}]{2018MNRAS.480.2846B}
{Boisvert}, J.~H., {Nelson}, B.~E., \& {Steffen}, J.~H. 2018, \mnras, 480, 2846

\bibitem[{Booth {et~al.}(2017)Booth, Clarke, Madhusudhan, \&
  Ilee}]{2017MNRAS.469.3994B}
Booth, R.~A., Clarke, C.~J., Madhusudhan, N., \& Ilee, J.~D. 2017, MNRAS, 469,
  p.3994

\bibitem[{{Bottke} {et~al.}(2005){Bottke}, {Durda}, {Nesvorn{\'y}}, {Jedicke},
  {Morbidelli}, {Vokrouhlick{\'y}}, \& {Levison}}]{2005Icar..175..111B}
{Bottke}, W.~F., {Durda}, D.~D., {Nesvorn{\'y}}, D., {et~al.} 2005, \icarus,
  175, 111

\bibitem[{Brauer {et~al.}(2008)Brauer, Dullemond, \&
  Henning}]{2008A&A...480..859B}
Brauer, F., Dullemond, C., \& Henning, T. 2008, A\&A, 480, pp.859

\bibitem[{{Bryan} {et~al.}(2019){Bryan}, {Knutson}, {Fulton}, {Lee}, {Batygin},
  {Ngo}, \& {Meshkat}}]{2018arXiv180608799B}
{Bryan}, M.~L., {Knutson}, H.~A., {Fulton}, B., {et~al.} 2019, AJ, 157, id. 52

\bibitem[{{Buchhave} {et~al.}(2018){Buchhave}, {Bitsch}, {Johansen}, {Latham},
  {Bizzarro}, {Bieryla}, \& {Kipping}}]{2018arXiv180206794B}
{Buchhave}, L.~A., {Bitsch}, B., {Johansen}, A., {et~al.} 2018, ApJ, 856, id.37

\bibitem[{{Chambers}(2019)}]{2019ApJ...879...98C}
{Chambers}, J. 2019, \apj, 879, 98

\bibitem[{{Chambers} {et~al.}(1996){Chambers}, {Wetherill}, \&
  {Boss}}]{1996Icar..119..261C}
{Chambers}, J.~E., {Wetherill}, G.~W., \& {Boss}, A.~P. 1996, \icarus, 119, 261

\bibitem[{{Chrenko} {et~al.}(2017){Chrenko}, {Bro{\v z}}, \&
  {Lambrechts}}]{2017A&A...606A.114C}
{Chrenko}, O., {Bro{\v z}}, M., \& {Lambrechts}, M. 2017, \aap, 606, A114

\bibitem[{Cossou {et~al.}(2014)Cossou, Raymond, Hersant, \&
  Pierens}]{2014arXiv1407.6011C}
Cossou, C., Raymond, S.~N., Hersant, F., \& Pierens, A. 2014, A\&A, 569, id.
  A56

\bibitem[{Cresswell \& Nelson(2008)}]{2008A&A...482..677C}
Cresswell, P. \& Nelson, R.~P. 2008, A\&A, 482, pp.677

\bibitem[{{Crida} {et~al.}(2006){Crida}, {Morbidelli}, \&
  {Masset}}]{2006Icar..181..587C}
{Crida}, A., {Morbidelli}, A., \& {Masset}, F. 2006, Icarus, 181, 587

\bibitem[{{D'Angelo} \& {Bodenheimer}(2013)}]{2013ApJ...778...77D}
{D'Angelo}, G. \& {Bodenheimer}, P. 2013, \apj, 778, 77

\bibitem[{{Dawson} \& {Johnson}(2018)}]{2018ARA&A..56..175D}
{Dawson}, R.~I. \& {Johnson}, J.~A. 2018, \araa, 56, 175

\bibitem[{{Dawson} \& {Murray-Clay}(2013)}]{2013ApJ...767L..24D}
{Dawson}, R.~I. \& {Murray-Clay}, R.~A. 2013, \apjl, 767, L24

\bibitem[{{Dr{\c a}{\.z}kowska} \& {Alibert}(2017)}]{2017A&A...608A..92D}
{Dr{\c a}{\.z}kowska}, J. \& {Alibert}, Y. 2017, \aap, 608, A92

\bibitem[{{Duffell} \& {MacFadyen}(2013)}]{2013ApJ...769...41D}
{Duffell}, P.~C. \& {MacFadyen}, A.~I. 2013, \apj, 769, 41

\bibitem[{{Dullemond} {et~al.}(2018){Dullemond}, {Birnstiel}, {Huang},
  {Kurtovic}, {Andrews}, {Guzm{\'a}n}, {P{\'e}rez}, {Isella}, {Zhu}, {Benisty},
  {Wilner}, {Bai}, {Carpenter}, {Zhang}, \& {Ricci}}]{2018ApJ...869L..46D}
{Dullemond}, C.~P., {Birnstiel}, T., {Huang}, J., {et~al.} 2018, \apjl, 869,
  L46

\bibitem[{Fischer \& Valenti(2005)}]{2005ApJ...622.1102F}
Fischer, D.~A. \& Valenti, J. 2005, ApJ, 622, pp. 1102

\bibitem[{{Flaherty} {et~al.}(2018){Flaherty}, {Hughes}, {Teague}, {Simon},
  {Andrews}, \& {Wilner}}]{2018ApJ...856..117F}
{Flaherty}, K.~M., {Hughes}, A.~M., {Teague}, R., {et~al.} 2018, \apj, 856, 117

\bibitem[{{Flock} {et~al.}(2015){Flock}, {Ruge}, {Dzyurkevich}, {Henning},
  {Klahr}, \& {Wolf}}]{2015A&A...574A..68F}
{Flock}, M., {Ruge}, J.~P., {Dzyurkevich}, N., {et~al.} 2015, \aap, 574, A68

\bibitem[{{Flock} {et~al.}(2019){Flock}, {Turner}, {Mulders}, {Hasegawa},
  {Nelson}, \& {Bitsch}}]{2019A&A...630A.147F}
{Flock}, M., {Turner}, N.~J., {Mulders}, G.~D., {et~al.} 2019, \aap, 630, A147

\bibitem[{{Ford} {et~al.}(2001){Ford}, {Havlickova}, \&
  {Rasio}}]{2001Icar..150..303F}
{Ford}, E.~B., {Havlickova}, M., \& {Rasio}, F.~A. 2001, Icarus, 150, 303

\bibitem[{{Ford} {et~al.}(2005){Ford}, {Lystad}, \&
  {Rasio}}]{2005Natur.434..873F}
{Ford}, E.~B., {Lystad}, V., \& {Rasio}, F.~A. 2005, \nat, 434, 873

\bibitem[{Fressin {et~al.}(2013)Fressin, Torres, Charbonneau, Bryson,
  Christiansen, Dressing, Jenkins, Walkowicz, \& Batalha}]{2013ApJ...766...81F}
Fressin, F., Torres, G., Charbonneau, D., {et~al.} 2013, ApJ, 766, id.81

\bibitem[{{Fung} {et~al.}(2014){Fung}, {Shi}, \&
  {Chiang}}]{2014ApJ...782...88F}
{Fung}, J., {Shi}, J.-M., \& {Chiang}, E. 2014, \apj, 782, 88

\bibitem[{{Ginzburg} \& {Chiang}(2020)}]{2020arXiv200612500G}
{Ginzburg}, S. \& {Chiang}, E. 2020, arXiv e-prints, arXiv:2006.12500

\bibitem[{{Gladman}(1993)}]{1993Icar..106..247G}
{Gladman}, B. 1993, Icarus, 106, 247

\bibitem[{{Hara} {et~al.}(2019){Hara}, {Bou{\'e}}, {Laskar}, {Delisle}, \&
  {Unger}}]{2019MNRAS.489..738H}
{Hara}, N.~C., {Bou{\'e}}, G., {Laskar}, J., {Delisle}, J.~B., \& {Unger}, N.
  2019, \mnras, 489, 738

\bibitem[{Howard(2013)}]{2013Sci...340..572H}
Howard, A.~W. 2013, Science, 340, pp.572

\bibitem[{{Ida} \& {Guillot}(2016)}]{2016A&A...596L...3I}
{Ida}, S. \& {Guillot}, T. 2016, \aap, 596, L3

\bibitem[{{Ida} {et~al.}(2016){Ida}, {Guillot}, \&
  {Morbidelli}}]{2016A&A...591A..72I}
{Ida}, S., {Guillot}, T., \& {Morbidelli}, A. 2016, \aap, 591, A72

\bibitem[{Ikoma {et~al.}(2000)Ikoma, Nakazawa, \& Emori}]{2000ApJ...537.1013I}
Ikoma, M., Nakazawa, K., \& Emori, H. 2000, ApJ, 537, pp. 1013

\bibitem[{{Izidoro} {et~al.}(2019){Izidoro}, {Bitsch}, {Raymond}, {Johansen},
  {Morbidelli}, {Lambrechts}, \& {Jacobson}}]{2019arXiv190208772I}
{Izidoro}, A., {Bitsch}, B., {Raymond}, S.~N., {et~al.} 2019, arXiv e-prints
  [\eprint[arXiv]{1902.08772}]

\bibitem[{Izidoro {et~al.}(2015)Izidoro, Morbidelli, Raymond, Hersant, \&
  Pierens}]{2015A&A...582A..99I}
Izidoro, A., Morbidelli, A., Raymond, S.~N., Hersant, F., \& Pierens, A. 2015,
  A\&A, 582, id.A99

\bibitem[{Izidoro {et~al.}(2017)Izidoro, Ogihara, Raymond, Morbidelli, Pierens,
  Bitsch, Cossou, \& Hersant}]{2017MNRAS.470.1750I}
Izidoro, A., Ogihara, M., Raymond, S.~N., {et~al.} 2017, MNRAS, 470, pp. 1750

\bibitem[{{Izidoro} {et~al.}(2015){Izidoro}, {Raymond}, {Morbidelli},
  {Hersant}, \& {Pierens}}]{2015ApJ...800L..22I}
{Izidoro}, A., {Raymond}, S.~N., {Morbidelli}, A., {Hersant}, F., \& {Pierens},
  A. 2015, \apjl, 800, L22

\bibitem[{{Jim{\'e}nez} \& {Masset}(2017)}]{2017MNRAS.471.4917J}
{Jim{\'e}nez}, M.~A. \& {Masset}, F.~S. 2017, \mnras, 471, 4917

\bibitem[{{Johansen} \& {Bitsch}(2019)}]{2019A&A...631A..70J}
{Johansen}, A. \& {Bitsch}, B. 2019, \aap, 631, A70

\bibitem[{Johansen \& Lacerda(2010)}]{2010MNRAS.404..475J}
Johansen, A. \& Lacerda, P. 2010, MNRAS, 404, pp. 475

\bibitem[{Johansen \& Lambrechts(2017)}]{Johansen2017}
Johansen, A. \& Lambrechts, M. 2017, AREP, 45

\bibitem[{Johansen {et~al.}(2015)Johansen, {Mac Low}, Lacerda, \&
  Bizzarro}]{Johansen2015}
Johansen, A., {Mac Low}, M.~M., Lacerda, P., \& Bizzarro, M. 2015, Science
  Advances, Vol.1, id. 1500109

\bibitem[{{Johnson} {et~al.}(2010){Johnson}, {Aller}, {Howard}, \&
  {Crepp}}]{J2010}
{Johnson}, J.~A., {Aller}, K.~M., {Howard}, A.~W., \& {Crepp}, J.~R. 2010,
  \pasp, 122, 905

\bibitem[{{Juri{\'c}} \& {Tremaine}(2008)}]{2008ApJ...686..603J}
{Juri{\'c}}, M. \& {Tremaine}, S. 2008, \apj, 686, 603

\bibitem[{{Kanagawa} {et~al.}(2015){Kanagawa}, {Tanaka}, {Muto}, {Tanigawa}, \&
  {Takeuchi}}]{2015MNRAS.448..994K}
{Kanagawa}, K.~D., {Tanaka}, H., {Muto}, T., {Tanigawa}, T., \& {Takeuchi}, T.
  2015, \mnras, 448, 994

\bibitem[{{Kanagawa} {et~al.}(2018){Kanagawa}, {Tanaka}, \&
  {Szuszkiewicz}}]{2018arXiv180511101K}
{Kanagawa}, K.~D., {Tanaka}, H., \& {Szuszkiewicz}, E. 2018, ApJ, 861, id.140

\bibitem[{{Keppler} {et~al.}(2018){Keppler}, {Benisty}, {M{\"u}ller},
  {Henning}, {van Boekel}, {Cantalloube}, {Ginski}, {van Holstein}, {Maire},
  {Pohl}, {Samland }, {Avenhaus}, {Baudino}, {Boccaletti}, {de Boer},
  {Bonnefoy}, {Chauvin}, {Desidera}, {Langlois}, {Lazzoni}, {Marleau},
  {Mordasini}, {Pawellek}, {Stolker}, {Vigan}, {Zurlo}, {Birnstiel},
  {Brandner}, {Feldt}, {Flock}, {Girard}, {Gratton}, {Hagelberg}, {Isella},
  {Janson}, {Juhasz}, {Kemmer}, {Kral}, {Lagrange}, {Launhardt}, {Matter},
  {M{\'e}nard}, {Milli}, {Molli{\`e}re}, {Olofsson}, {P{\'e}rez}, {Pinilla},
  {Pinte}, {Quanz}, {Schmidt}, {Udry}, {Wahhaj}, {Williams}, {Buenzli},
  {Cudel}, {Dominik}, {Galicher}, {Kasper}, {Lannier}, {Mesa}, {Mouillet},
  {Peretti}, {Perrot}, {Salter}, {Sissa}, {Wildi}, {Abe}, {Antichi},
  {Augereau}, {Baruffolo}, {Baudoz}, {Bazzon}, {Beuzit}, {Blanchard}, {Brems},
  {Buey}, {De Caprio}, {Carbillet}, {Carle}, {Cascone}, {Cheetham}, {Claudi},
  {Costille}, {Delboulb{\'e}}, {Dohlen}, {Fantinel}, {Feautrier}, {Fusco},
  {Giro}, {Gluck}, {Gry}, {Hubin}, {Hugot}, {Jaquet}, {Le Mignant}, {Llored},
  {Madec}, {Magnard}, {Martinez}, {Maurel}, {Meyer}, {M{\"o}ller-Nilsson},
  {Moulin}, {Mugnier}, {Orign{\'e}}, {Pavlov}, {Perret}, {Petit}, {Pragt},
  {Puget}, {Rabou}, {Ramos}, {Rigal}, {Rochat}, {Roelfsema}, {Rousset}, {Roux},
  {Salasnich}, {Sauvage}, {Sevin}, {Soenke}, {Stadler}, {Suarez}, {Turatto}, \&
  {Weber}}]{2018A&A...617A..44K}
{Keppler}, M., {Benisty}, M., {M{\"u}ller}, A., {et~al.} 2018, \aap, 617, A44

\bibitem[{{Kimmig} {et~al.}(2020){Kimmig}, {Dullemond}, \&
  {Kley}}]{2020A&A...633A...4K}
{Kimmig}, C.~N., {Dullemond}, C.~P., \& {Kley}, W. 2020, \aap, 633, A4

\bibitem[{{Kley} \& {Dirksen}(2006)}]{2006A&A...447..369K}
{Kley}, W. \& {Dirksen}, G. 2006, \aap, 447, 369

\bibitem[{{Kokaia} {et~al.}(2020){Kokaia}, {Davies}, \&
  {Mustill}}]{2020MNRAS.492..352K}
{Kokaia}, G., {Davies}, M.~B., \& {Mustill}, A.~J. 2020, \mnras, 492, 352

\bibitem[{{K{\"u}rster} {et~al.}(2015){K{\"u}rster}, {Trifonov}, {Reffert},
  {Kostogryz}, \& {Rodler}}]{2015A&A...577A.103K}
{K{\"u}rster}, M., {Trifonov}, T., {Reffert}, S., {Kostogryz}, N.~M., \&
  {Rodler}, F. 2015, \aap, 577, A103

\bibitem[{{Kutra} \& {Wu}(2020)}]{2020arXiv200308431K}
{Kutra}, T. \& {Wu}, Y. 2020, arXiv e-prints, arXiv:2003.08431

\bibitem[{{Lambrechts} \& {Johansen}(2012)}]{2012A&A...544A..32L}
{Lambrechts}, M. \& {Johansen}, A. 2012, A\&A, 544, id.A32

\bibitem[{Lambrechts \& Johansen(2014)}]{2014A&A...572A.107L}
Lambrechts, M. \& Johansen, A. 2014, A\&A, 572, id.A107

\bibitem[{Lambrechts {et~al.}(2014)Lambrechts, Johansen, \&
  Morbidelli}]{2014A&A...572A..35L}
Lambrechts, M., Johansen, A., \& Morbidelli, A. 2014, A\&A, 572, id. A35

\bibitem[{{Lambrechts} \& {Lega}(2017)}]{2017A&A...606A.146L}
{Lambrechts}, M. \& {Lega}, E. 2017, \aap, 606, A146

\bibitem[{{Lambrechts} {et~al.}(2019){Lambrechts}, {Morbidelli}, {Jacobson},
  {Johansen}, {Bitsch}, {Izidoro}, \& {Raymond}}]{2019arXiv190208694L}
{Lambrechts}, M., {Morbidelli}, A., {Jacobson}, S.~A., {et~al.} 2019, A\&A,
  627, id. A83

\bibitem[{{Lee} \& {Peale}(2002)}]{2002ApJ...567..596L}
{Lee}, M.~H. \& {Peale}, S.~J. 2002, \apj, 567, 596

\bibitem[{{Lega} {et~al.}(2014){Lega}, {Crida}, {Bitsch}, \&
  {Morbidelli}}]{2014MNRAS.440..683L}
{Lega}, E., {Crida}, A., {Bitsch}, B., \& {Morbidelli}, A. 2014, MNRAS, 440,
  p.683

\bibitem[{{Lega} {et~al.}(2013){Lega}, {Morbidelli}, \&
  {Nesvorn{\'y}}}]{2013MNRAS.431.3494L}
{Lega}, E., {Morbidelli}, A., \& {Nesvorn{\'y}}, D. 2013, \mnras, 431, 3494

\bibitem[{{Lenz} {et~al.}(2019){Lenz}, {Klahr}, \&
  {Birnstiel}}]{2019ApJ...874...36L}
{Lenz}, C.~T., {Klahr}, H., \& {Birnstiel}, T. 2019, \apj, 874, 36

\bibitem[{Levison {et~al.}(2015)Levison, Kretke, \&
  Duncan}]{2015Natur.524..322L}
Levison, H.~F., Kretke, K., \& Duncan, M.~J. 2015, Nature, 524, pp. 322

\bibitem[{Lubow \& D'Angelo(2006)}]{2006ApJ...641..526L}
Lubow, S.~H. \& D'Angelo, G. 2006, ApJ, 641, pp.526

\bibitem[{{Luque} {et~al.}(2019){Luque}, {Trifonov}, {Reffert}, {Quirrenbach},
  {Lee}, {Albrecht}, {Fredslund Andersen}, {Antoci}, {Grundahl}, {Schwab}, \&
  {Wolthoff}}]{2019A&A...631A.136L}
{Luque}, R., {Trifonov}, T., {Reffert}, S., {et~al.} 2019, \aap, 631, A136

\bibitem[{Machida {et~al.}(2010)Machida, Kokubo, Inutsuka, \&
  Matsumoto}]{2010MNRAS.405.1227M}
Machida, M.~N., Kokubo, E., Inutsuka, S.~I., \& Matsumoto, T. 2010, MNRAS, 405,
  pp. 1227

\bibitem[{Mamajek(2009)}]{2009AIPC.1158....3M}
Mamajek, E.~E. 2009, AIP Conference Proceedings, 1158, pp.3

\bibitem[{{Masset} \& {Snellgrove}(2001)}]{2001MNRAS.320L..55M}
{Masset}, F. \& {Snellgrove}, M. 2001, \mnras, 320, L55+

\bibitem[{{Masset} \& {Papaloizou}(2003)}]{2003ApJ...588..494M}
{Masset}, F.~S. \& {Papaloizou}, J.~C.~B. 2003, \apj, 588, 494

\bibitem[{{Matsumoto} \& {Ogihara}(2020)}]{2020ApJ...893...43M}
{Matsumoto}, Y. \& {Ogihara}, M. 2020, \apj, 893, 43

\bibitem[{{Mayor} {et~al.}(2011){Mayor}, {Marmier}, {Lovis}, {Udry},
  {S{\'e}gransan}, {Pepe}, {Benz}, {Bertaux}, {Bouchy}, {Dumusque}, {Lo Curto},
  {Mordasini}, {Queloz}, \& {Santos}}]{2011arXiv1109.2497M}
{Mayor}, M., {Marmier}, M., {Lovis}, C., {et~al.} 2011, ArXiv e-prints
  [\eprint[arXiv]{1109.2497}]

\bibitem[{Mayor \& Queloz(1995)}]{1995Natur.378..355M}
Mayor, M. \& Queloz, D. 1995, Nature, 378, pp.355

\bibitem[{Morbidelli {et~al.}(2016)Morbidelli, Bitsch, Crida, Gounelle,
  Guillot, Jacobson, Johansen, Lambrechts, \& Lega}]{2016Icar..267..368M}
Morbidelli, A., Bitsch, B., Crida, A., {et~al.} 2016, Icarus, 267, p. 368

\bibitem[{Morbidelli {et~al.}(2009)Morbidelli, Bottke, Nesvorny, \&
  Levison}]{2009Icar..204..558M}
Morbidelli, A., Bottke, W.~F., Nesvorny, D., \& Levison, H.~F. 2009, Icarus,
  204, p.558

\bibitem[{Morbidelli {et~al.}(2015)Morbidelli, Lambrechts, Jacobson, \&
  Bitsch}]{2015Icar..258..418M}
Morbidelli, A., Lambrechts, M., Jacobson, S.~A., \& Bitsch, B. 2015, Icarus,
  258, p. 418

\bibitem[{{Morbidelli} \& {Nesvorny}(2012)}]{2012A&A...546A..18M}
{Morbidelli}, A. \& {Nesvorny}, D. 2012, A\&A, 546, id.A18

\bibitem[{Movshovitz \& Podolak(2008)}]{2008Icar..194..368M}
Movshovitz, N. \& Podolak, M. 2008, Icarus, 194, p.368

\bibitem[{{Mulders} {et~al.}(2018){Mulders}, {Pascucci}, {Apai}, \&
  {Ciesla}}]{2018AJ....156...24M}
{Mulders}, G.~D., {Pascucci}, I., {Apai}, D., \& {Ciesla}, F.~J. 2018, \aj,
  156, 24

\bibitem[{Mustill {et~al.}(2015)Mustill, Davies, \&
  Johansen}]{2015ApJ...808...14M}
Mustill, A.~J., Davies, M.~B., \& Johansen, A. 2015, ApJ, 808, id. 14

\bibitem[{{Narang} {et~al.}(2018){Narang}, {Manoj}, {Furlan}, {Mordasini},
  {Henning}, {Mathew}, {Banyal}, \& {Sivarani}}]{2018AJ....156..221N}
{Narang}, M., {Manoj}, P., {Furlan}, E., {et~al.} 2018, \aj, 156, 221

\bibitem[{{Ndugu} {et~al.}(2018){Ndugu}, {Bitsch}, \&
  {Jurua}}]{2018MNRAS.474..886N}
{Ndugu}, N., {Bitsch}, B., \& {Jurua}, E. 2018, \mnras, 474, 886

\bibitem[{Ormel \& Klahr(2010)}]{2010A&A...520A..43O}
Ormel, C.~W. \& Klahr, H.~H. 2010, A\&A, 520, id.A43

\bibitem[{{Paardekooper} {et~al.}(2011){Paardekooper}, {Baruteau}, \&
  {Kley}}]{2011MNRAS.410..293P}
{Paardekooper}, S.~J., {Baruteau}, C., \& {Kley}, W. 2011, MNRAS, 410, 293

\bibitem[{Paardekooper \& Mellema(2006)}]{2006A&A...453.1129P}
Paardekooper, S.-J. \& Mellema, G. 2006, A\&A, 453, pp.1129

\bibitem[{{Papaloizou} {et~al.}(2001){Papaloizou}, {Nelson}, \&
  {Masset}}]{2001A&A...366..263P}
{Papaloizou}, J.~C.~B., {Nelson}, R.~P., \& {Masset}, F. 2001, \aap, 366, 263

\bibitem[{{Pfeil} \& {Klahr}(2019)}]{2019ApJ...871..150P}
{Pfeil}, T. \& {Klahr}, H. 2019, \apj, 871, 150

\bibitem[{{Pierens} {et~al.}(2013){Pierens}, {Cossou}, \&
  {Raymond}}]{2013A&A...558A.105P}
{Pierens}, A., {Cossou}, C., \& {Raymond}, S.~N. 2013, A\&A, 558, id.A105

\bibitem[{Pierens {et~al.}(2014)Pierens, Raymond, Nesvorny, \&
  Morbidelli}]{2014ApJ...795L..11P}
Pierens, A., Raymond, S.~N., Nesvorny, D., \& Morbidelli, A. 2014, ApJL, 795,
  id. L11

\bibitem[{Pinilla {et~al.}(2012)Pinilla, Benisty, \&
  Birnstiel}]{2012A&A...545A..81P}
Pinilla, P., Benisty, M., \& Birnstiel, T. 2012, A\&A, 545, id. A81

\bibitem[{{Pinilla} {et~al.}(2012){Pinilla}, {Birnstiel}, {Ricci}, {Dullemond},
  {Uribe}, {Testi}, \& {Natta}}]{2012A&A...538A.114P}
{Pinilla}, P., {Birnstiel}, T., {Ricci}, L., {et~al.} 2012, \aap, 538, A114

\bibitem[{{Pinte} {et~al.}(2016){Pinte}, {Dent}, {M{\'e}nard}, {Hales}, {Hill},
  {Cortes}, \& {de Gregorio-Monsalvo}}]{2016ApJ...816...25P}
{Pinte}, C., {Dent}, W.~R.~F., {M{\'e}nard}, F., {et~al.} 2016, \apj, 816, 25

\bibitem[{{Pinte} {et~al.}(2018){Pinte}, {Price}, {M{\'e}nard}, {Duch{\^e}ne},
  {Dent}, {Hill}, {de Gregorio-Monsalvo}, {Hales}, \&
  {Mentiplay}}]{2018ApJ...860L..13P}
{Pinte}, C., {Price}, D.~J., {M{\'e}nard}, F., {et~al.} 2018, \apjl, 860, L13

\bibitem[{Piso \& Youdin(2014)}]{2014ApJ...786...21P}
Piso, A.~M.~A. \& Youdin, A. 2014, ApJ, 786, id. 21

\bibitem[{{Pollack} {et~al.}(1996){Pollack}, {Hubickyj}, {Bodenheimer},
  {Lissauer}, {Podolak}, \& {Greenzweig}}]{1996Icar..124...62P}
{Pollack}, J.~B., {Hubickyj}, O., {Bodenheimer}, P., {et~al.} 1996, Icarus,
  124, 62

\bibitem[{{Raymond} {et~al.}(2009{\natexlab{a}}){Raymond}, {Armitage}, \&
  {Gorelick}}]{2009ApJ...699L..88R}
{Raymond}, S.~N., {Armitage}, P.~J., \& {Gorelick}, N. 2009{\natexlab{a}},
  \apjl, 699, L88

\bibitem[{{Raymond} {et~al.}(2008){Raymond}, {Barnes}, \&
  {Mandell}}]{2008MNRAS.384..663R}
{Raymond}, S.~N., {Barnes}, R., \& {Mandell}, A.~M. 2008, \mnras, 384, 663

\bibitem[{{Raymond} {et~al.}(2009{\natexlab{b}}){Raymond}, {O'Brien},
  {Morbidelli}, \& {Kaib}}]{2009Icar..203..644R}
{Raymond}, S.~N., {O'Brien}, D.~P., {Morbidelli}, A., \& {Kaib}, N.~A.
  2009{\natexlab{b}}, Icarus, 203, p. 644

\bibitem[{{Ros} \& {Johansen}(2013)}]{2013A&A...552A.137R}
{Ros}, K. \& {Johansen}, A. 2013, A\&A, 552, id.A137

\bibitem[{{Ros} {et~al.}(2019){Ros}, {Johansen}, {Riipinen}, \&
  {Schlesinger}}]{2019A&A...629A..65R}
{Ros}, K., {Johansen}, A., {Riipinen}, I., \& {Schlesinger}, D. 2019, \aap,
  629, A65

\bibitem[{Santos {et~al.}(2004)Santos, Israelia, \&
  Mayor}]{2004A&A...415.1153S}
Santos, N.~C., Israelia, G., \& Mayor, M. 2004, A\&A, 415, p.1153

\bibitem[{{Savvidou} {et~al.}(2020){Savvidou}, {Bitsch}, \&
  {Lambrechts}}]{2020arXiv200514097S}
{Savvidou}, S., {Bitsch}, B., \& {Lambrechts}, M. 2020, A\&A, 640, id. A63

\bibitem[{{Schlaufman}(2018)}]{2018ApJ...853...37S}
{Schlaufman}, K.~C. 2018, \apj, 853, 37

\bibitem[{{Schlecker} {et~al.}(2020){Schlecker}, {Mordasini}, {Emsenhuber},
  {Klahr}, {Henning}, {Burn}, {Alibert}, \& {Benz}}]{2020arXiv200705563S}
{Schlecker}, M., {Mordasini}, C., {Emsenhuber}, A., {et~al.} 2020, arXiv
  e-prints, arXiv:2007.05563

\bibitem[{{Schreiber} \& {Klahr}(2018)}]{2018ApJ...861...47S}
{Schreiber}, A. \& {Klahr}, H. 2018, \apj, 861, 47

\bibitem[{{Schulik} {et~al.}(2019){Schulik}, {Johansen}, {Bitsch}, \&
  {Lega}}]{2019A&A...632A.118S}
{Schulik}, M., {Johansen}, A., {Bitsch}, B., \& {Lega}, E. 2019, \aap, 632,
  A118

\bibitem[{Simon {et~al.}(2016)Simon, Armitage, Li, \&
  Youdin}]{2016ApJ...822...55S}
Simon, J., Armitage, P.~J., Li, R., \& Youdin, A. 2016, ApJ, 822, id. 55

\bibitem[{{Singer} {et~al.}(2019){Singer}, {McKinnon}, {Gladman},
  {Greenstreet}, {Bierhaus}, {Stern}, {Parker}, {Robbins}, {Schenk}, {Grundy},
  {Bray}, {Beyer}, {Binzel}, {Weaver}, {Young}, {Spencer}, {Kavelaars},
  {Moore}, {Zangari}, {Olkin}, {Lauer}, {Lisse}, {Ennico}, {New Horizons
  Geology}, Team, {New Horizons Surface Composition Science Theme Team}, \&
  {New Horizons Ralph and LORRI Teams}}]{2019Sci...363..955S}
{Singer}, K.~N., {McKinnon}, W.~B., {Gladman}, B., {et~al.} 2019, Science, 363,
  955

\bibitem[{{Sotiriadis} {et~al.}(2017){Sotiriadis}, {Libert}, {Bitsch}, \&
  {Crida}}]{2017A&A...598A..70S}
{Sotiriadis}, S., {Libert}, A.-S., {Bitsch}, B., \& {Crida}, A. 2017, \aap,
  598, A70

\bibitem[{{Stern} {et~al.}(2019){Stern}, {Weaver}, {Spencer}, {Olkin},
  {Gladstone}, {Grundy}, {Moore}, {Cruikshank}, {Elliott}, {McKinnon},
  {Parker}, {Verbiscer}, {Young}, {Aguilar}, {Albers}, {Andert}, {Andrews},
  {Bagenal}, {Banks}, {Bauer}, {Bauman}, {Bechtold}, {Beddingfield}, {Behrooz},
  {Beisser}, {Benecchi}, {Bernardoni}, {Beyer}, {Bhaskaran}, {Bierson},
  {Binzel}, {Birath}, {Bird}, {Boone}, {Bowman}, {Bray}, {Britt}, {Brown},
  {Buckley}, {Buie}, {Buratti}, {Burke}, {Bushman}, {Carcich}, {Chaikin},
  {Chavez}, {Cheng}, {Colwell}, {Conard}, {Conner}, {Conrad}, {Cook}, {Cooper},
  {Custodio}, {Dalle Ore}, {Deboy}, {Dharmavaram}, {Dhingra}, {Dunn}, {Earle},
  {Egan}, {Eisig}, {El-Maarry}, {Engelbrecht}, {Enke}, {Ercol}, {Fattig},
  {Ferrell}, {Finley}, {Firer}, {Fischetti}, {Folkner}, {Fosbury}, {Fountain},
  {Freeze}, {Gabasova}, {Glaze}, {Green}, {Griffith}, {Guo}, {Hahn}, {Hals},
  {Hamilton}, {Hamilton}, {Hanley}, {Harch}, {Harmon}, {Hart}, {Hayes},
  {Hersman}, {Hill}, {Hill}, {Hofgartner}, {Holdridge}, {Hor{\'a}nyi},
  {Hosadurga}, {Howard}, {Howett}, {Jaskulek}, {Jennings}, {Jensen}, {Jones},
  {Kang}, {Katz}, {Kaufmann}, {Kavelaars}, {Keane}, {Keleher}, {Kinczyk},
  {Kochte}, {Kollmann}, {Krimigis}, {Kruizinga}, {Kusnierkiewicz}, {Lahr},
  {Lauer}, {Lawrence}, {Lee}, {Lessac-Chenen}, {Linscott}, {Lisse}, {Lunsford},
  {Mages}, {Mallder}, {Martin}, {May}, {McComas}, {McNutt}, {Mehoke}, {Mehoke},
  {Nelson}, {Nguyen}, {N{\'u}{\~n}ez}, {Ocampo}, {Owen}, {Oxton}, {Parker},
  {P{\"a}tzold}, {Pelgrift}, {Pelletier}, {Pineau}, {Piquette}, {Porter},
  {Protopapa}, {Quirico}, {Redfern}, {Regiec}, {Reitsema}, {Reuter},
  {Richardson}, {Riedel}, {Ritterbush}, {Robbins}, {Rodgers}, {Rogers}, {Rose},
  {Rosendall}, {Runyon}, {Ryschkewitsch}, {Saina}, {Salinas}, {Schenk},
  {Scherrer}, {Schlei}, {Schmitt}, {Schultz}, {Schurr}, {Scipioni}, {Sepan},
  {Shelton}, {Showalter}, {Simon}, {Singer}, {Stahlheber}, {Stanbridge},
  {Stansberry}, {Steffl}, {Strobel}, {Stothoff}, {Stryk}, {Stuart}, {Summers},
  {Tapley}, {Taylor}, {Taylor}, {Tedford}, {Throop}, {Turner}, {Umurhan}, {Van
  Eck}, {Velez}, {Versteeg}, {Vincent}, {Webbert}, {Weidner}, {Weigle},
  {Wendel}, {White}, {Whittenburg}, {Williams}, {Williams}, {Williams},
  {Winters}, {Zangari}, \& {Zurbuchen}}]{2019Sci...364.9771S}
{Stern}, S.~A., {Weaver}, H.~A., {Spencer}, J.~R., {et~al.} 2019, Science, 364,
  aaw9771

\bibitem[{Suzuki {et~al.}(2016)Suzuki, Ogihara, Morbidelli, Crida, \&
  Guillot}]{2016arXiv160900437S}
Suzuki, T.~K., Ogihara, M., Morbidelli, A., Crida, A., \& Guillot, T. 2016,
  A\&A, 596, id.A74

\bibitem[{{Teague} {et~al.}(2018){Teague}, {Bae}, {Bergin}, {Birnstiel}, \&
  {Foreman-Mackey}}]{2018ApJ...860L..12T}
{Teague}, R., {Bae}, J., {Bergin}, E.~A., {Birnstiel}, T., \& {Foreman-Mackey},
  D. 2018, \apjl, 860, L12

\bibitem[{Thorngren {et~al.}(2016)Thorngren, Fortney, Murray-Clay, \&
  Lopez}]{2016ApJ...831...64T}
Thorngren, D., Fortney, J., Murray-Clay, R.~A., \& Lopez, E. 2016, ApJ, 831,
  id. 64

\bibitem[{{Trifonov} {et~al.}(2014){Trifonov}, {Reffert}, {Tan}, {Lee}, \&
  {Quirrenbach}}]{2014A&A...568A..64T}
{Trifonov}, T., {Reffert}, S., {Tan}, X., {Lee}, M.~H., \& {Quirrenbach}, A.
  2014, \aap, 568, A64

\bibitem[{Tsiganis {et~al.}(2005)Tsiganis, Gomes, Morbidelli, \&
  Levison}]{2005Natur.435..459T}
Tsiganis, K., Gomes, R., Morbidelli, A., \& Levison, H.~F. 2005, Nature, 435,
  pp.459

\bibitem[{{Visser} \& {Ormel}(2016)}]{2016A&A...586A..66V}
{Visser}, R.~G. \& {Ormel}, C.~W. 2016, \aap, 586, A66

\bibitem[{{Voelkel} {et~al.}(2020){Voelkel}, {Klahr}, {Mordasini},
  {Emsenhuber}, \& {Lenz}}]{2020arXiv200403492V}
{Voelkel}, O., {Klahr}, H., {Mordasini}, C., {Emsenhuber}, A., \& {Lenz}, C.
  2020, arXiv e-prints, arXiv:2004.03492

\bibitem[{{Walsh} {et~al.}(2011){Walsh}, {Morbidelli}, {Raymond}, {O'Brien}, \&
  {Mandell}}]{2011Natur.475..206W}
{Walsh}, K.~J., {Morbidelli}, A., {Raymond}, S.~N., {O'Brien}, D.~P., \&
  {Mandell}, A.~M. 2011, Nature, 475, 206

\bibitem[{{Wittenmyer} {et~al.}(2019{\natexlab{a}}){Wittenmyer}, {Bergmann},
  {Horner}, {Clark}, \& {Kane}}]{2019MNRAS.484.4230W}
{Wittenmyer}, R.~A., {Bergmann}, C., {Horner}, J., {Clark}, J., \& {Kane},
  S.~R. 2019{\natexlab{a}}, \mnras, 484, 4230

\bibitem[{{Wittenmyer} {et~al.}(2019{\natexlab{b}}){Wittenmyer}, {Clark},
  {Zhao}, {Horner}, {Wang}, \& {Johns}}]{2019MNRAS.484.5859W}
{Wittenmyer}, R.~A., {Clark}, J.~T., {Zhao}, J., {et~al.} 2019{\natexlab{b}},
  \mnras, 484, 5859

\bibitem[{{Wittenmyer} {et~al.}(2013){Wittenmyer}, {Wang}, {Horner}, {Tinney},
  {Butler}, {Jones}, {O'Toole}, {Bailey}, {Carter}, {Salter}, {Wright}, \&
  {Zhou}}]{2013ApJS..208....2W}
{Wittenmyer}, R.~A., {Wang}, S., {Horner}, J., {et~al.} 2013, \apjs, 208, 2

\bibitem[{{Wu}(2019)}]{2019ApJ...874...91W}
{Wu}, Y. 2019, \apj, 874, 91

\bibitem[{{Yang} {et~al.}(2017){Yang}, {Johansen}, \&
  {Carrera}}]{2017A&A...606A..80Y}
{Yang}, C.-C., {Johansen}, A., \& {Carrera}, D. 2017, \aap, 606, A80

\bibitem[{Youdin \& Lithwick(2007)}]{2007Icar..192..588Y}
Youdin, A. \& Lithwick, Y. 2007, Icarus, 192, p. 588

\bibitem[{{Zhang} {et~al.}(2015){Zhang}, {Blake}, \&
  {Bergin}}]{2015ApJ...806L...7Z}
{Zhang}, K., {Blake}, G.~A., \& {Bergin}, E.~A. 2015, \apjl, 806, L7

\bibitem[{{Zhang} {et~al.}(2018){Zhang}, {Zhu}, {Huang}, {Guzm{\'a}n},
  {Andrews}, {Birnstiel}, {Dullemond}, {Carpenter}, {Isella}, {P{\'e}rez},
  {Benisty}, {Wilner}, {Baruteau}, {Bai}, \& {Ricci}}]{2018ApJ...869L..47Z}
{Zhang}, S., {Zhu}, Z., {Huang}, J., {et~al.} 2018, \apjl, 869, L47

\bibitem[{{Zhu} \& {Wu}(2018)}]{2018arXiv180502660Z}
{Zhu}, W. \& {Wu}, Y. 2018, AJ, 156, id.92

\end{thebibliography}
\end{document}